\documentclass[12pt,preprint]{aastex}

\shorttitle{Heavy Element Abundances in the Bulge}
\shortauthors{Johnson et al.}

\newcommand\iso[2]{$^{\rm #1}$#2}

\begin{document}

\title{Constraints on the Formation of the Galactic Bulge from Na, Al, and 
Heavy Element Abundances in Plaut's Field}

\author{
Christian I. Johnson\altaffilmark{1,5},
R. Michael Rich\altaffilmark{1},
Chiaki Kobayashi\altaffilmark{2}, and
Jon P. Fulbright\altaffilmark{3,4}
}

\altaffiltext{1}{Department of Physics and Astronomy, UCLA, 430 Portola Plaza,
Box 951547, Los Angeles, CA 90095-1547, USA; cijohnson@astro.ucla.edu;
rmr@astro.ucla.edu}

\altaffiltext{2}{Centre for Astrophysics Research, University of Hertfordshire,
Hatfield  AL10 9AB, UK; c.kobayashi@herts.ac.uk}

\altaffiltext{3}{Department of Physics and Astronomy, Johns Hopkins University,
Baltimore, MD 21218, USA; jfulb@skysrv.pha.jhu.edu}

\altaffiltext{4}{Visiting astronomer, Cerro Tololo Inter--American
Observatory, National Optical Astronomy Observatory, which are operated by the
Association of Universities for Research in Astronomy, under contract with the
National Science Foundation.}

\altaffiltext{5}{National Science Foundation Astronomy and Astrophysics
Postdoctoral Fellow}

\begin{abstract}

We report chemical abundances of Na, Al, Zr, La, Nd, and Eu for 39 red 
giant branch (RGB) stars and 23 potential inner disk red clump stars located 
in Plaut's low extinction window.  We also measure lithium for a super Li--rich
RGB star.  The abundances were determined by spectrum synthesis of high 
resolution (R$\approx$25,000), high signal--to--noise (S/N$\sim$50--100
pixel$^{\rm -1}$) spectra obtained with the Blanco 4m telescope and Hydra 
multifiber spectrograph.  For the bulge RGB stars, we find a general increase 
in the [Na/Fe] and [Na/Al] ratios with increasing metallicity, and a similar 
decrease in [La/Fe] and [Nd/Fe].  Additionally, the [Al/Fe] and [Eu/Fe] 
abundance trends almost identically follow those of the $\alpha$--elements, and
the [Zr/Fe] ratios exhibit relatively little change with [Fe/H].  
The consistently low [La/Eu] ratios of the RGB stars indicate that at least a
majority of bulge stars formed rapidly ($\la$1 Gyr) and before the main 
s--process could become a significant pollution source.  In contrast, we find 
that the potential inner disk clump stars exhibit abundance patterns more 
similar to those of the thin and thick disks.  Comparisons between the 
abundance trends at different bulge locations suggest that the inner and outer 
bulge formed on similar timescales.  However, we find evidence of some 
abundance differences between the most metal--poor and metal--rich stars in 
various bulge fields.  The data also indicate that the halo may have had a more
significant impact on the outer bulge initial composition than the inner
bulge composition.  The [Na/Fe] and to a lesser extent [La/Fe] abundances 
further indicate that the metal--poor bulge, at least at $\sim$1 kpc from the 
Galactic center, and thick disk may not share an identical chemistry.

\end{abstract}


\section{INTRODUCTION}

It has long been known that stars in the Galactic bulge exhibit an interesting
chemical composition.  Bulge stars tend to be relatively metal--rich 
([Fe/H]\footnote{We use of the standard spectroscopic notation where 
[A/B]$\equiv$log(N$_{\rm A}$/N$_{\rm B}$)$_{\rm star}$--log(N$_{\rm A}$/N$_{\rm B}$)$_{\sun}$ and log $\epsilon$(A)$\equiv$log(N$_{\rm A}$/N$_{\rm H}$)+12.0 
for elements A and B.}$\ga$--0.6) like the thin and thick disk, but exhibit
larger [$\alpha$/Fe] ratios than similar metallicity thin disk stars (e.g., 
McWilliam \& Rich 1994; Fulbright et al. 2007; Lecureur et al. 2007; Melendez 
et al. 2008; Alves--Brito et al. 2010).  The first paper in this series 
(Johnson et al. 2011) confirmed that the [$\alpha$/Fe] enhancements extend out
to at least the Plaut field (b=--8$\degr$) we consider here, and that this
field follows the declining metallicity gradient with increasing Galactic 
latitude found by Zoccali et al. (2008).  Gonzalez et al. (2011) also find
that the uniform [$\alpha$/Fe] enhancements extend from at least b=--4$\degr$ 
to b=--12$\degr$ along the minor--axis.  The pervasive trend of 
$\alpha$--enhancement in the bulge is classically considered to reflect early, 
rapid enrichment due to massive star supernovae (SNe).

Dynamically, the Plaut field stars considered here exhibit cylindrical rotation
that is well--modeled by the N--body bar model of Shen et al. (2010).  This
work suggests that the bulge has its origin in a primordial massive disk, 
and thus to some extent the bar and inner thick disk may share a largely 
similar formation history.  Recent thick disk and bulge abundance analyses 
(e.g., Melendez et al. 2008; Alves--Brito et al. 2010; Bensby et al. 2010a; 
2011) support this scenario and indicate that the thick disk and bulge may in
fact share very similar chemical compositions.  However, we note that the 
bulge may actually be composed of two separate populations,
with one similar in metallicity and composition to the thick disk and the 
other considerably more metal--rich (Babusiaux et al. 2010; Bensby et al. 2010a;
2011; Hill et al. 2011).  Additionally, in Johnson et al. (2011) we found that 
a population of red clump stars, likely belonging to an inner disk population, 
are as $\alpha$--enhanced as the bulge red giant branch (RGB) stars but have 
metallicities and radial velocities consistent with the thin disk.  All of 
these findings raise the question of whether a unique chemical ``tag" for 
bulge stars exists.

Although the $\alpha$--element abundance trends seem to indicate that the thick
disk and bulge share very similar chemical enrichment histories,
the lighter odd--Z and heavier neutron--capture elements offer the possibility 
of refining our inferences concerning the bulge's chemical evolution timescale 
relative to other populations.  Unlike the $\alpha$--elements, which are almost
solely produced in massive stars, light odd--Z elements like Na and Al can
be produced over longer timescales in both massive and intermediate mass 
stars.  Furthermore, their unique abundance signatures in other stellar 
populations, such as globular clusters, may make it possible to derive 
constraints on the bulge's merger history.  In a similar fashion, the elements
heavier than the Fe--peak are mostly produced in either the rapid (r--process) 
or slow (s--process) neutron--capture processes.  While the r--process is 
thought to be active in core--collapse SNe and thus traces rapid chemical 
enrichment, the s--process is mostly active in low and intermediate mass 
thermally pulsing asymptotic giant branch (AGB) stars living $\sim$500 Myr to 
several Gyr (e.g., see recent review by Sneden et al. 2008 and references 
therein).  Since the production of light odd--Z, $\alpha$, and heavy elements 
traces a wide range of timescales, the bulge and thick disk should exhibit 
comparable abundance trends for all elements if they truly experienced similar 
formation timescales and enrichment histories.

Therefore, we have measured Na, Al, Zr, La, Nd, and Eu 
abundances for 39 RGB stars in a single field located near Plaut's low 
extinction window.  We have also analyzed these elements in a similar sample
of 23 inner disk red clump stars identified in Johnson et al. (2011).  We
compare the abundance trends measured here to those available in other bulge 
fields located at different Galactic latitudes.  These comparisons will 
provide insight into whether the inner and outer bulge experienced any
significant formation timescale and/or composition differences.  We then 
compare our results to those available in the literature for the Galactic halo,
thick disk, and thin disk.  This analysis will help us understand whether 
the thick disk and bulge are truly chemically indistinguishable, and also yield
insight into how the various populations may have influenced the bulge's 
chemical enrichment.

\section{OBSERVATIONS AND DATA ANALYSES}

Details of the target selection, observations, data reductions, and previous
abundance determinations are provided in Johnson et al. (2011).  However, 
here we provide a brief summary of the critical observation and analysis 
information necessary for interpreting the results presented in this paper.

We have previously obtained high resolution (R$\approx$25,000), high 
signal--to--noise (S/N$\sim$50--100 pixel$^{\rm -1}$) spectra of a combined 92 
stars in two separate fields located near Plaut's low--extinction window.  All
data were obtained using the Blanco 4m telescope instrumented with the Hydra
multifiber spectrograph at Cerro Tololo Inter--American Observatory.  In 
Johnson et al. (2011) we derived [Fe/H], [Si/Fe], and [Ca/Fe] ratios via
equivalent width (EW) and spectrum synthesis analyses for 61 giants in 
``Field 1" (l=--1$\degr$, b=--8.5$\degr$) and [Fe/H] ratios for an additional 
31 giants in ``Field 2" (l=0$\degr$, b=--8$\degr$).  While we concluded that 
38 of the targets observed in Field 1 and all of the targets observed in 
Field 2 are likely RGB stars belonging to the bulge population, we 
determined that 23 of the stars in Field 1 are likely foreground red clump 
stars belonging to an inner disk population.  Note however that the population
assignments for these 23 stars are not robust, and are based solely 
on tentative identifications from literature sources (e.g., Zoccali et al. 
2003; Vieira et al. 2007; Rangwala et al. 2009).

Although we observed Field 1 in four separate wavelength regions (6000--6250, 
6150--6400, 6500--6800, and 7650--7950 \AA), Field 2 was only observed in 
two wavelength regions (6150--6400 and 6500--6800 \AA).  Given this, and the 
fact that the S/N ratios for the Field 1 spectra were significantly higher 
than for Field 2, we only report the additional abundances for Field 1 in this
paper.  However, there is one star in Field 2 that has been identified as a 
Li--rich giant (2MASS 18183679--3251454) and we have determined abundances of
additional elements, including lithium, for this star.

The model atmosphere parameters for the stars presented here are the same as 
those listed in Johnson et al. (2011; their Table 1).  The effective 
temperatures (T$_{\rm eff}$) for all stars were derived from dereddened 
V--K$_{\rm s}$ colors, and the surface gravities (log g) for the RGB stars were
calculated based on each star's bolometric magnitude, assuming a distance of 
8 kpc.  Since the distances to the foreground red clump stars are not known, 
their surface gravity values were estimated from the Padova stellar 
evolutionary tracks (Girardi et al. 2000).  Model atmosphere metallicities 
were set at the derived [Fe/H] ratio for each star, and microturbulence 
(V$_{\rm t}$) was determined by removing trends in Fe I abundance as a function
of line strength.  The final model atmosphere parameters were then used to 
interpolate within the $\alpha$--rich ODFNEW ATLAS9 grid (Castelli et al. 
1997\footnote{The model atmosphere grid can be downloaded from http://wwwuser.oat.ts.astro.it/castelli/grids.html.}).

\subsection{Spectrum Synthesis Abundance Determinations}

The abundances of all elements analyzed here were determined via spectrum
synthesis using a newly modified parallel version of the 2010 LTE line 
analysis code MOOG\footnote{MOOG can be downloaded at 
http://www.as.utexas.edu/~chris/moog.html.} (Sneden 1973).  The modified code, 
developed for this project, provides the same interface and abundance 
algorithms as the original MOOG \emph{synth} driver, but the calculations for 
different chemical compositions are evenly distributed among individual 
processing cores via the open message passing interface (Open MPI\footnote{Open
MPI is freely available and can be downloaded at http://www.open-mpi.org/.}) 
communication protocol.  The effective result of this parallelization scheme is
an overall calculation speed increase to the end user that scales nearly 
linearly with the number of processing cores being utilized.  As in Johnson et 
al. (2011), all abundance ratios reported here have been calculated relative to 
Arcturus.  A summary of the lines used here and the adopted Arcturus [X/Fe] 
ratios is provided in Table 1.

\subsubsection{Sodium and Aluminum}

The abundances of sodium and aluminum were determined using the 6154/6160 \AA\ 
Na I lines and the 6696/6698 \AA\ Al I lines.  The synthesized regions for
both line sets spanned 6150--6170 \AA\ for Na and 6690--6705 \AA\ for Al.  The
log gf values for all significant atomic and molecular lines in these regions
were first set in Arcturus by forcing the synthetic spectra to match the 
high resolution, high S/N Arcturus atlas\footnote{The Arcturus atlas can be
downloaded at: http://www.noao.edu/archives.html.}.  All Arcturus abundances
were set by adopting the [X/Fe] ratios in Fulbright et al. (2007; their Table 
2).  Elements not analyzed in Fulbright et al. (2007) were set at the values
given in Peterson et al. (1993; their Table 3), and elements not listed in
Peterson et al. (1993) were set at [X/Fe]=0.  Therefore, the [Na/Fe] and 
[Al/Fe] ratios listed in Table 2 are measured relative to the Arcturus
abundances of [Na/Fe]=$+$0.09 and [Al/Fe]=$+$0.38, and the [Fe/H] values for 
each star are the same as those listed in Johnson et al. (2011; their Table 1).

Given the large metallicity, and to a lesser extent T$_{\rm eff}$ and log(g),
range in our sample, it is likely that the Na and Al line profiles in our 
spectra reflect moderate departures from LTE.  However, the subordinate lines
used here typically have NLTE corrections $\la$0.10--0.15 dex (e.g., Gratton et 
al. 1999; Mashonkina et al. 2000; Gehren et al. 2004; Andrievsky et al. 2008; 
Lind et al. 2011).  Unfortunately, there are no ``standard" NLTE corrections
employed in the literature, in part because the various calculations often
utilize different input physics and model atmospheres.  Therefore, the 
[Na/Fe] and [Al/Fe] abundances provided in Table 2 do not include any NLTE
corrections, but we caution the reader that any metallicity and/or temperature
dependent abundance trends seen in our data, especially those at the level 
$\la$0.10 dex, could be affected by strictly assuming LTE\footnote{For 
similar reasons we did not apply NLTE corrections to any other elements analyzed
here except Li.}.  Although we have attempted to cancel out many of the model 
atmosphere dependencies by anchoring our abundance scale to Arcturus, truly 
self--consistent 3D and NLTE calculations will be needed in order to validate 
any subtle abundance trends.

\subsubsection{Zirconium and Neodymium}

The zirconium and neodymium abundances listed in Table 2 are based on an
average of the 6134, 6140, and 6143 \AA\ Zr I lines and the 6740 \AA\ Nd II 
line.  Although both of these elements have multiple, stable isotopes of both
even and odd mass numbers, we are not aware of any existing hyperfine and/or
isotope dependent linelists available in the literature for the transitions 
used here.  However, as discussed below, we have concluded that neglecting 
these issues should not lead to any significant errors in the derived 
abundances for either element.  

For the case of the Zr I lines, the solar system isotopic ratios presented in 
Anders \& Grevesse (1989; their Table 3) indicate that 51.45$\%$ of the total 
Zr abundance should reside as the even isotope \iso{90}{Zr}, which will not be 
subject to hyperfine broadening, and the remaining even isotopes (\iso{92}{Zr},
\iso{94}{Zr}, and \iso{96}{Zr}) make up a combined 37.33$\%$ of the total Zr 
abundance.  The lone stable odd isotope (\iso{91}{Zr}) makes up only 11.22$\%$ 
of the total Zr abundance and likely does not contribute a strong broadening 
effect.  We therefore conclude that because the \iso{90}{Zr}, \iso{92}{Zr},
\iso{94}{Zr}, and \iso{96}{Zr} isotopes dominate the line profile no additional
corrections are required.

A similar argument can be made for the Nd II measurements.  Like zirconium, 
neodymium is made up of multiple long--lived isotopes, and in the solar system
the total neodymium abundance is distributed into \iso{142}{Nd}, \iso{143}{Nd},
\iso{144}{Nd}, \iso{145}{Nd}, \iso{146}{Nd}, \iso{148}{Nd}, and \iso{150}{Nd}
in the proportions 27.13$\%$, 12.18$\%$, 23.80$\%$, 8.30$\%$, 17.19$\%$, 
5.76$\%$, and 5.64$\%$, respectively (Anders \& Grevesse 1989).  However, Aoki
et al. (2001) and Den Hartog et al. (2003) point out that the odd isotopes 
constitute little more than 20$\%$ of the total Nd abundance and therefore 
their hyperfine broadening effects are negligible.  Similarly, the isotopic
broadening for the blue transitions listed in Aoki et al. (2001; their Table 
A3) do not exceed $\sim$0.004 \AA.  Therefore, we assume that the isotope 
broadening in the 6740 \AA\ Nd II line used here is of the same order of 
magnitude and can therefore also be ignored.

Since neither of our adopted ``standard" Arcturus abundance scales (Peterson 
et al. 1993 and Fulbright et al. 2007) report [Zr/Fe] or [Nd/Fe] ratios, we 
employed a slightly different method for determining the log gf
values of the transitions used for these elements.  In both cases we started 
with the log gf values provided by the VALD\footnote{The Vienna Atomic Line 
Database can be accessed at http://vald.astro.univie.ac.at/~vald/php/vald.php.}
compilation (Kupka et al. 2000) and set the oscillator strengths based on an 
inverse solar analysis (i.e., the oscillator strengths were adjusted until the 
synthetic line profiles matched the observed line profiles in the high 
resolution, high S/N solar spectrum provided with the Arcturus atlas).  The 
adopted solar abundances are log $\epsilon$(Zr)=$+$2.60 and 
log $\epsilon$(Nd)=$+$1.45 (Anders \& Grevesse 1989), which are nearly 
identical to the recommended values provided in Asplund et al. (2009; their 
Table 1).  Finally, we used these log gf values to determine the Zr and Nd 
abundances for each line in the Arcturus atlas in order to perform a 
line--by--line differential abundance analysis for the program stars.  The 
average abundance ratios derived for Arcturus were [Zr/Fe]=$+$0.00 and 
[Nd/Fe]=$+$0.05.  

\subsubsection{Lanthanum and Europium}

Unlike the elements listed in $\S$2.1.1--2.1.2, the lines used to derive the
[La/Fe] (6262 \AA) and [Eu/Fe] (6645 \AA) ratios can be
affected significantly by hyperfine structure and/or isotopic broadening.  The 
total lanthanum abundance is almost entirely made up of a single 
stable isotope (\iso{139}{La}) so isotopic broadening is not an issue.  
However, the absorption lines arising from this odd mass number isotope are 
often strongly affected by hyperfine structure broadening.  In order to 
properly account for this effect in the 6262 \AA\ La II line, we used the 
laboratory derived hyperfine linelist provided by Lawler et al. (2001a).  The 
reference Arcturus [La/Fe] abundance was set by taking the laboratory log gf 
values and fitting the 6262 \AA\ La II line profile in the Arcturus atlas.  
This yielded an Arcturus abundance of [La/Fe]=--0.06, assuming the solar 
log $\epsilon$(La)=$+$1.13, as determined in Lawler et al. (2001a).  

Europium is one of the more complex elements to analyze because many of its 
transitions are significantly affected both by isotopic and hyperfine structure
broadening.  Although the 6645 \AA\ Eu II line used here is usually relatively
weak ($\la$50 mA), it is still important to account for these two effects 
because failing to do so can lead to systematic errors exceeding 0.1 dex.  In
order to properly fit the 6645 \AA\ Eu II line we used the linelist provided
by Lawler et al. (2001b) and assumed a solar isotopic mix
(\iso{151}{Eu}=47.8$\%$, \iso{153}{Eu}=52.2$\%$).  Similar to the case of La
mentioned above, we determined the reference Arcturus [Eu/Fe] abundance by 
taking the laboratory log gf values and fitting a synthetic line profile to
the 6645 \AA\ Eu II line in the Arcturus atlas.  This provided an abundance of
[Eu/Fe]=$+$0.29, assuming the solar log $\epsilon$(Eu)=$+$0.52 (Lawler et al. 
2001b).  The final [La/Fe] and [Eu/Fe] ratios for each star are listed in 
Table 2.

\subsubsection{The Li--Rich Giant 2MASS 18183679--3251454}

\begin{figure}
\epsscale{1.00}
\plotone{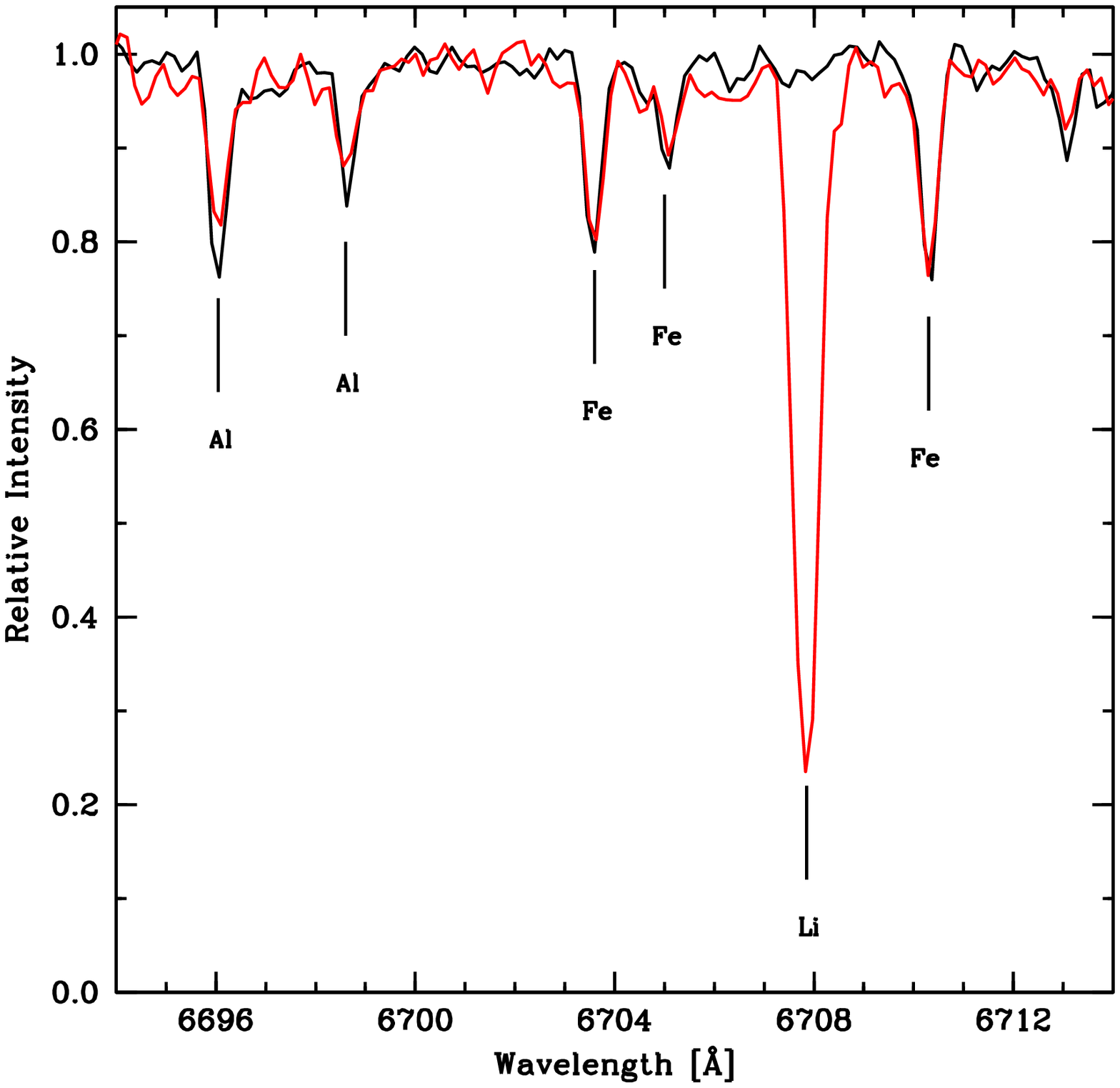}
\caption{This figure illustrates the line strength difference between the 
Li--rich giant 2MASS 18183679--3251454 (solid red line) and a similar 
``Li--normal" giant 2MASS 18181929--3404128 (solid black line).  Both stars 
have nearly identical temperatures, surface gravities, and metallicities, but 
the Li lines are drastically different.  Note also the line strength 
differences in the two Al lines.}
\label{f1}
\end{figure}

Although the 6707 \AA\ Li I resonance line is typically too weak to be 
measured in the spectra of evolved RGB stars, one star in our sample (2MASS
18183679--3251454) exhibited a very strong Li line.  In order to highlight
the unusually strong Li line in this star, which has an EW of 504.3 m\AA, 
Figure \ref{f1} illustrates the stark difference in Li line strengths between 
2MASS 18183679--3251454 and another star in our sample with similar 
T$_{\rm eff}$, log(g), and [Fe/H].  Unlike the elements mentioned above, we 
did not measure the Li abundance relative to Arcturus because the 6707 
\AA\ line strengths between 2MASS 18183679--3251454 and Arcturus are 
significantly different.  Instead, we adopted the linelist of Hobbs et al. 
(1999).  While lines of both \iso{6}{Li} and \iso{7}{Li} are blended in the 
spectra of dwarf stars, the Li present in this star is not primordial and is 
likely dominated by \iso{7}{Li}.  We therefore derived a Li abundance 
from spectrum synthesis assuming that \iso{7}{Li} is the dominant isotope. 
The final derived LTE abundance is log $\epsilon$(Li)=$+$3.56.  Fortunately, 
the NLTE correction is small (Lind et al. 2009), and the NLTE Li abundance 
decreases only slightly to log $\epsilon$(Li)=$+$3.51.

Given the lower S/N and limited spectral coverage for this star, it is 
difficult to determine whether 2MASS 18183679--3251454 exhibits any other 
unusual spectral features.  However, examination of Table 2 (and the subsequent
figures) suggests it may have slightly lower [Al/Fe] and [Eu/Fe] ratios than 
the other Plaut field stars but normal [La/Fe].  Interestingly, the combined 
datasets of McWilliam \& Rich (1994), Gonzalez et al. (2009), Lebzelter et al.
(2012), and the present study have found seven Li--rich giants out of a 
sample of $\sim$850 bulge stars.  This suggests that the fraction of 
Li--rich giants in the bulge is $\sim$1$\%$, and is comparable to what is found
in other stellar populations (e.g., Brown et al. 1989; Kraft et al. 1999; de 
La Reza et al. 1997; Ruchti et al. 2011).

\subsection{Abundance Error Estimates}

In Table 3 we list the estimated log $\epsilon$(X) abundance uncertainties
for each element (except Li) measured in all stars.  We provide individual 
abundance sensitivities to model atmosphere parameter 
uncertainties of $\Delta$T$_{\rm eff}$$+$100 K, $\Delta$log(g)$+$0.3 dex, 
$\Delta$[M/H]$+$0.3 dex, $\Delta$V$_{\rm t}$$+$0.3 km s$^{\rm -1}$, and also 
include the measured line--to--line dispersion for each element in each star.  
Note that for elements and/or stars where only a single line was available to 
measure we have assigned a default line--to--line dispersion of 0.07 dex.  This
value is equal to the average line--to--line dispersion of all instances where 
more than one line was measured.  A final estimate of the total uncertainty 
for each element is also included in Table 3.

Although the average total abundance uncertainties tend to range from 
$\sim$0.15--0.20 dex, it is likely that these represent conservative upper 
limits.  In Johnson et al. (2011) we concluded that values of 
$\Delta$T$_{\rm eff}$$+$50 K and $\Delta$[M/H]$+$0.16 dex might be more 
appropriate.  The T$_{\rm eff}$ uncertainty was estimated by assuming that
the E(B--V) values were accurate to within $\pm$15$\%$, which equaled the 
E(B--V) dispersion across the full Hydra field--of--view.  We also note that
the dispersion in the Alonso et al. (1999) V--K$_{\rm s}$ color--temperature
relation used here is only 25 K.  The [M/H] uncertainty of 0.16 dex was based
on the average log $\epsilon$(Fe) line--to--line dispersion.  However, it 
seems likely that a microturbulence uncertainty of 0.3 km s$^{\rm -1}$, 
especially for more metal--rich stars with stronger lines, is a reasonable 
assumption.  For bulge field stars, the surface gravity is often the most
difficult model parameter to constrain.  While we assumed a fixed distance of 
8 kpc, it is likely that an individual bulge star's distance may range 
anywhere from $\sim$6--10 kpc (e.g., see Zoccali et al. 2008, their Figure 10).
This corresponds to a change in derived log(g) of $\sim$0.2--0.25 dex, which 
makes the $\Delta$log(g) of 0.3 dex used here an appropriate (though possibly 
conservative) choice.  On the other hand, distances to the foreground red 
clump stars are almost completely unconstrained.  However, assuming the stars
actually belong to the red clump means they should have surface gravities 
close to log(g)$\sim$2.3, and therefore their true surface gravities probably
do not exceed the 0.3 dex range estimated in Table 3.

Examination of Table 3 reveals that not all elements are equally affected by
model atmosphere parameter uncertainties.  While changes in T$_{\rm eff}$
noticeably affect the abundances derived from neutral species (Na, Al, and 
especially Zr), the abundances derived from singly ionized species (La, Nd, and
Eu) are affected at $\la$0.03 dex level.  In contrast, changes to the 
model atmosphere surface gravity and metallicity can strongly affect the 
singly ionized species but have a nearly negligible influence on the abundances
derived from neutral species.  Note that the [La/Fe], [Nd/Fe], and [Eu/Fe]
ratios provided in Table 2 are measured relative to Fe I rather than Fe II.
While this is not necessarily a problem as long as the surface gravities are 
accurate, the increased sensitivity to surface gravity means that one should 
use caution when interpreting small abundance differences related to the La, Nd,
and Eu trends.  Microturbulence uncertainties have a very small
affect on most lines in metal--poor stars because the lines are weak and on
the linear portion of the curve--of--growth.  However, Table 3 illustrates the
growing role microturbulence plays with increasing metallicity and line
strength.  In the most metal--rich stars, abundances of Na, Al, and Zr can be
affected at the $>$0.1 dex level while the influence on singly ionized 
species is typically $<$0.05 dex in magnitude.

\section{RESULTS}

\subsection{The Light Elements: Sodium and Aluminum}

In addition to the often used [$\alpha$/Fe] ratio, the two light odd--Z 
elements Na and Al can be exploited as sensitive tracers of a stellar 
population's chemical enrichment history.  Na and Al production can have 
multiple sources tracing timescales ranging from $\la$30 Myr to more than 
several Gyr.  While most of the Na and Al in the Galaxy is produced during the 
hydrostatic helium, carbon, and neon burning stages of massive stars (e.g., 
Arnett \& Thielemann 1985; Thielemann \& Arnett 1985; Woosley \& Weaver 1995), 
these elements can also be created via the NeNa and MgAl proton--capture cycles
active in the envelopes of massive and intermediate mass stars (e.g., Decressin
et al. 2007; Ventura \& D'Antona 2009; Karakas 2010) and near the 
hydrogen--burning shells of some lower mass RGB stars (e.g., Sweigart \& Mengel
1979; Denisenkov \& Denisenkova 1990; D'Antona \& Ventura 2007).  However,
the photospheric [Na/Fe] and [Al/Fe] ratios measured in bulge RGB stars are
likely not significantly altered by \emph{in situ} processing because their
relatively high metallicity ([Fe/H]$\ga$--0.6, on average) does not provide 
favorable conditions for efficient NeNa and/or MgAl cycling and deep \emph{in 
situ} mixing (e.g., D'Antona \& Ventura 2007).  Therefore, it is probably a 
safe assumption that our derived photospheric [Na/Fe] and [Al/Fe] ratios 
reflect the stars' original composition, which resulted purely from external 
pollution.

\begin{figure}
\epsscale{1.00}
\plotone{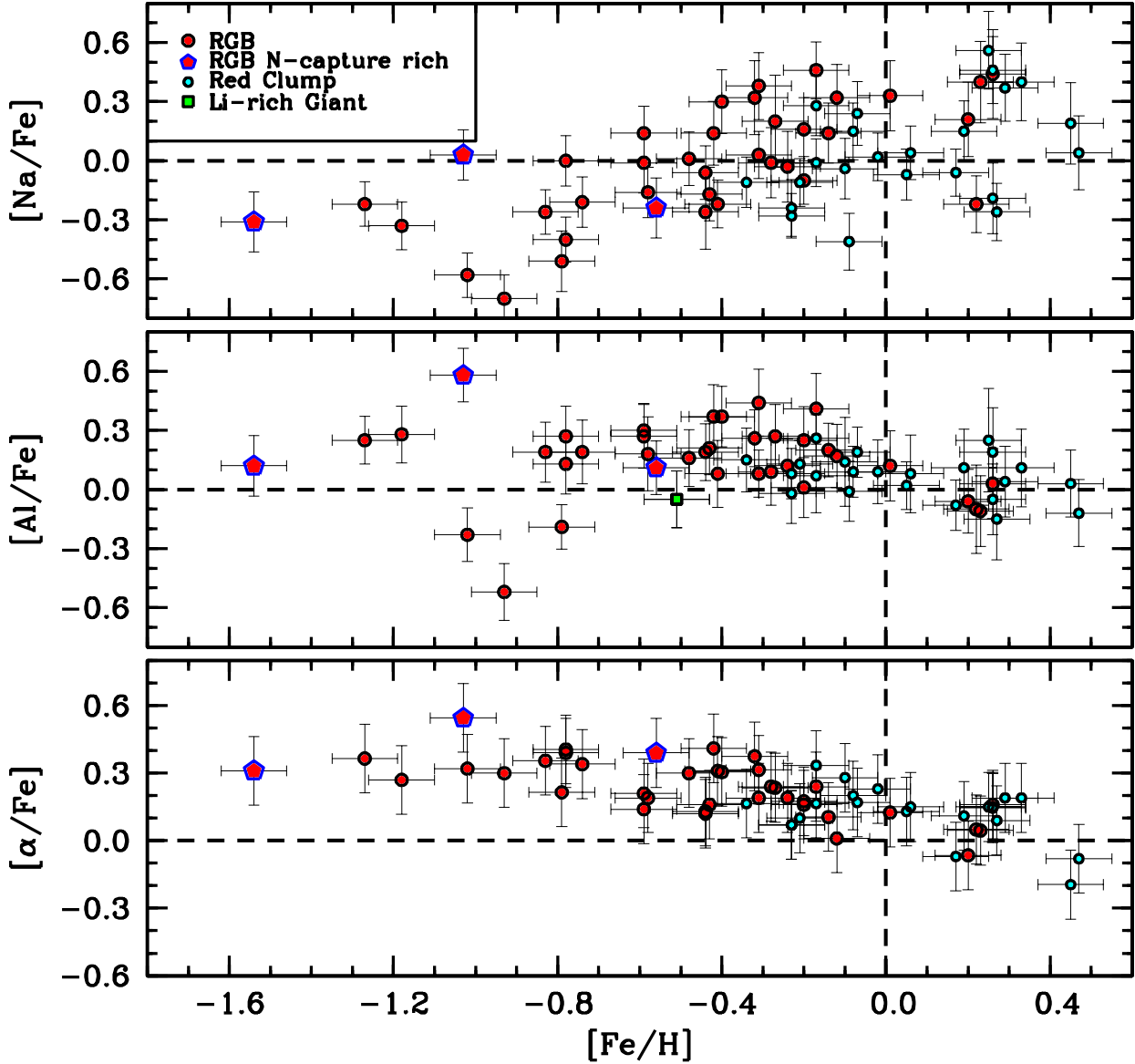}
\caption{Plots of [Na/Fe], [Al/Fe], and [$\alpha$/Fe] are shown as a function
of [Fe/H].  The [$\alpha$/Fe] abundances are from Johnson et al. (2011) and
[$\alpha$/Fe] is defined as $\onehalf$([Si/Fe]$+$[Ca/Fe]).  The red filled 
circles are the bulge RGB stars, and the filled cyan circles are the 
foreground red clump stars in our sample.  The blue outlined pentagons are the
stars with anomalously high neutron--capture element abundances, and the filled
green box is the Li--rich giant.}
\label{f2}
\end{figure}

In Figure \ref{f2} we plot the abundance trends of [Na/Fe], [Al/Fe], and (for 
reference) the [$\alpha$/Fe] abundances from Johnson et al. (2011) as a 
function of [Fe/H].  At the low metallicity end ([Fe/H]$\la$--0.8), Na and Al
are generally under-- and over--abundant at [Na/Fe]$\sim$--0.3 and 
[Al/Fe]$\sim$$+$0.2, respectively.  If the three anomalously low Na and Al 
stars (2MASS 18182256--3401248, 2MASS 18174351--3401412, and 2MASS 
18183521--3344124) are excluded, then the [Na/Fe] and [Al/Fe] abundance trends
are essentially flat between --1.5$\la$[Fe/H]$\la$--0.8.  However, the origin 
of these three low Na and Al stars is unclear, but each could have been 
captured from the Galactic halo or a 
disrupted stellar system, such as a globular cluster or dwarf galaxy (see also 
$\S$4.2.1).  Although these three stars span a relatively narrow metallicity 
range ([Fe/H]=--1.02 to --0.79), they exhibit a large radial velocity range from
--25 to $+$49 km s$^{\rm -1}$.  This suggests that the stars probably did not
all originate from a single globular cluster, assuming the initial kinematics
are preserved.  Interestingly, the other peculiar metal--poor stars in 
Figure \ref{f2} (blue outlined pentagons), which have large neutron--capture 
element enhancements, mostly follow the same [Na/Fe], [Al/Fe], and 
[$\alpha$/Fe] trends as the other bulge giants.

While the general [Na/Fe] and [Al/Fe] trends are relatively flat at 
[Fe/H]$\la$--0.8, the behavior of [Na/Fe] and [Al/Fe] begins to differentiate 
at higher metallicities.  This is naturally expected from the metallicity
dependence of yields for odd--Z elements, which depend on the surplus of 
neutrons from \iso{22}{Ne} and therefore on the initial CNO abundance.  As can 
be seen in Figure \ref{f2}, Al tends to mimic 
the $\alpha$--element trend such that the average [Al/Fe] ratio remains both 
moderately enhanced and constant until [Fe/H]$\sim$--0.2, where it then begins 
to decline.  The only exception appears to be the lone Li--rich giant (filled
green box), which exhibits a moderately low [Al/Fe]=--0.05.  On the other hand,
the average Na abundance increases from [Na/Fe]$\sim$--0.3 to $+$0.3 over that 
same range in [Fe/H], and then remains roughly constant at [Fe/H]$\ga$--0.2.  

Although the targets identified as possible inner disk red clump stars (filled 
cyan circles) show a nearly identical [$\alpha$/Fe] trend as the bulge RGB 
stars, it is difficult to determine whether their [Na/Fe] and [Al/Fe] 
distributions are unique.  Visual inspection of Figure \ref{f2} indicates that 
there may be a tendency for the clump stars to have systematically lower 
[Na/Fe] but similar [Al/Fe] ratios compared to bulge RGB stars of the same
metallicity.  A quantitative comparison between the two samples shows that the
bulge RGB stars with [Fe/H]$>$--0.4 have $\langle$[Na/Fe]$\rangle$=$+$0.19
($\sigma$=0.20) and $\langle$[Al/Fe]$\rangle$=$+$0.15 ($\sigma$=0.17) while
the clump stars have $\langle$[Na/Fe]$\rangle$=$+$0.05 ($\sigma$=0.25) and 
$\langle$[Al/Fe]$\rangle$=$+$0.07 ($\sigma$=0.11).  However, two--sided
Kolmogorov--Smirnov tests (Press et al. 1992) comparing the [X/Fe] 
distributions of the RGB and clump stars are unable to confidently reject
the null hypothesis that the [Na/Fe] and [Al/Fe] samples are drawn from the 
same parent population.\footnote{We adopt the notion that the null hypothesis 
(i.e., that the two distributions are the same) can be rejected if the p value 
is ``small" ($<$0.05).  For both [Na/Fe] and [Al/Fe] distributions the p values
are $>$0.20.}

\begin{figure}
\epsscale{1.00}
\plotone{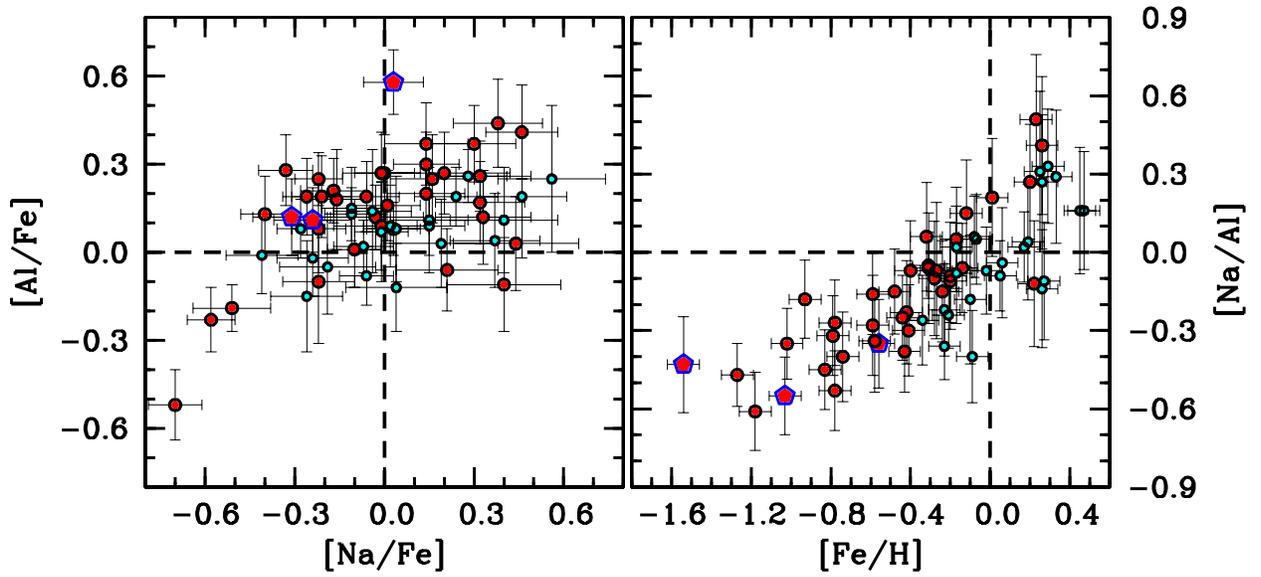}
\caption{The left panel shows [Na/Fe] vs [Al/Fe] for all stars in our sample.
The right panel illustrates the change in the [Na/Al] ratio as a function of
[Fe/H].  The symbols are the same as those in Figure \ref{f2}.}
\label{f3}
\end{figure}

In Figure \ref{f3} we show [Na/Fe] versus [Al/Fe] and the [Na/Al] ratio as a 
function of [Fe/H].  A well--known correlation exists between [Na/Fe] and 
[Al/Fe] in Galactic globular clusters (e.g., see review by Gratton et al. 
2004), and it is believed that this correlation is driven primarily by the 
mixing of pristine gas with material that has been enriched with the 
by--products of the NeNa and MgAl proton--capture cycles.  However, as can be
seen in Figure \ref{f3}, the RGB and clump stars analyzed here do not show
significant evidence for the existence of a Na--Al correlation.  This is 
supported by the relatively weak Pearson (R$_{\rm p}$) and Spearman 
(R$_{\rm s}$) correlation coefficients for the data (R$_{\rm p}$=0.298 and 
R$_{\rm s}$=0.252), excluding obvious outliers in the distribution.  Although
there does not appear to be a strong correlation between [Na/Fe] and [Al/Fe],
the [Na/Al] ratio plotted as a function of [Fe/H] shows a relatively steep
trend such that [Na/Al] increases with metallicity.  Both the bulge giants and
the foreground clump stars appear to share this chemical trait.  Note that
the three metal--poor stars with low [Na/Fe] and [Al/Fe] ratios no longer 
appear as outliers in the distribution when Na is normalized to Al.  This 
suggests that at least the Na and Al in these stars may share a common 
production origin with the rest of the metal--poor bulge giants.

\subsection{The Neutron--Capture Elements: Zirconium--Europium}

Unlike the lighter elements, which are produced predominantly through 
charged--particle reactions, most of the isotopes of elements heavier than
the Fe--peak are produced through successive neutron captures on preexisting 
seed nuclei (e.g., see the recent review by Sneden et al. 2008).  Heavy element
production typically occurs through one of two pathways: the slow 
neutron--capture process (s--process), which is active at low neutron exposure 
rates, and the rapid neutron--capture process (r--process), which is active
at high neutron exposure rates.  In general, each process is responsible
for synthesizing about half of the stable heavy element isotopes, but many
isotopes can also be produced through both processes (e.g., Burris et al. 2000; 
their Table 5).  Furthermore, each general production mechanism can be 
deconvolved into subprocesses that predominantly produce a set mass range of 
nuclei.  

The s--process is typically divided into three subcomponents: the ``main", 
``weak", and ``strong" s--processes.  The products of the main component,
nuclei in the range 88$\la$A$\la$208 (e.g., K{\"a}ppeler 1989; Arlandini et al. 
1999), are the most commonly studied in the literature.  This s--process 
mechanism is believed to be active in $\sim$1.5--4.0 M$_{\rm \sun}$ thermally 
pulsing AGB stars (e.g., Busso et al. 1999; Herwig 2005; Straniero et al. 
2006), and may be responsible for some fraction of the elements Zr, La, and
Nd analyzed here.  While the main component is thought to become active over
relatively long timescales ($\ga$500 Myr), the weak s--process may be active
as little as $\sim$10 Myr after the onset of star formation.  The environmental
and nucleosynthesis details for the weak s--process are not as well understood 
as those of the main s--process, but it is believed to be active in 
$>$10 M$_{\rm \sun}$ stars during the He--burning (e.g., Peters 1968; Lamb et 
al. 1977; Prantzos et al. 1987; Langer et al. 1989; Prantzos 1990; Raiteri et 
al. 1991a) and C--burning stages (e.g., Arcoragi et al. 1991; Raiteri et al. 
1991b; Raiteri et al. 1993; The et al. 2007; Pignatari et al. 
2010).\footnote{Note that low metallicity, rapidly rotating stars may also
be able to selectively produce the ``light" s--process elements Y, Sr, and 
(possibly) Zr, due to enhanced rotational mixing (Chiappini et al. 2011).}  
This production mechanism is typically used to explain the enhanced abundances
of isotopes in the mass range 60$\la$A$\la$90, found in some metal--poor stars 
(e.g., Burris et al. 2000).  The Zr analyzed here may have some
production via the weak s--process.  Lastly, the strong s--process component is
invoked to explain approximately half of the production of \iso{208}{Pb}, and 
is also thought to be active in low and intermediate mass, low metallicity AGB 
stars (e.g., Clayton \& Rassbach 1967; Beer \& Macklin 1985; Gallino et al. 
1998).  However, it is unlikely that any of the elements analyzed here have 
been significantly affected by the strong s--process.

In a similar fashion to the s--process, the r--process can be differentiated
into two different mechanisms: the ``main" and ``weak" processes.  While the 
exact sources of these two r--process components are not currently known, both
are strongly suspected to be associated with the final stages of massive stars
exploding as core--collapse SNe (e.g., Lattimer et al. 1977; Truran 1981; 
Mathews \& Cowan 1990; Cowan et al. 1991; Takahashi et al. 1994; Woosley et al.
1994; Freiburghaus et al. 1999; Truran et al. 2002; Arnould et al. 2007).  
Although the main r--process is responsible for the production of many isotopes
with A$\ga$130--140, the weak r--process component is believed to only produce 
isotopes with A$\la$130--140.  In particular, the weak r--process mechanism 
is typically invoked in order to explain the existence of certain metal--poor
stars exhibiting preferential enhancement of only the r--process nuclei near
the Sr--Y--Zr peak (e.g., McWilliam 1998; Burris et al. 2000; Johnson \& Bolte 
2002; Travaglio et al. 2004; Fran{\c c}ois et al. 2007; Honda et al. 2007).  
For the present work the main r--process may affect the abundances of La, Nd, 
and especially Eu, and the weak r--process may contribute to the production of 
Zr.

\subsubsection{Zirconium}

\begin{figure}
\epsscale{1.00}
\plotone{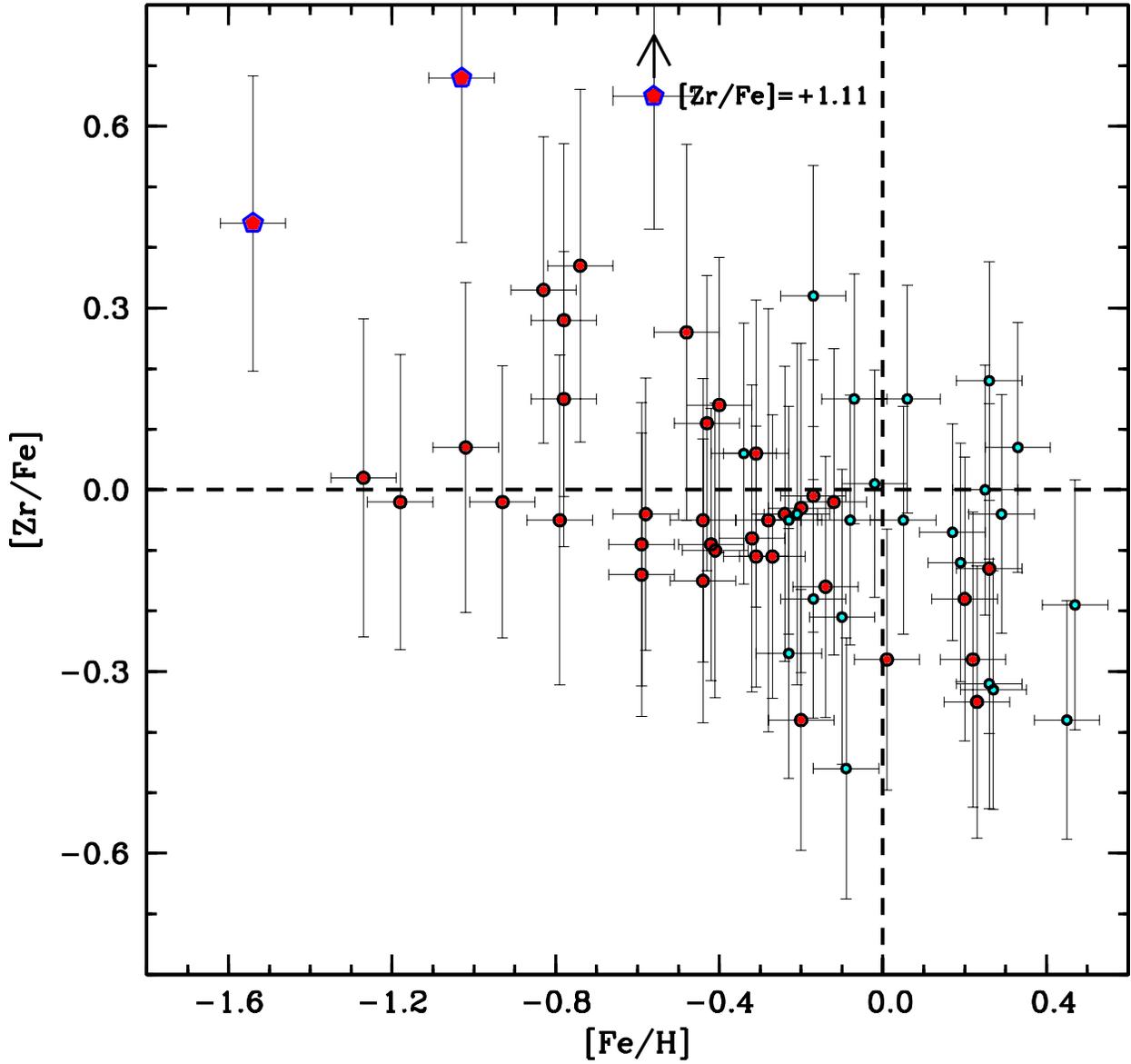}
\caption{The [Zr/Fe] ratios are plotted as a function of [Fe/H].
The symbols are the same as those in Figure \ref{f2}.}
\label{f4}
\end{figure}

In Figure \ref{f4} we show a plot of [Zr/Fe] as a function of [Fe/H] for all 
stars analyzed here.  If we ignore the three super Zr--rich stars, then the 
remaining stars in our sample with [Fe/H]$<$--0.9 have a nearly uniform Zr
abundance of $\langle$[Zr/Fe]$\rangle$=$+$0.01 ($\sigma$=0.04).  In the range
[Fe/H]=--0.8 to --0.4 there is some evidence of two separate sequences, where
one group of stars is enhanced at [Zr/Fe]$\sim$$+$0.25 and the other is
moderately Zr--poor at [Zr/Fe]$\sim$--0.10.  Interestingly, these two
distributions merge together at [Fe/H]$\approx$--0.4 and continue a trend of
decreasing [Zr/Fe] abundance with increasing [Fe/H] down to [Zr/Fe]$\sim$--0.3
at [Fe/H]$\sim$$+$0.3.  However, the larger errors associated with the Zr I
measurements makes it difficult to assess how robust the difference is between
the ``Zr--enhanced" and ``Zr--poor" stars.  The clump stars do not appear to 
exhibit any strong trends in [Zr/Fe] as a function of [Fe/H] and instead have 
an average abundance of $\langle$[Zr/Fe]$\rangle$=--0.08 but with a relatively 
large star--to--star scatter ($\sigma$=0.19).  A two--sided KS test cannot rule
out that the RGB and clump [Zr/Fe] distributions are significantly different 
at [Fe/H]$>$--0.4.

\subsubsection{Lanthanum, Neodymium, and Europium}

\begin{figure}
\epsscale{1.00}
\plotone{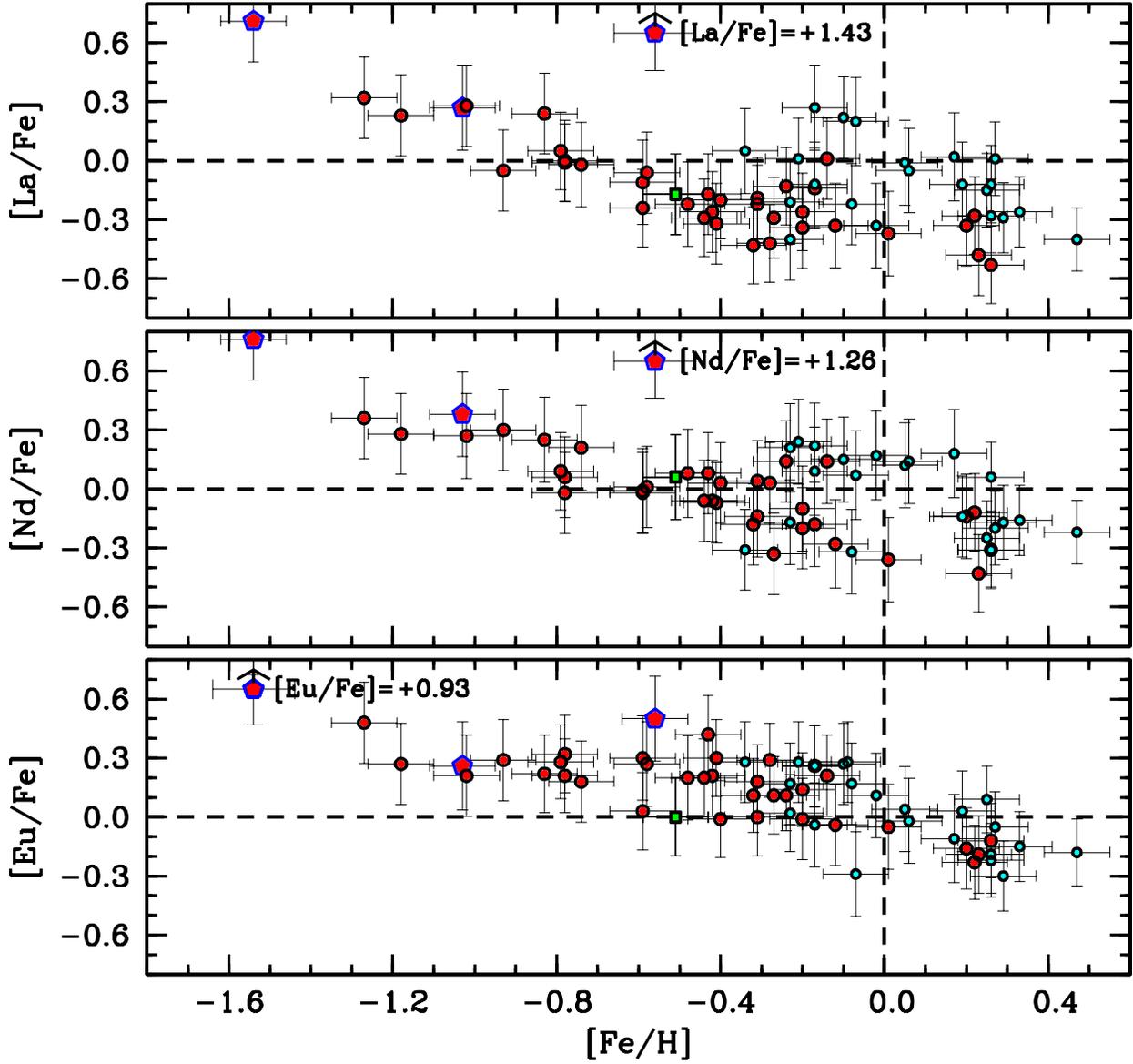}
\caption{[La/Fe], [Nd/Fe], and [Eu/Fe] are plotted as a function of [Fe/H].  
The symbols are the same as those in Figure \ref{f2}.}
\label{f5}
\end{figure}

In Figure \ref{f5} we plot the [La/Fe], 
[Nd/Fe], and [Eu/Fe] ratios for all stars as a function of [Fe/H].  First 
examining La and Nd, it is clear from Figure \ref{f5} that both elements 
exhibit very similar abundance trends (with a $\sim$0.1--0.2 dex systematic 
offset).  On the metal--poor end, nearly all stars are enhanced at 
[La,Nd/Fe]$\approx$$+$0.30.  At metallicities greater 
than [Fe/H]=--0.80, the [La/Fe] and [Nd/Fe] trends decline with increasing 
metallicity until [Fe/H]$\sim$--0.4.  Beyond [Fe/H]=--0.4, the [La,Nd/Fe] 
ratios remain roughly constant at [La,Nd/Fe]$\sim$--0.30.  Although the 
star--to--star scatter is larger for the clump stars, there is 
some indication that the clump stars may have higher [La/Fe] and [Nd/Fe] 
abundances than the bulge giants.  In the metallicity regime covered by the 
clump stars ([Fe/H]$\ga$--0.4), the bulge stars have 
$\langle$[La/Fe]$\rangle$=--0.29 ($\sigma$=0.14) and 
$\langle$[Nd/Fe]$\rangle$=--0.14 ($\sigma$=0.17) while the
clump stars have $\langle$[La/Fe]$\rangle$=--0.10 ($\sigma$=0.19) and 
$\langle$[Nd/Fe]$\rangle$=--0.03 ($\sigma$=0.20).  Interestingly, a two--sided
KS test suggests that the clump and RGB [La/Fe] distributions might be drawn
from different parent populations, but the case for [Nd/Fe] is not as strong
with a p--value of 0.09 (compared to 0.01 for La).  If larger sample sizes 
confirm this result then it may be possible that the [La/Fe] and [Nd/Fe] ratios
could be used as sensitive discriminators between the bulge and inner disk 
populations, assuming the clump stars identified here are actually part of the
inner disk.

\begin{figure}
\epsscale{1.00}
\plotone{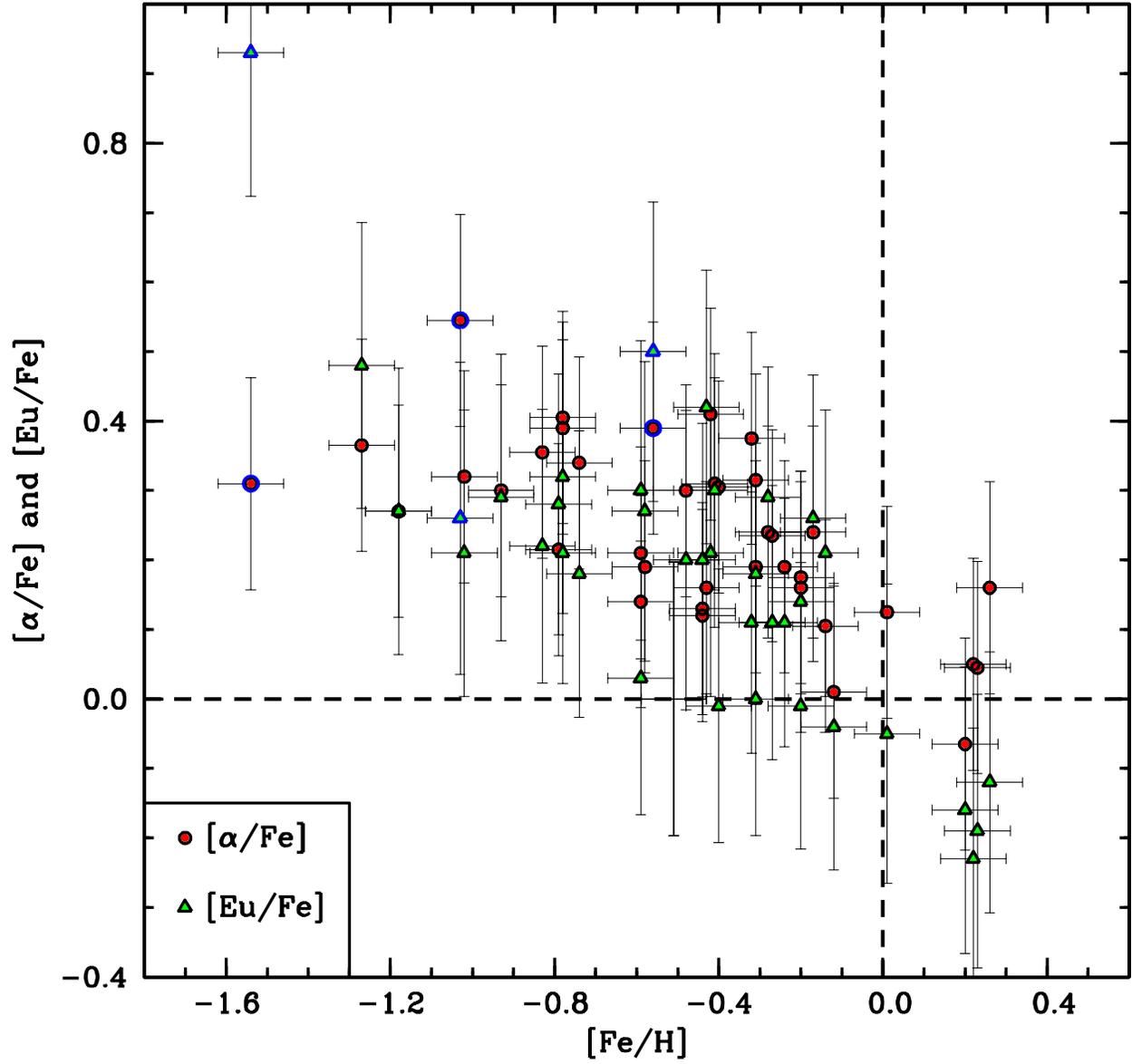}
\caption{[$\alpha$/Fe] (filled red circles) and [Eu/Fe] (filled green 
triangles) are plotted as a function of [Fe/H].  Note that the objects plotted 
here are only the bulge RGB stars, and the blue outlined points are the same
as those in Figures \ref{f2}--\ref{f5}.}
\label{f6}
\end{figure}

For Eu, inspection of Figure \ref{f5} reveals that [Eu/Fe] exhibits 
a different trend than La and Nd.  Although Eu is also enhanced in the 
metal--poor stars at [Eu/Fe]$\sim$$+$0.30, the overall shape of the [Eu/Fe]
versus [Fe/H] trend follows a much more ``$\alpha$--like" distribution.  Like
the [$\alpha$/Fe] distribution shown in Figure \ref{f2}, [Eu/Fe] remains 
enhanced at [Eu/Fe]$\sim$$+$0.30 until [Fe/H]$\sim$--0.4 where [Eu/Fe] then
begins to decline at higher metallicities.  In fact, overplotting [Eu/Fe]
and [$\alpha$/Fe], as is shown in Figure \ref{f6}, illustrates the remarkable
agreement between the distributions of these two elements.  The similar 
behavior between the $\alpha$ elements and Eu is not unexpected since both
groups are believed to be produced by massive stars.  It is
also worth noting that while the clump stars may potentially have different 
[La/Fe] and [Nd/Fe] abundances than the bulge RGB stars, there does not appear 
to be any significant difference between the two populations with respect to 
the [Eu/Fe] (or [$\alpha$/Fe]) distributions.

\begin{figure}
\epsscale{1.00}
\plotone{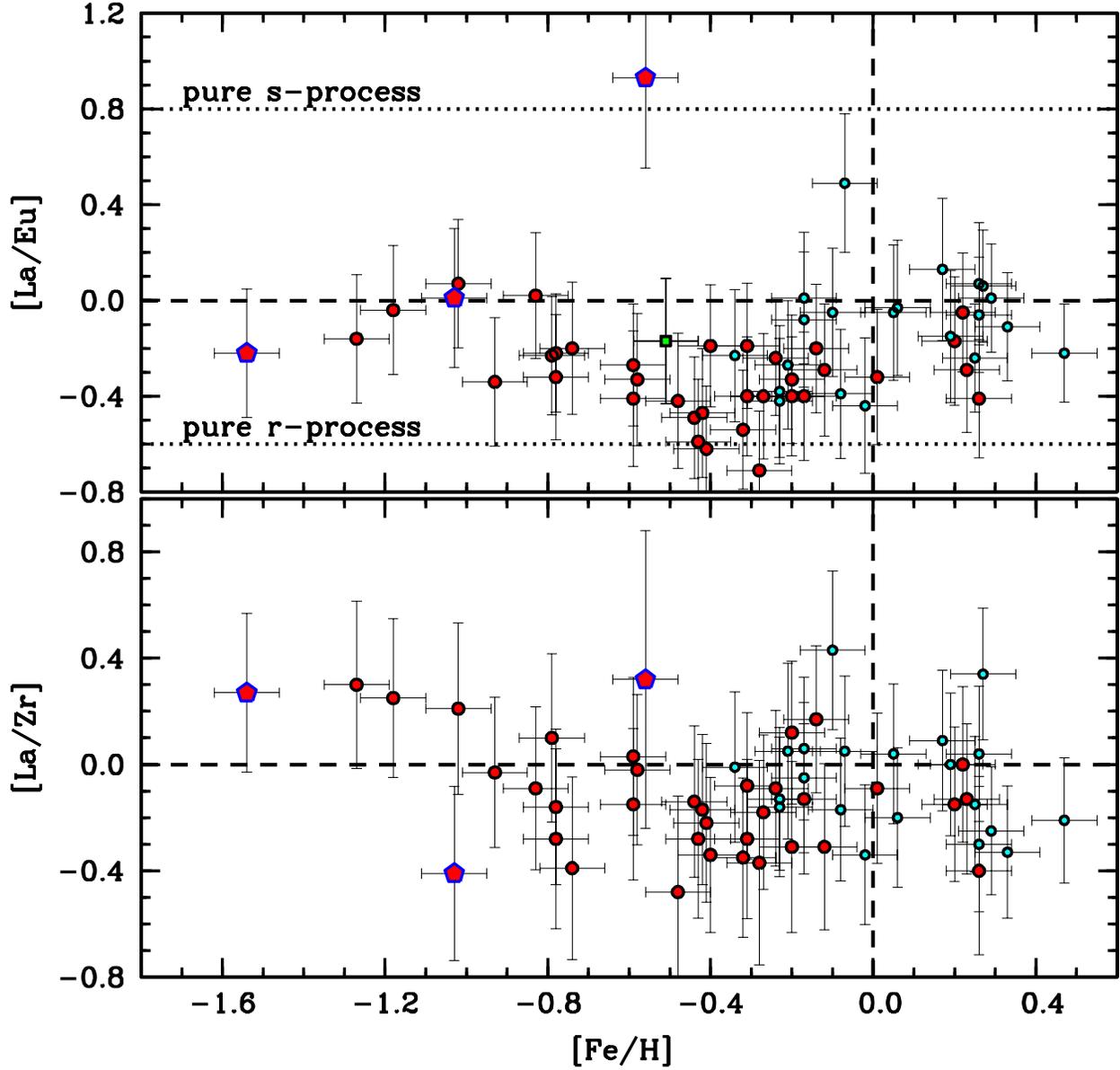}
\caption{[La/Eu] and [La/Zr] are plotted as a function of [Fe/H].  The symbols 
are the same as those in Figure \ref{f2}.  The dotted lines in the left panel 
indicate the abundance ratios expected for pure r-- and s--process enrichment 
by Kappeler et al. (1989) and the ``standard pocket" model of Bisterzo et 
al. (2010), respectively.}
\label{f7}
\end{figure}

The [La/Eu] ratio is often used as a tracer of the relative contributions the
r-- and s--processes have made to a stellar population.  This indicator is 
useful because La can be made in both processes, but Eu is made almost entirely
by the r--process ($\sim$97$\%$; e.g., Burris et al. 2000).  Similarly, when
the s--process is active plots of the ``heavy" (e.g., Ba and La) versus 
``light" (e.g., Y and Zr) s--process elements can yield insight into the 
metallicity of the production site.  The s--process in a metal--poor 
environment is expected to produce a higher heavy/light ratio because there are
more neutrons available per seed nucleon and the neutron--capture chain 
proceeds to heavier nuclei.  In Figure \ref{f7} we plot the [La/Eu] and [La/Zr]
ratios as a function of [Fe/H] for all program stars.  We find that the [La/Eu]
trend, especially for [Fe/H]$>$--0.8, is nearly flat at 
$\langle$[La/Eu]$\rangle$$\sim$--0.3 and exhibits a relatively 
consistent star--to--star scatter ($\sigma$=0.15).  The consistency of the 
[La/Eu] magnitude and scatter across a wide range in [Fe/H] suggests that the 
abundance variations are likely dominated by measurement errors.  The low 
[La/Eu] ratios are consistent with the r--process being the dominant 
neutron--capture production process for nearly all of the bulge RGB 
stars\footnote{Note that star 2MASS 18174742--3348098 exhibits a [La/Eu] ratio 
that is consistent with pure s--process production.  However, given the apparent
dominance of the r--process in bulge stars, it seems likely that this star
is the result of mass transfer from an AGB companion or was captured from an
external stellar system.}.  However, several of the most metal--poor stars 
appear to have [La/Eu] ratios that are higher than the bulk of the RGB 
population.  This suggests that at least some of the most metal--poor bulge 
stars may have experienced enrichment by the s--process, and is supported by 
the enhanced [La/Zr] ratios in these same stars.  In a similar fashion, the 
relatively flat [La/Zr] trend of the more metal--rich RGB stars may be 
consistent with the absence of a significant s--process contribution.  However,
we note again that the larger errors on the Zr abundance may be masking any 
subtle, underlying trends.  Interestingly, while the clump stars do not seem to
significantly differentiate themselves from the bulge RGB population in the 
[La/Zr] plot, these stars appear to exhibit a non--negligible rise in the 
[La/Eu] ratio at [Fe/H]$>$0.  If confirmed, then this may be evidence for a 
longer formation timescale in the inner disk compared to the Galactic bulge.

\section{DISCUSSION}

\subsection{Comparison to Other Bulge Field Locations}

\subsubsection{The Light Elements}

\begin{figure}
\epsscale{1.00}
\plotone{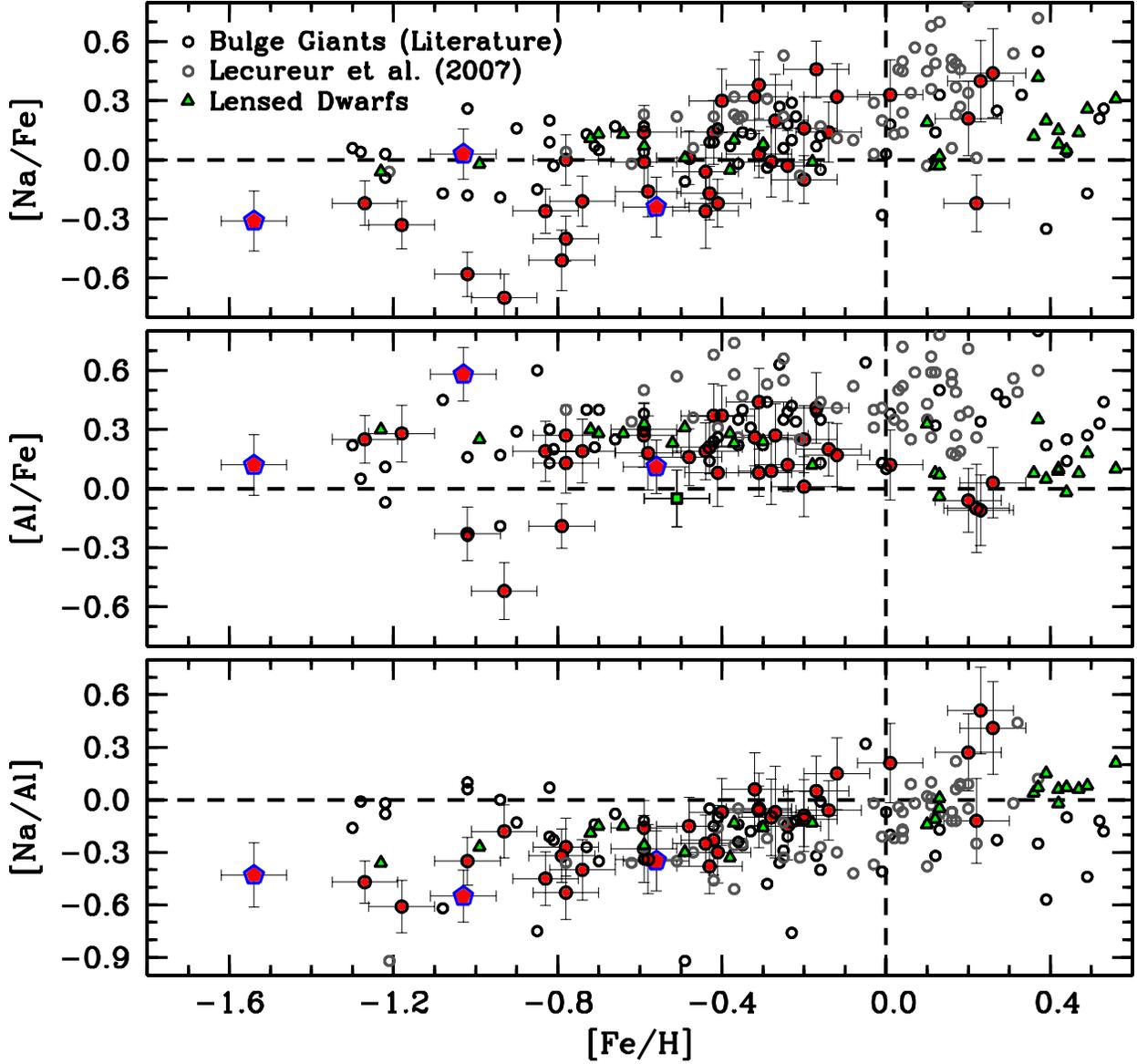}
\caption{[Na/Fe], [Al/Fe], and [Na/Al] plotted as a function of [Fe/H] for 
bulge stars only.  The symbols for our data are the same as those in 
Figure \ref{f2}.  The open black circles represent bulge RGB data from 
McWilliam \& Rich (1994), Fulbright et al. (2007), and Alves--Brito et al.
(2010).  The open grey circles are the bulge RGB data from Lecureur et al.
(2007), and the filled green triangles are from the microlensed dwarf studies
by Bensby et al. (2010a; 2011).}
\label{f8}
\end{figure}

In Figure \ref{f8} we compare the [Na/Fe], [Al/Fe], and [Na/Al] ratios as a 
function of [Fe/H] for our data and those available in the literature.  First
examining [Na/Fe], all regions of the bulge represented in Figure \ref{f8}
(essentially the minor--axis from b=--4$\degr$ to b=--12$\degr$) generally 
follow the same abundance pattern.  In particular, the average [Na/Fe] 
ratio increases as a function of metallicity with little change in the 
star--to--star dispersion.  On average, the [Na/Fe] abundance appears to rise 
from [Na/Fe]$\sim$--0.20 to $+$0.00 at low metallicity and increase up to 
[Na/Fe]$\sim$$+$0.20 to $+$0.40 at the high metallicity end.  Even considering
the differences in measurement techniques, data quality, and 
temperature/gravity regimes covered by the points in Figure \ref{f8}, we find
no evidence supporting a strong minor--axis abundance gradient in [Na/Fe].  
This finding is consistent with the lack of any gradient in [$\alpha$/Fe].

However, there could be subtle differences at the metal--poor and metal--rich 
ends of the distribution.  Although the number of abundance measurements for 
stars with [Fe/H]$<$--1 is small, Figure \ref{f8} indicates that the [Na/Fe] 
ratios for the most metal--poor Plaut field stars (ignoring the three very low 
Na and Al stars) may be 0.1--0.3 dex lower than those found at 
b$\la$--4$\degr$.  If this result is confirmed by larger sample sizes then it 
could be an indication of composition inhomogeneities in the early bulge.  The 
cause of the discrepancies at the metal--rich end of the [Na/Fe] distribution,
noting especially the large range of derived [Na/Fe] ratios, is unclear.
Combining all literature data with the present study indicates that stars with
[Fe/H]$>$0 fall into the range --0.3$\la$[Na/Fe]$\la$$+$1.0, though we note 
that only Lecureur et al. (2007) find a significant number of stars with 
[Na/Fe]$>$+0.5.  It may be the case that the larger dispersion seen at 
[Fe/H]$>$0 simply reflects the difficulty in accurately analyzing cool, 
metal--rich giants.

The [Al/Fe] distribution is nearly identical for all bulge stars shown in 
Figure \ref{f8}, at least in the range --1.5$\la$[Fe/H]$\la$--0.2.  
Specifically, the stars tend to be enhanced at [Al/Fe]$\sim$$+$0.2, regardless 
of metallicity and bulge location.  Although the Lecureur et al. (2007) [Al/Fe]
data again show higher abundances and increased star--to--star scatter, 
the general trend of enhanced [Al/Fe] with no strong metallicity dependence 
appears to be the same.  However, two different [Al/Fe] trends emerge at 
[Fe/H]$\ga$--0.2.  The Plaut field and microlensed dwarfs exhibit a decrease in
[Al/Fe] toward solar values, but the data from McWilliam \& Rich (1994), 
Fulbright et al. (2007), Lecureur et al. (2007), and Alves--Brito et al. (2010)
consistently show enhanced [Al/Fe] at all metallicities.  This is a 
particularly puzzling problem because the stars contributing to the two 
separate trends essentially sample the same regions of the bulge.

As is noted in Bensby et al. (2011) and can be seen in Figure \ref{f8}, two of
the metal--rich microlensed bulge dwarfs exhibit higher [Al/Fe] (and 
[Na/Fe]) abundances than the other dwarfs of similar metallicity.  
Interestingly, the higher [Al/Fe] ratios of these two stars match the 
enhancements observed in bulge giants by McWilliam \& Rich (1994), Fulbright et
al. (2007), Lecureur et al. (2007), and Alves--Brito et al. (2010).  These
two dwarfs, coupled with the literature data, might hint that two 
different populations of metal--rich bulge stars exist.  Alternatively, the true
[Al/Fe] distribution for the metal--rich bulge may span a range from roughly 
[Al/Fe]=--0.10 to $+$0.40, and the microlensed dwarf and Plaut data happen to 
predominantly sample the low--Al stars.  There could also be age and/or
initial mass function (IMF) variations that contributed to the different 
abundance trends.  

Most analyses tend to find that the bulge is at least 10 Gyr old with
a $\sim$1 Gyr age spread (e.g., Ortolani et al. 1995; Feltzing \& Gilmore 2000;
Kuijken \& Rich 2002; Zoccali et al. 2003; Clarkson et al. 2008), but the data 
do not fully rule out the existence of a young, metal--rich bulge population
(see also Nataf \& Gould 2011).  In fact, several of the microlensed dwarfs are
estimated to be only $\sim$3--4 Gyr old.  While we do not have explicit age 
estimates for the Plaut field stars, in Johnson et al. (2011) we used old 
globular cluster isochrones to photometrically estimate the metallicity 
distribution function in Plaut's field from RGB--tip stars and found excellent 
agreement with the spectroscopic metallicities.  Therefore, we assume that 
most, if not all, of the RGB stars are old.  It is worth noting that the two 
metal--rich [Al/Fe] enhanced dwarfs seen in Figure \ref{f8} have estimated ages
of $\sim$3 and 13 Gyr, and plotting [Al/Fe] versus age for all of the data in 
Bensby et al. (2010a; 2011) indicates that the two parameters are not strongly
correlated.  This suggests that age is not a primary cause of the metal--rich 
[Al/Fe] discrepancy.

\begin{figure}
\epsscale{1.00}
\plotone{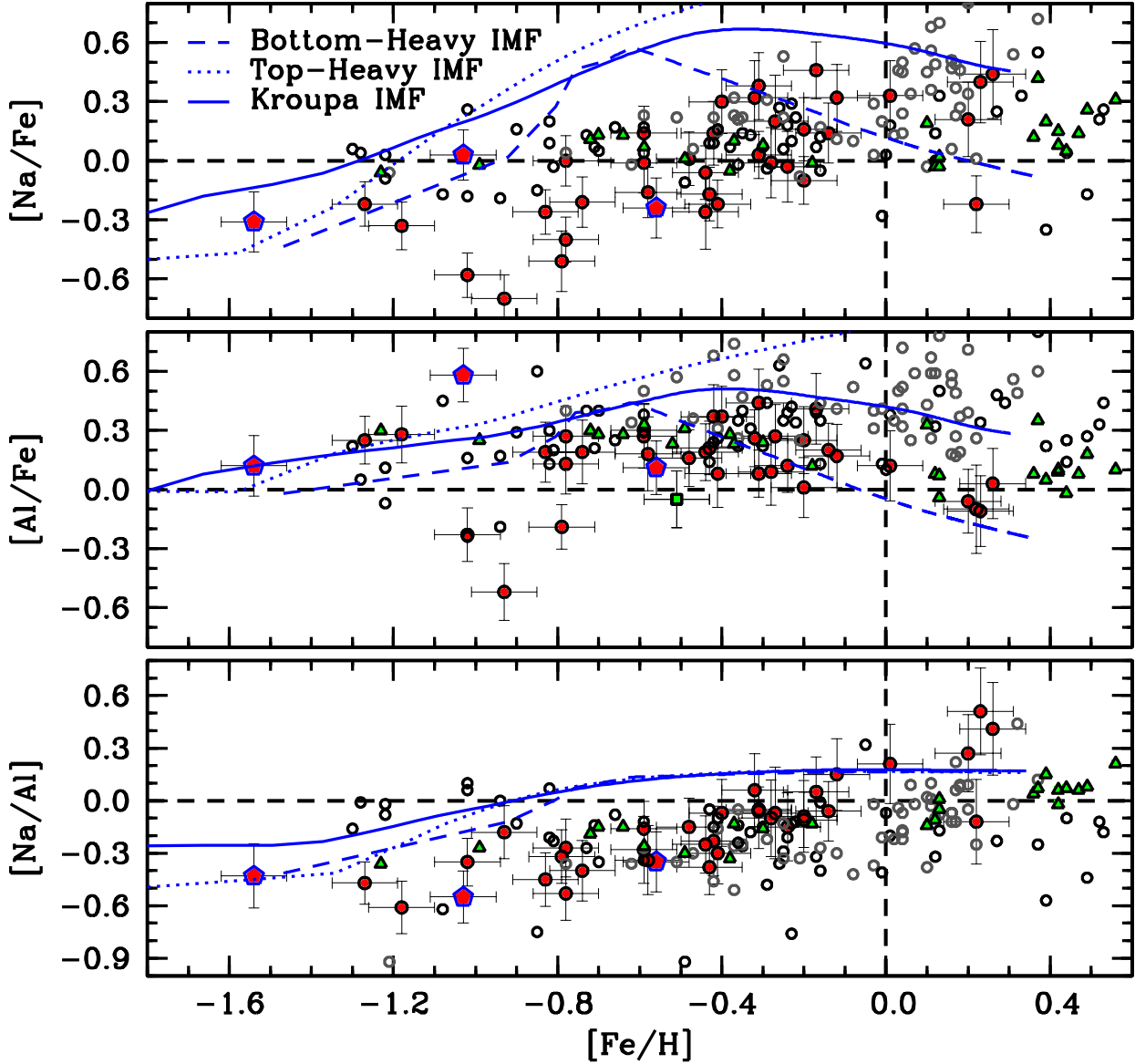}
\caption{The same plot as Figure \ref{f8} with bulge chemical enrichment 
models of varying IMF overplotted.  The models follow the prescriptions 
outlined in Kobayashi et al. (2006; 2010; see also $\S$4 for more details).
The solid blue lines illustrate the models based on a Kroupa IMF (x=1.3; 
Kroupa 2008), the dotted blue lines indicate models based on an extreme 
top--heavy IMF (x=0.3), and the dashed blue lines indicate models based on an
extreme bottom--heavy IMF (x=1.6).  Note that Ballero et al. (2007) find an
IMF slope x$\leq$0.95 with a high star formation efficiency to best fit the 
bulge.}
\label{f9}
\end{figure}

\begin{figure}
\epsscale{1.00}
\plotone{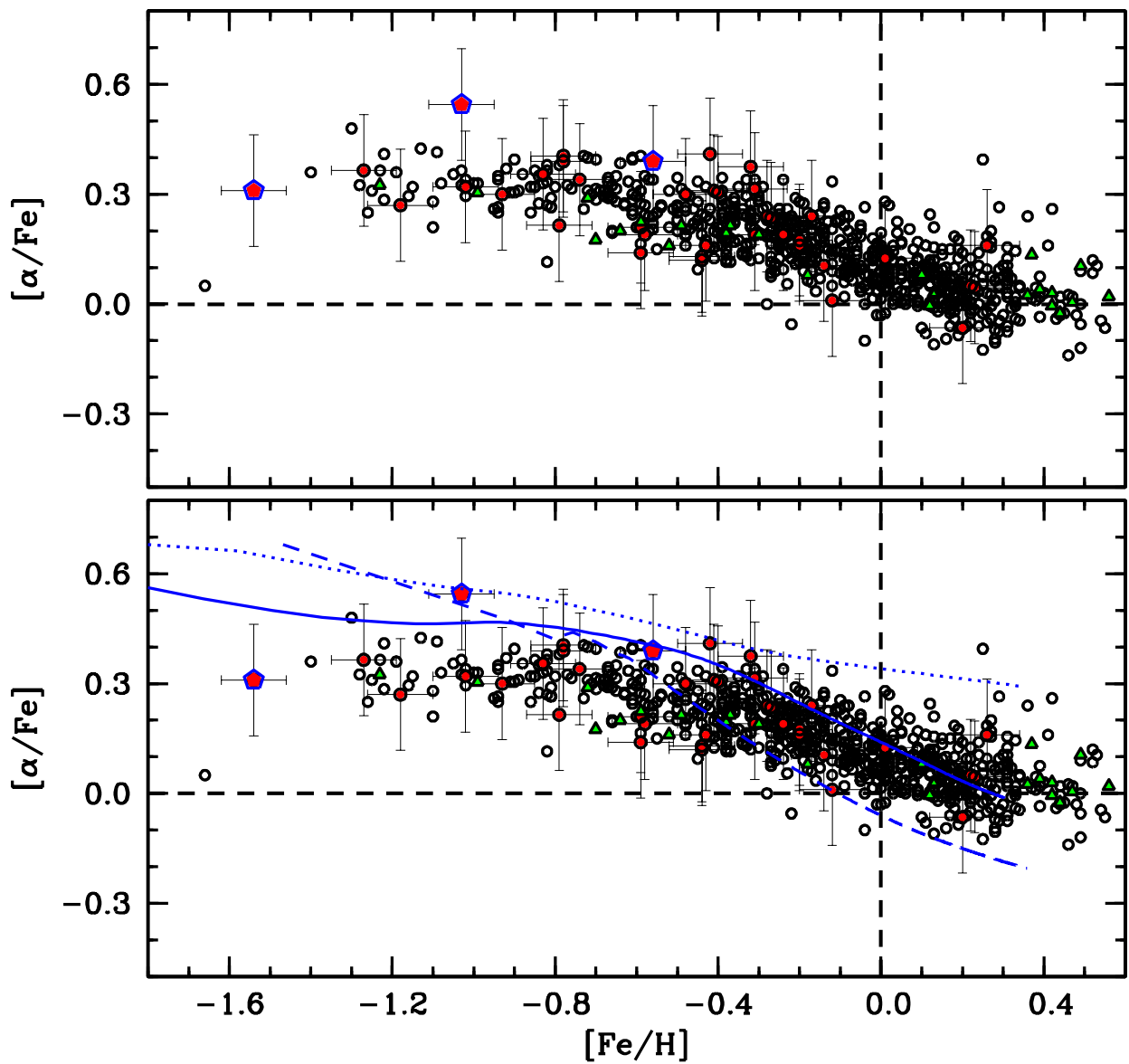}
\caption{Similar to Figures \ref{f8}--\ref{f9}, the [$\alpha$/Fe] ratio is 
plotted as a function of [Fe/H] for bulge RGB and dwarf stars.  The symbols 
are the same as those in Figure \ref{f8}, and [$\alpha$/Fe] is defined 
as $\onehalf$([Si/Fe]$+$[Ca/Fe]).  The literature RGB data are from 
McWilliam \& Rich (1994), Fulbright et al. (2007), Alves--Brito et al. (2010),
Johnson et al. (2011), and Gonzalez et al. (2011).  The microlensed dwarf
data are from Bensby et al. (2010a; 2011).  The bottom panel contains the same
data as the top panel, but with the same bulge chemical enrichment models from
Figure \ref{f9} overplotted.}
\label{f10}
\end{figure}

In order to investigate whether IMF variations could cause [Al/Fe] differences,
we have plotted the data from Figure \ref{f8}, along with three chemical 
enrichment models for the Galactic bulge, in Figure \ref{f9}.  The star
formation rates are modified from Kobayashi et al. (2006; 2011) to meet the
observed metallicity distribution function and include updated super--solar 
metallicity yields (up to Z=0.05) and nuclear reaction rates.  The three models
are calculated for a Kroupa (2008) IMF (x=1.3), an extreme top--heavy IMF 
(x=0.3), and an extreme bottom--heavy IMF (x=1.6).  The star formation, infall,
and galactic wind timescales are (0.2 Gyr, 5 Gyr, 3 Gyr) for x=1.3, (0.01 Gyr, 
5 Gyr, Inf.) for x=0.3, and (0.01 Gyr, 5 Gyr, 1 Gyr) for x=1.6, respectively.  
As can clearly be seen in Figure \ref{f9}, the models predict that stars with 
[Fe/H]$>$--0.2 originating from populations with different IMFs should exhibit
significantly different light element abundances.  However, in Figure \ref{f10}
we plot [$\alpha$/Fe] as a function of [Fe/H] from several studies spanning the
same regions of the bulge as those in Figures \ref{f8}--\ref{f9}.  
Figure \ref{f10} reiterates the conclusions from Johnson et al. (2011) and 
Gonzalez et al. (2011) that the [$\alpha$/Fe] ratio trends are essentially 
identical for all bulge stars.  Since the $\alpha$--elements are produced 
almost exclusively in massive stars, these data suggest that there 
were not any significant IMF variations in the bulge.  Additionally, formation 
timescale differences should also be ruled out from the [$\alpha$/Fe] trends.  
In $\S$4.1.2 we show that the heavy element abundances indicate that the 
enrichment timescale was probably very similar throughout the bulge.

Since the [Na/Fe] and [Al/Fe] ratios alone may be sensitive to parameters such
as model atmosphere deficiencies and 3D/NLTE corrections, we have also plotted
the [Na/Al] ratio in Figures \ref{f8}--\ref{f9} in an attempt to mitigate
these effects.  Similar to what was observed in Figure \ref{f3}, we find that
normalizing Na by Al generally decreases the star--to--star scatter in a 
given metallicity bin.  While the agreement is very good in the range 
--0.8$\la$[Fe/H]$\la$--0.2, the slope of the [Na/Al] 
increase is steepest for the Plaut field data followed by the Lecureur et al.
(2007), Bensby et al. (2010a; 2011), and finally the Baade's window data by 
McWilliam \& Rich (1994), Fulbright et al. (2007), and Alves--Brito et al. 
(2010).  The reason for the [Na/Al] differences is not immediately clear, and 
the Lecureur et al. (2007) data, which span multiple bulge fields at various 
Galactic latitudes, do not show any correlations between [Na/Al] ratio and 
location.  Interestingly, the chemical enrichment models shown in Figure 
\ref{f9} have two important predictions: (1) the [Na/Al] ratio should rise 
with metallicity and (2) this ratio should be nearly independent of the IMF.  
While the data do clearly exhibit an increase in [Na/Al] with [Fe/H], all of 
the models predict that the rise should occur at [Fe/H]$\la$--1; however, this 
is not observed until at least [Fe/H]=--0.4.  Additionally, the model [Na/Al]
ratios are typically at least 0.2 dex higher than the observed values.  
However, the yields of these elements are subject to significant changes as the
models improve to include newer/updated input physics, nuclear reaction rates,
and employ fully self--consistent 3D treatments of parameters such as 
rotation and convection.  Although the [Na/Al] distribution seems to reinforce 
the previously stated notion that IMF variations are not the cause of the 
metal--rich abundance differences, other factors such as inflow/outflow of gas 
could play an important and unaccounted for role.  On the other hand, these 
differences may simply reflect the difficulty in analyzing metal--rich spectra.

\subsubsection{The Heavy Elements}

\begin{figure}
\epsscale{1.00}
\plotone{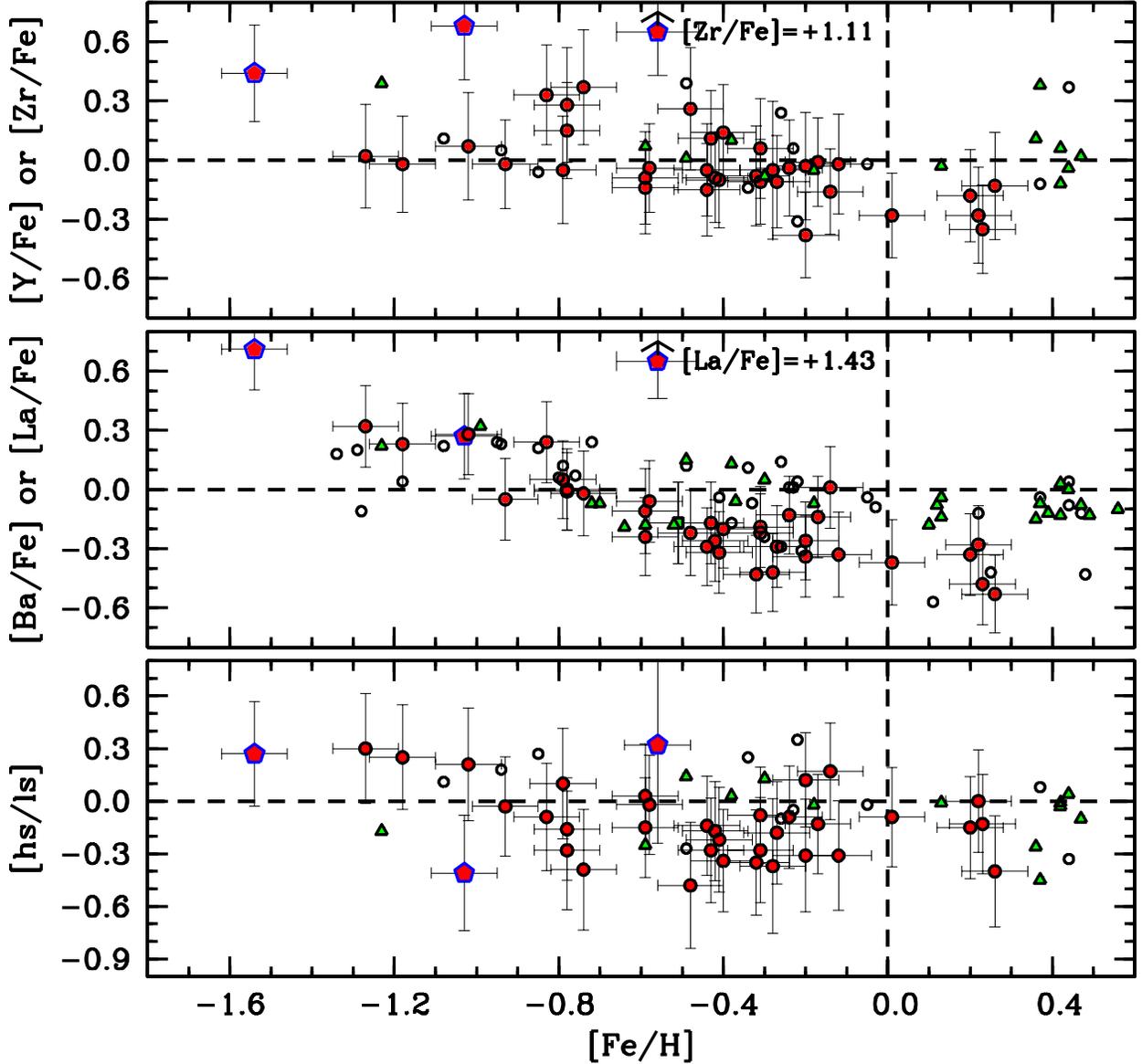}
\caption{The top panel shows the abundance trends of [Y/Fe] (literature data) 
or [Zr/Fe] (our data) as a function of [Fe/H].  The symbols
are the same as those in Figures \ref{f2} and \ref{f8}.  The literature data
are from Bensby et al. (2010a; 2011) for the dwarfs and McWilliam \& Rich (1994)
for the giants.  The middle panel plots [Ba/Fe] or
[La/Fe] as a function of [Fe/H].  [Ba/Fe] data are shown for Bensby et al.
(2010a; 2011) and [La/Fe] data are shown for our data, McWilliam \& Rich (1994),
and McWilliam et al. (2010).  The bottom panel plots the ``heavy--to--light"
neutron--capture ratio ([hs/ls], where heavy is defined as either [Ba/Fe] or
[La/Fe] and light is defined as either [Y/Fe] or [Zr/Fe]), as a function of 
[Fe/H].  The literature data are from Bensby et al. (2010a; 2011) and 
McWilliam \& Rich (1994).}
\label{f11}
\end{figure}

In Figure \ref{f11} we compare the abundances of [Zr/Fe], [La/Fe], and the 
``heavy--to--light" ratio ([hs/ls]) versus [Fe/H] between our data and those
available in the literature.\footnote{A detailed discussion of Nd 
abundance trends is omitted because very few Nd measurements are available
in the literature.}  Since few [Zr/Fe] and [La/Fe] measurements exist 
for bulge stars in the literature, we have substituted [Y/Fe] and [Ba/Fe] 
abundances when appropriate.  While there is likely to be some systematic 
offset between [Y/Fe] and [Zr/Fe] or [Ba/Fe] and [La/Fe], these element pairs 
are neighbors in atomic number and are believed to trace very similar 
production mechanisms.  

We generally find good agreement between all of the data sets plotted in 
Figure \ref{f11}.  Specifically, it appears that the heavier neutron--capture
elements Ba and La exhibit a general trend of decreasing abundance as a 
function of increasing metallicity, but the Y and Zr trends are much more
shallow.  The literature and Plaut field data all include a handful of 
stars with enhanced [Y,Zr/Fe] ratios that deviate from the otherwise relatively
flat distribution, but the small sample sizes per metallicity bin, especially
on the metal--poor end, make it difficult to understand the origin of these
enhanced stars.  Many of the stars that appear to be enhanced in Y and/or Zr 
do not stand out as outliers in the Ba and La plot.  It is possible that the 
Y/Zr--enhanced stars could have been stochastically affected by the weak
s-- and/or r--processes, which are believed to selectively produce elements 
near the Sr--Y--Zr peak without affecting heavier elements like Ba and La (see 
the brief discussion in $\S$3.2).  The multiple possible production mechanisms
for Y and Zr compared to Ba and La, which are only thought to be produced in
the main s-- and r--processes, may be an explanation for the slightly 
different abundance trends.  Similarly, since Y and Zr may be produced in
a wider range of stars than Ba and La, particularly in the weak s--process by
massive stars also producing Fe, this may explain why the [Y,Zr/Fe] ratio 
remains roughly constant over a large metallicity range while the average 
[Ba,La/Fe] ratio experiences a steady decline at [Fe/H]$>$--0.8.\footnote{Note
that this may not be true if the weak s--process is mostly active in lower
mass SNe, which do not produce much Fe.}  However, more observations are 
needed and we note that many of the Zr and Y abundances shown in 
Figure \ref{f11} have uncertainties $>$0.2 dex.

While more than 70$\%$ of the Ba and La in the solar system is thought to have 
been produced by the main s--process in $\sim$1.5--4 M$_{\sun}$ thermally 
pulsing AGB stars (e.g., Burris et al. 2000; Bisterzo et al. 2010), a 
non--negligible 15--30$\%$ is still produced by the r--process.  If the bulge 
formed in $\la$1 Gyr then it is possible that not enough time was available for
significant s--process enrichment.  Therefore, we may be able to assume that 
the [Ba,La/Fe] trend shown in Figure \ref{f11} mostly reflects metallicity 
dependent r--process yields.  Alternatively, it is possible that only a 
specific mass range of stars produces Ba and La via the r--process, and the 
declining trend is a result of Fe but not Ba or La being produced.  
Interestingly, nearly all of the data points agree that the [Ba,La/Fe] ratios 
in the most metal--poor bulge stars are enhanced at [Ba,La/Fe]$\sim$$+$0.2 dex.
Could the initially high [Ba,La/Fe] ratios reflect pre--enrichment of the early
bulge gas?  As we will show in $\S$4.2.2, the most metal--poor bulge stars do 
in fact fall in the range of observed [La/Fe] abundances for the most 
metal--rich Galactic halo stars.  Furthermore, the slightly enhanced [hs/ls] 
ratios in the most metal--poor stars suggest that some s--process enrichment 
occurred.  Unfortunately, little is quantitatively known about the pure 
r--process yields at [Fe/H]$>$--1.5.  However, the main s--process is predicted
to produce significant variations in the [hs/ls] ratio as a function of 
metallicity (e.g., Bisterzo et al. 2010; their Figure 11).  Therefore, we can 
speculate that the general invariance of the [hs/ls] ratio at [Fe/H]$\ga$--1 
may be additional evidence that the main s--process did not play a major role 
in the bulge's chemical enrichment.

\begin{figure}
\epsscale{1.00}
\plotone{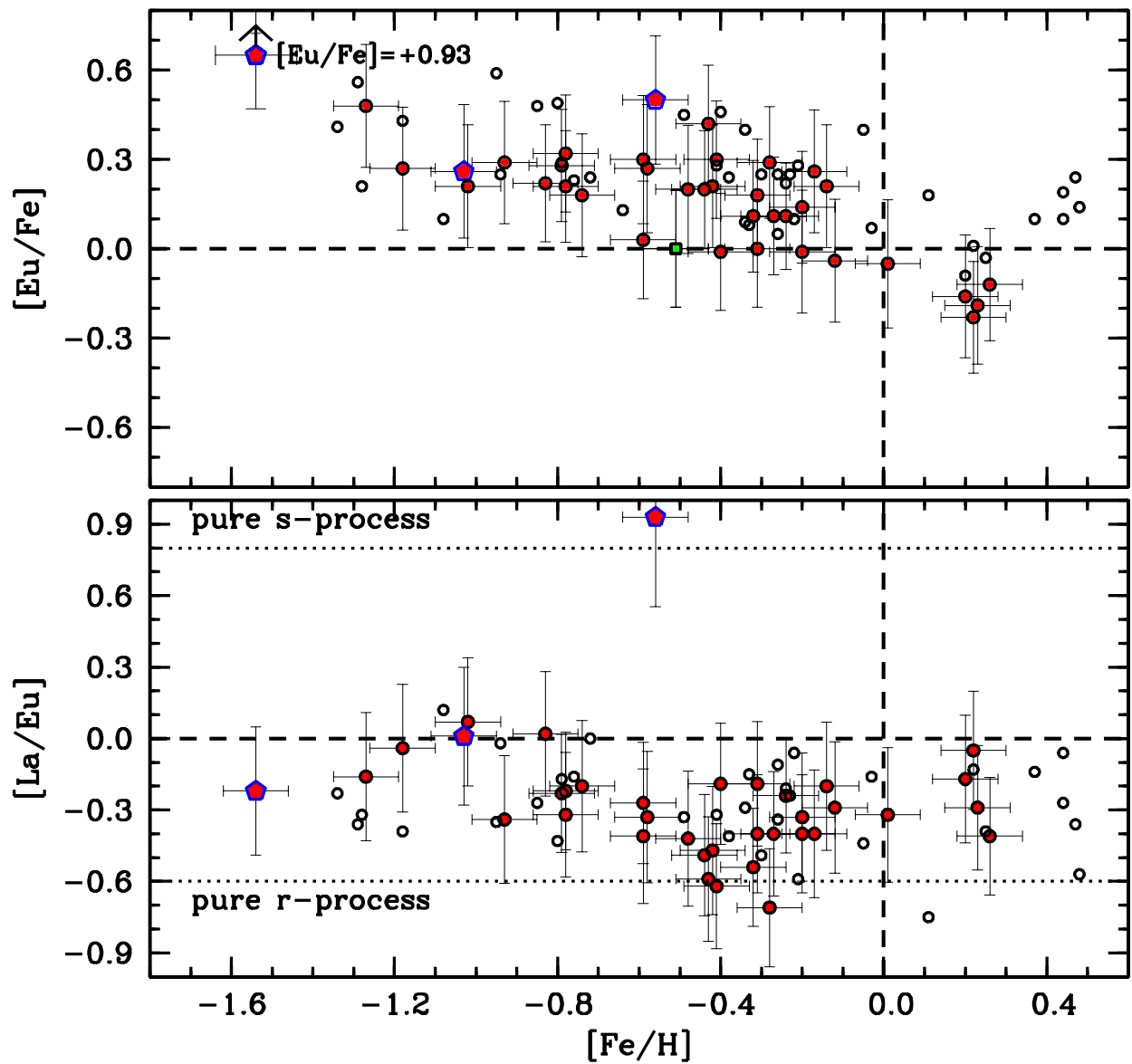}
\caption{In the top panel the [Eu/Fe] ratio is plotted as a function of [Fe/H].
The symbols are the same as those in Figure \ref{f8}.  The literature data
are from McWilliam \& Rich (1994) and McWilliam et al. (2010).  The bottom
panel shows the [La/Eu] ratio as a function of [Fe/H], using our data set 
along with the McWilliam \& Rich (1994) and McWilliam et al. (2010) data.
The pure r-- and s--process lines are the same as those in Figure \ref{f7}.}
\label{f12}
\end{figure}

Since Eu is believed to be produced almost exclusively in massive stars, the 
naive expectation is that the trend of [Eu/Fe] versus [Fe/H] should closely 
mimic that of [$\alpha$/Fe].  In fact, at least in the Plaut field,
we find that the [Eu/Fe] and [$\alpha$/Fe] trends are nearly indistinguishable
(see Figure \ref{f6}).  In Figure \ref{f12} we plot [Eu/Fe] and [La/Eu] versus
[Fe/H] for the Plaut field and the Baade's window data from McWilliam \& Rich
(1994) and McWilliam et al. (2010), and note that all three data sets exhibit
nearly identical [Eu/Fe] trends following the general [$\alpha$/Fe] 
distribution.  While there is some discrepancy at [Fe/H]$>$0, we believe this 
may be related to different studies normalizing Eu II with Fe I (our data) or 
Fe II (McWilliam \& Rich 1994; McWilliam et al. 2010) and/or the small sample 
size of our data at high metallicity.  As was pointed out in McWilliam et al. 
(2010), the standard Tinsley (1979) chemical enrichment picture implies that 
the downturn in [$\alpha$/Fe] and [Eu/Fe] are due to the addition of Fe from 
Type Ia SNe, and suggests a $\ga$1 Gyr formation timescale for the 
bulge.  However, very little is known about the exact production location of 
the r--process, and detailed yield calculations do not exist.  Therefore, we 
reach the same conclusion as McWilliam et al. (2010) that the downturn in 
[Eu/Fe] is likely related more to the metallicity dependent yields of the 
r--process than the Type II/Type Ia SN ratio.\footnote{If the bulge stars, even
at high metallicity, are in fact dominated by the r--process, then the bulge 
may prove to be a useful population for examining the r--process at high 
metallicity.  In principle, this is difficult to do with
thin and thick disk stars because of contamination by the s--process.}  

While the [La/Fe] and [Eu/Fe] ratios may not individually reveal the strength
of s-- and r--process contributions, the [La/Eu] ratio is predicted to be a 
strong indicator of the main r--process versus the main s--process 
contributions.  Additionally, normalizing La II by Eu II should cancel out any 
significant errors associated with measuring La II and Eu II relative to Fe I 
or Fe II.  As can be seen in Figure \ref{f12}, a majority of bulge stars 
exhibit [La/Eu] ratios that are consistent with the r--process being the 
dominant neutron--capture production mechanism.  Additionally, the [La/Eu] 
trends at b=--4$\degr$ and b=--8$\degr$ are nearly identical and do not show 
any indication of an increase in [La/Eu] with rising [Fe/H].  These 
observations suggest that the enrichment timescales for the inner and outer 
bulge were very similar and that this enrichment occurred on a timescale rapid 
enough (probably $\la$1 Gyr) to avoid significant contributions from the 
main s--process.  The nearly indistinguishable [La/Fe] and [Nd/Fe] trends shown
in Figure \ref{f5} also support the lack of a main s--process contribution.  In 
the solar system, roughly 50$\%$ of the Nd is produced by the main s--process
(compared to $\sim$75$\%$ for La), and if it were contributing to the bulge's 
chemical enrichment then we might expect to observe a change in the [Nd/Fe] 
compared to [La/Fe] trends.  

At first glance, it seems as if the [La/Eu] and [$\alpha$/Fe] data produce 
conflicting views on the bulge's formation.  The [La/Eu] ratios indicate that 
star formation in the bulge was relatively rapid, but the [$\alpha$/Fe] data
suggest a more extended star formation period that allowed Type Ia SNe to 
contribute.  Unfortunately, the actual timescale over which Type Ia SNe become 
important is not well constrained and may range from as little as a few hundred
Myr (e.g., Matteucci \& Recchi 2001; Raskin et al. 2009) to a few Gyr (e.g., 
Yoshii et al. 1996).  Additionally, there is likely to be a non--negligible 
metallicity dependence in the yields of $\alpha$--elements from Type II SNe.  
In fact, most Type II SN models show a trend of decreasing [$\alpha$/Fe] yields
with increasing metallicity (e.g., Kobayashi et al. 2006; their Figure 5).
However, it remains to be seen what role the Type II/Ia SN ratio played in 
the bulge's chemical enrichment.

A remaining problem with the rapid formation scenario is the previously 
mentioned discovery by Bensby et al. (2010a; 2011) of a significant number of 
young ($\sim$3--4 Gyr) and relatively metal--rich bulge dwarfs.  This contrasts
with the [La/Eu] data presented here and in the literature that seem to 
rule out a chemical enrichment timescale (and thus age spread) of several Gyr.
At the moment there is no obvious solution to this problem, but it could be 
possible that all previous analyses of bulge RGB stars have selectively chosen 
targets belonging to the dominant old population.  Much larger sample sizes, 
[La/Eu] determinations for the young and old bulge dwarf populations, and a 
better understanding of the r--process at solar and super--solar metallicities 
are clearly needed.

\subsection{Comparison to the Halo, Thick Disk, and Thin Disk}

\subsubsection{The Light Elements}

\begin{figure}
\epsscale{0.75}
\plotone{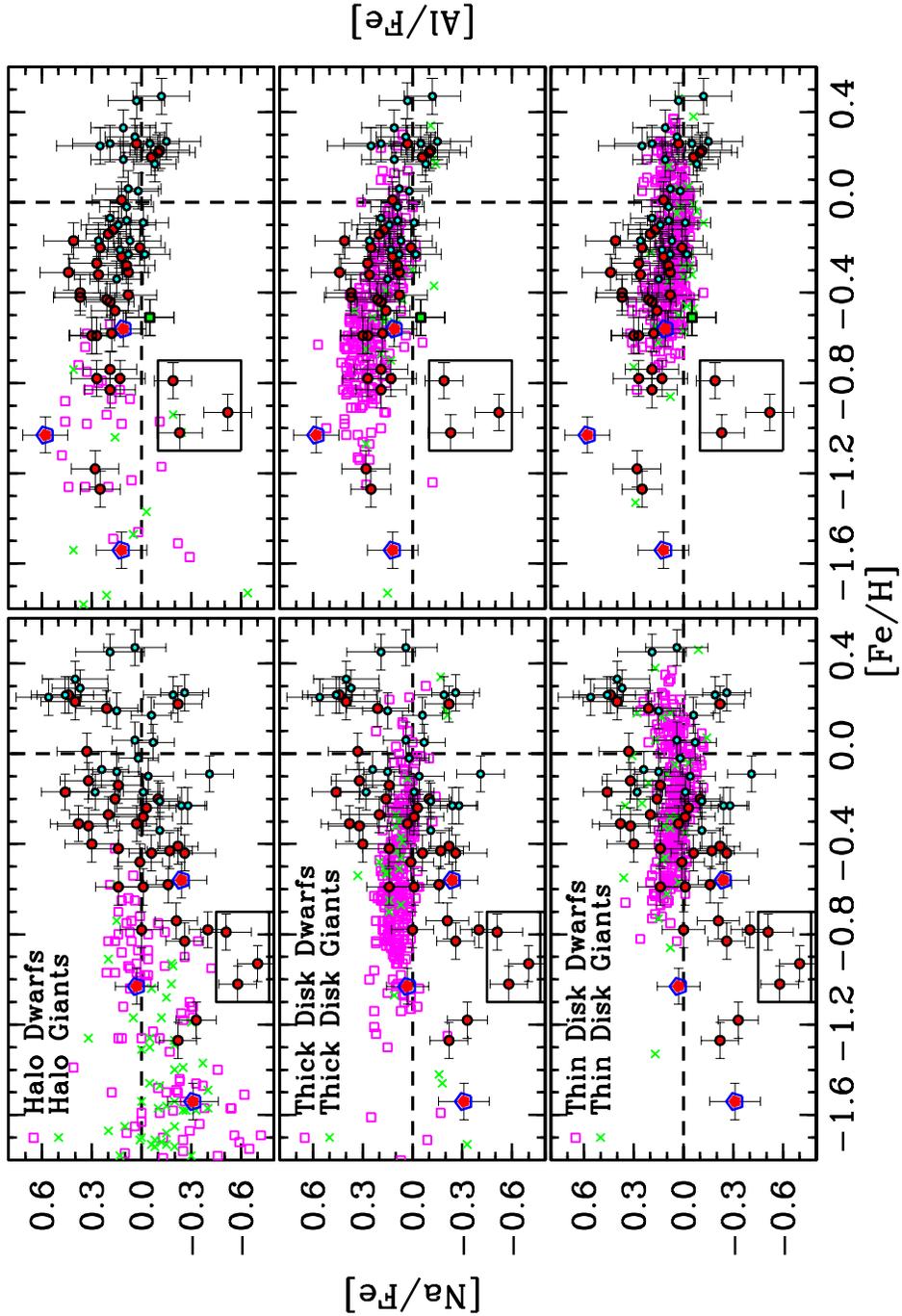}
\caption{[Na/Fe] and [Al/Fe] are plotted as a function of [Fe/H] for our sample
and the literature.  The red, cyan, and blue outlined symbols are the same as
those in Figure \ref{f2}, and a box has been placed around the three low Na/Al
stars for guidance.  The magenta open boxes represent literature dwarf
abundances for the halo (top panels), thick disk (middle panels), and thin
disk (bottom panels).  Similarly, the green crosses represent literature 
giant abundances.  The halo data are from McWilliam et al. (1995), Ryan et al.
(1996), Nissen \& Schuster (1997), Hanson et al. (1998), Fulbright (2000),
Johnson (2002), Stephens \& Boesgaard (2002), Reddy et al. (2006), and
Alves--Brito et al. (2010).  The thin and thick disk data are from Edvardsson
et al. (1993), Nissen \& Schuster (1997), Hanson et al. (1998), Fulbright 
(2000), Prochaska et al. (2000), Stephens \& Boesgaard (2002), Bensby et al. 
(2003; 2005), Reddy et al. (2003; 2006), and Brewer \& Carney (2006).}
\label{f13}
\end{figure}

\begin{figure}
\epsscale{1.00}
\plotone{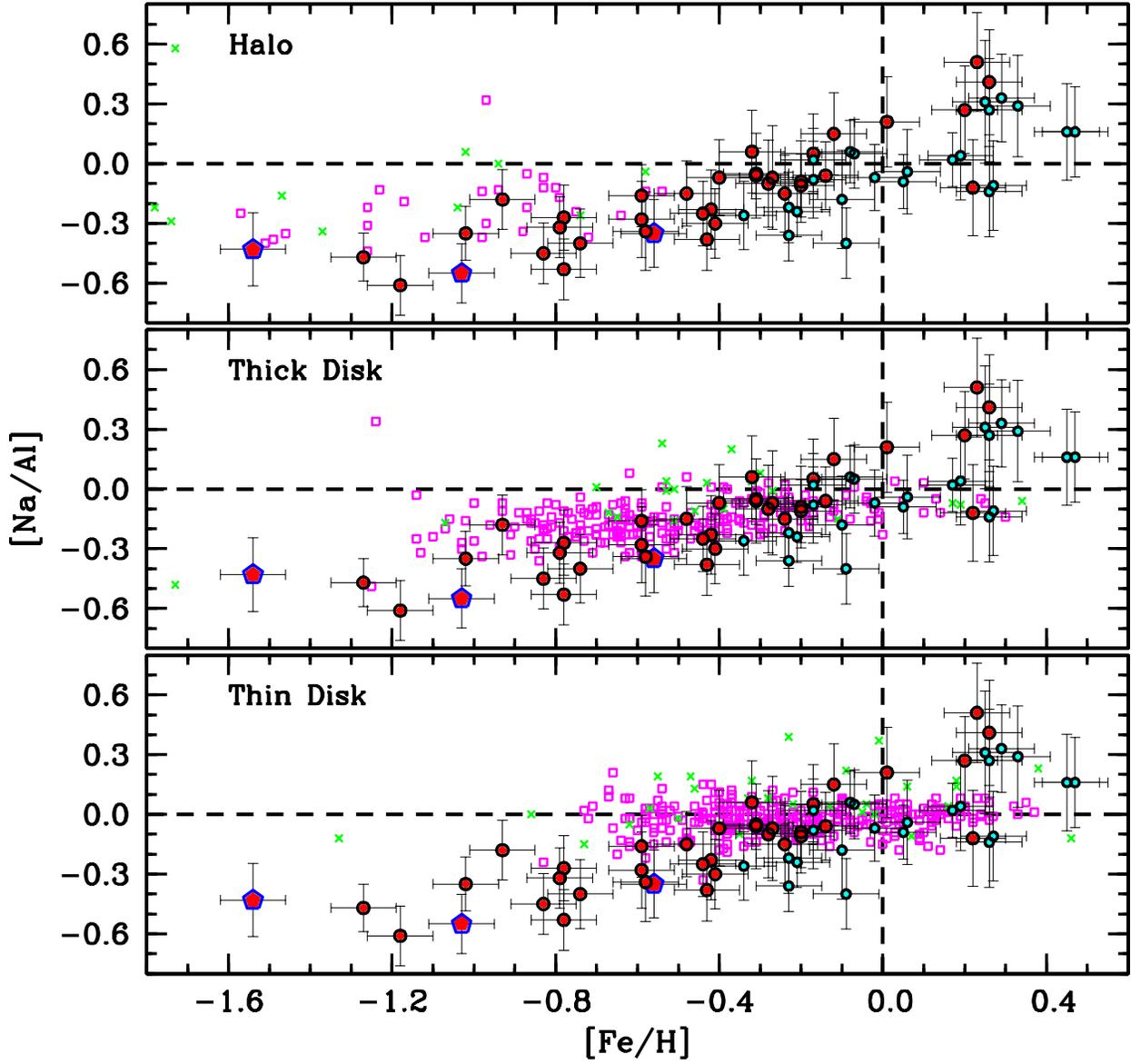}
\caption{A similar plot to Figure \ref{f13} comparing our bulge [Na/Al] 
abundances as a function of [Fe/H] to the halo, thick disk, and thin disk.}
\label{f14}
\end{figure}

In Figures \ref{f13}--\ref{f14} we plot dwarf and giant [Na/Fe], [Al/Fe], and 
[Na/Al] abundances as a function of [Fe/H] for the Plaut field, Galactic halo, 
thick disk, and thin disk.  The halo and disk data are compiled from the 
literature, and we have checked to ensure that the abundance ratios from 
different studies are in agreement, on average.  While we generally find very 
good agreement between the different studies without applying any corrections, 
the [Al/Fe] ratios for stars with [Fe/H]$<$--0.4 in the Edvardsson et al. 
(1993) data set were systematically increased by $+$0.1 dex to match the other 
literature sources.  Since there may be non--negligible systematic differences
between abundance ratios derived from dwarfs and giants, in Figure \ref{f13}
and subsequent figures we differentiate between dwarf and giant data for 
comparison.  As can be seen in Figures \ref{f13}--\ref{f14}, the giant and 
dwarf [Na/Fe], [Al/Fe], and [Na/Al] abundances typically agree to within 
$\sim$$\pm$0.1 dex, on average.

Although a handful of thick disk stars at [Fe/H]$<$--0.8 have [Na/Fe]$<$0,
a majority of the data points scatter between $+$0.0$\la$[Na/Fe]$\la$$+$0.2.
In contrast, nearly all of the similar metallicity halo stars fall in 
the range --0.6$\la$[Na/Fe]$\la$$+$0.2.  Therefore, our limited data seem to
suggest that the most metal--poor stars in Plaut's field exhibit more 
halo--like [Na/Fe] ratios, and may be distinctly different from the most 
metal--poor thick disk stars.  If confirmed in additional fields, the 
observation that outer bulge abundance ratios are more similar to the halo
and inner bulge abundances more similar to the thick disk would be evidence
of non--negligible composition differences in the early bulge.  However, we note
that very few Na abundance measurements are available for low metallicity thick
disk giants, which could be important given that NLTE and 3D effects tend to 
increase in magnitude with decreasing metallicity (e.g., Asplund 2005).  On the
other hand, the average [Na/Fe] difference between the low metallicity Plaut 
field giant and thick disk dwarf stars is $\sim$0.3--0.4 dex, which is 
significantly larger than the roughly 0.1--0.15 dex NLTE correction predicted 
for cool, metal--poor giants using the subordinate 6154/6160 Na I lines (e.g., 
see Lind et al. 2011, their Figure 4).

While the [Na/Fe] differences between the inner and outer bulge disappear at 
[Fe/H]$\ga$--0.4 (see Figures \ref{f8}--\ref{f9}), it is clear from 
Figure \ref{f13} that the thick disk and Plaut field differentiate at 
[Fe/H]$>$--0.2 (although they are roughly similar in the range 
--0.6$\la$[Fe/H]$\la$--0.2).  On average, the Plaut field stars are more 
enhanced in [Na/Fe] and continue a trend of increasing [Na/Fe] with [Fe/H].
In contrast, the thick (and thin) disk giant and dwarf data peak in [Na/Fe] at 
[Fe/H]$\sim$--0.6 and then gradually decline at solar and super--solar 
metallicities.  Interestingly, the thin disk data show an increase in [Na/Fe]
at [Fe/H]$>$0 that brings the bulge and thin disk populations into modest 
agreement for the most metal--rich stars.

\begin{figure}
\epsscale{1.00}
\plotone{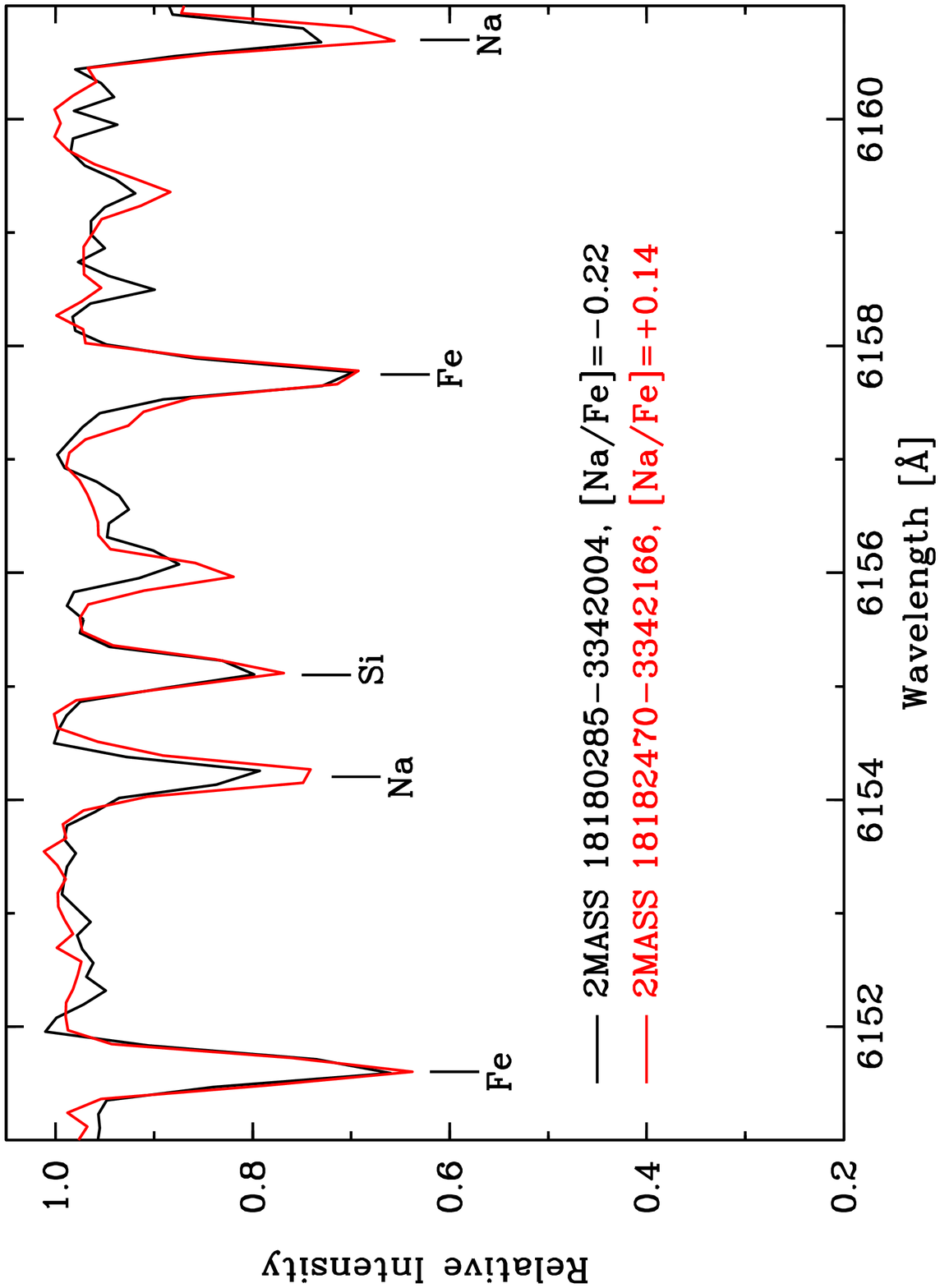}
\caption{Two Plaut field giants of similar temperature, gravity, and 
metallicity (T$_{\rm eff}$$\approx$4150 K, log(g)$\approx$1.25 (cgs), and
[Fe/H]$\approx$--0.40) but different [Na/Fe] abundances are overplotted.  This 
illustrates that at least part of the star--to--star dispersion in the Na 
abundances is real.}
\label{f15}
\end{figure}

An interesting question is whether the larger star--to--star 
scatter in the Plaut field compared to disk data has any significance.  
Although our bulge data and those from the literature share similar 
dispersions, this may be an artifact of the larger uncertainties associated 
with analyzing bulge, as opposed to local disk, stars.  However, in 
Figure \ref{f15} we overplot two Plaut field giants of similar T$_{\rm eff}$, 
log(g), and [Fe/H] but different [Na/Fe].  Note that the Na line strength 
differences indicate that at least part of our measured [Na/Fe] scatter is 
real.  We note also that Melendez et al. (2008) and Alves--Brito et al. (2010) 
measured bulge, halo, and disk stars using the same analysis techniques and 
found comparable star--to--star scatter among all of the populations.  
We also find that the possible inner disk red clump stars in our 
sample tend to follow the same [Na/Fe] pattern as the metal--rich thin and 
thick disk, albeit with larger scatter, which is consistent with the findings 
by Bensby et al. (2010b) that inner and local disk stars have similar chemical 
compositions.

Unlike the case for Na, the Plaut field [Al/Fe] abundances closely
follow the same trend as the halo and thick disk at all metallicities, and 
remain $\ga$0.1 dex more enhanced than the thin disk until [Fe/H]$>$0.  We note
also that for especially the thick disk there appears to be a roughly 0.1 dex 
offset between the literature disk and giant [Al/Fe] ratios, and the Plaut 
field abundances more closely follow those of the literature giant than 
dwarf trend.  Since the halo and metal--poor thick disk share nearly 
indistinguishable [Al/Fe] abundances, it is not surprising that there is a
similar degree of agreement between the different bulge fields.  The similarity 
in [Al/Fe] (but not [Na/Fe]) between these different populations supports our 
proposed scenario in which the halo more strongly influenced the initial 
composition of the outer bulge while the thick disk more strongly influenced
the inner bulge.  As mentioned in $\S$4.1.1 and seen in Figures 
\ref{f8}--\ref{f9}, the Plaut field and bulge dwarf data indicate a downturn 
in [Al/Fe] and follow the thick disk trend, but the other bulge literature 
abundances suggest that [Al/Fe] remains significantly enhanced.  Unfortunately,
adding the thin/thick disk data to this picture does not help solve the 
[Al/Fe] discrepancy at high metallicities.  Finally, in a similar 
fashion to the [Na/Fe] abundances, the [Al/Fe] ratios of the Plaut field clump 
stars once again follow the thin/thick disk distribution.

The origin of the three low [Na/Fe] and [Al/Fe] stars is not immediately clear.
We note that the stars are near the lower envelope of the halo [Na/Fe] and 
[Al/Fe] distributions, but it seems unlikely that three randomly chosen halo 
stars at [Fe/H]$\sim$--1 would all have [Na/Fe]$\la$--0.5 and 
[Al/Fe]$\la$--0.2.  Alternatively, one or all of these stars could have formed 
in the first generation of a now disrupted or tidally stripped globular 
cluster.  While first generation globular cluster stars are low in both 
[Na/Fe] and [Al/Fe], very few of these stars have [Al/Fe]$<$--0.1 and none 
really reach [Al/Fe]$\la$--0.2 (e.g., Carretta et al. 2009).  Interestingly, 
two of the stars have [La/Eu]$\approx$--0.30, which is consistent with the 
other bulge stars, the halo, and globular clusters.  However, the other star 
(2MASS 18182256--3401248) has [La/Eu]$\approx$$+$0.10, which suggests it may 
have a different origin than the other two low Na and Al stars.  While some 
dwarf galaxies also host stars with low [Na/Fe] and [Al/Fe] and enhanced 
[La/Eu] ratios (e.g., see review by Venn et al. 2004 and references therein), 
the enhanced [$\alpha$/Fe] ratios for these three stars makes
a dwarf galaxy origin unlikely.

The [Na/Al] plots in Figure \ref{f14} further suggest that there may be 
key composition differences between the thick disk and outer bulge.  For stars
with [Fe/H]$\la$--0.8, the [Na/Al] ratios of the Plaut field are clearly lower
than both the thick disk dwarf and giant data, and instead are more similar to 
those of the halo.  While there could be systematic differences affecting our 
data that could shift the points into better agreement with both the halo and 
thick disk, that is unlikely to flatten the [Na/Al] slope (though we note that
the rise in [Na/Al] is greater for the thick disk dwarf compared to giant 
data).  In contrast, the metal--poor inner bulge data from 
Figures \ref{f8}--\ref{f9} are a better match to the thick disk.  At higher 
metallicities ([Fe/H]$\ga$--0.2), the Plaut field data clearly extend to larger
[Na/Al] ratios than the average thin or thick disk star, but the inner disk red
clump stars generally follow the thick disk trend.  Interestingly, the shape 
of the [Na/Al] distribution may be sensitive to the stellar population.  In 
particular, the thin disk exhibits essentially no change in [Na/Al] as a 
function of metallicity, the thick disk and halo show steeper positive slopes, 
and the Plaut field has the steepest positive slope.  This may be a result
of increasing Na/Al yields in massive stars, with faster star formation and 
chemical enrichment producing a steeper relation versus iron.  It would be 
interesting if that were the case because the [Na/Al] ratio would be tracing a 
different timescale than the [La/Eu] ratio, which we have shown in 
Figure \ref{f12} as being identical at b=--4$\degr$ and b=--8$\degr$.

\subsubsection{The Heavy Elements}

\begin{figure}
\epsscale{0.80}
\plotone{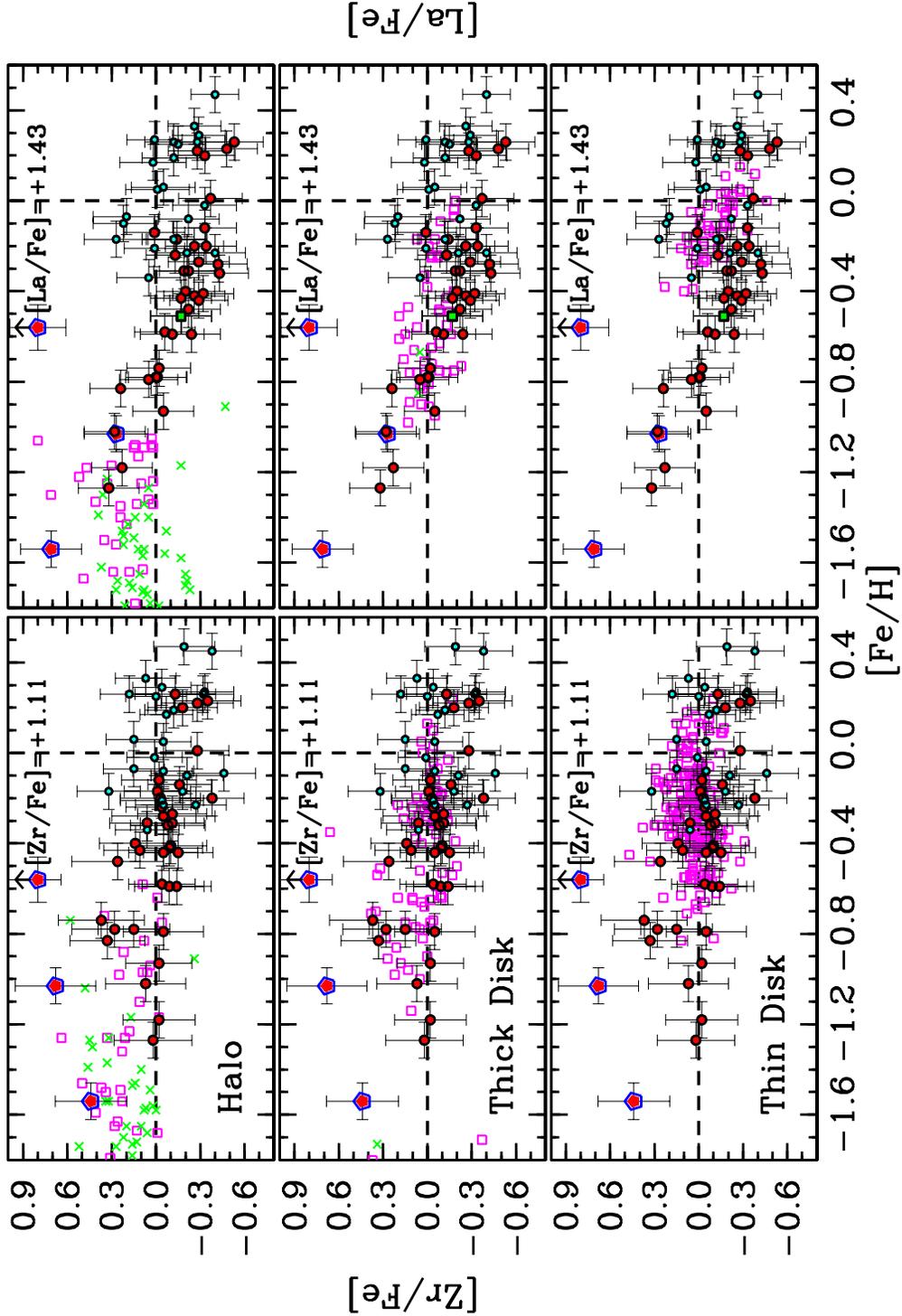}
\caption{Literature [Zr/Fe] and [La/Fe] data are plotted as a function of 
[Fe/H] in comparison with our bulge results.  The symbols are the same as those
in Figure \ref{f13}.  The halo data are from McWilliam et al. (1995), Ryan et 
al. (1996), Fulbright (2000), Johnson (2002), and Simmerer et al. (2004).  The 
thin and thick disk data are from Edvardsson et al. (1993), Reddy et al. 
(2003), and Brewer \& Carney (2006).}
\label{f16}
\end{figure}

In Figure \ref{f16} we plot [Zr/Fe] and [La/Fe] as a function of [Fe/H] for
the Plaut field, halo, thick, and thin disk.  While there is a global trend 
of decreasing [Zr/Fe] abundance with increasing [Fe/H], the various 
populations do not exhibit significantly different [Zr/Fe] trends.  This 
follows the results shown in Figure \ref{f11}, which indicate that the 
light neutron--capture element [Y/Fe] and [Zr/Fe] trends are nearly 
indistinguishable at different bulge locations.  Similarly, the inner disk
clump stars also exhibit little variation with [Fe/H] and agree with the thin 
and thick disk trends.  However, the halo stars at [Fe/H]$\la$--1 tend to have 
[Zr/Fe]$\ga$$+$0.1, on average, while the most metal--poor Plaut field stars 
have [Zr/Fe]$\sim$$+$0.00.  Given the small number of bulge [Zr/Fe] 
measurements at [Fe/H]$<$--1 and the larger errors associated with the measured
[Zr/Fe] ratios, it is difficult to determine whether this reflects a real 
difference.  Several of the bulge stars near [Fe/H]$\sim$--0.8
exhibit [Zr/Fe] enhancements that match the highest halo and thick disk values.

While there is some ambiguity regarding the [Zr/Fe] (and [Y/Fe]) trends, the
[La/Fe] distribution provides better differentiation among the various 
populations.  On average, the halo exhibits [La/Fe]$>$$+$0.00, but most of the 
thin and thick disk stars lie at [La/Fe]$\leq$$+$0.00.  Figure \ref{f16} shows 
that the most metal--poor Plaut field bulge giants are at least 
0.1--0.2 dex more La--enhanced than the thick disk and are instead much more 
similar to the halo composition.  Although the bulge [La/Fe] values overlap 
with the thick disk stars between --1.0$\la$[Fe/H]$\la$--0.6, the bulge stars 
with [Fe/H]$>$--0.6 are at least 0.1--0.3 dex lower than many of the thin and 
thick disk stars.  However, this discrepancy is smaller for the Baade's window 
data and may only be present at the 0.1--0.2 dex level.  While the [La/Fe] trend
for the thick disk slowly declines as a function of increasing [Fe/H], the 
bulge fields exhibit a sharper decline in [La/Fe] at [Fe/H]$>$--0.8.  In 
contrast, the inner disk clump stars once again exhibit large 
star--to--star abundance variations, but their [La/Fe] ratios are roughly 
similar to the thin and thick disk, on average.

\begin{figure}
\epsscale{0.80}
\plotone{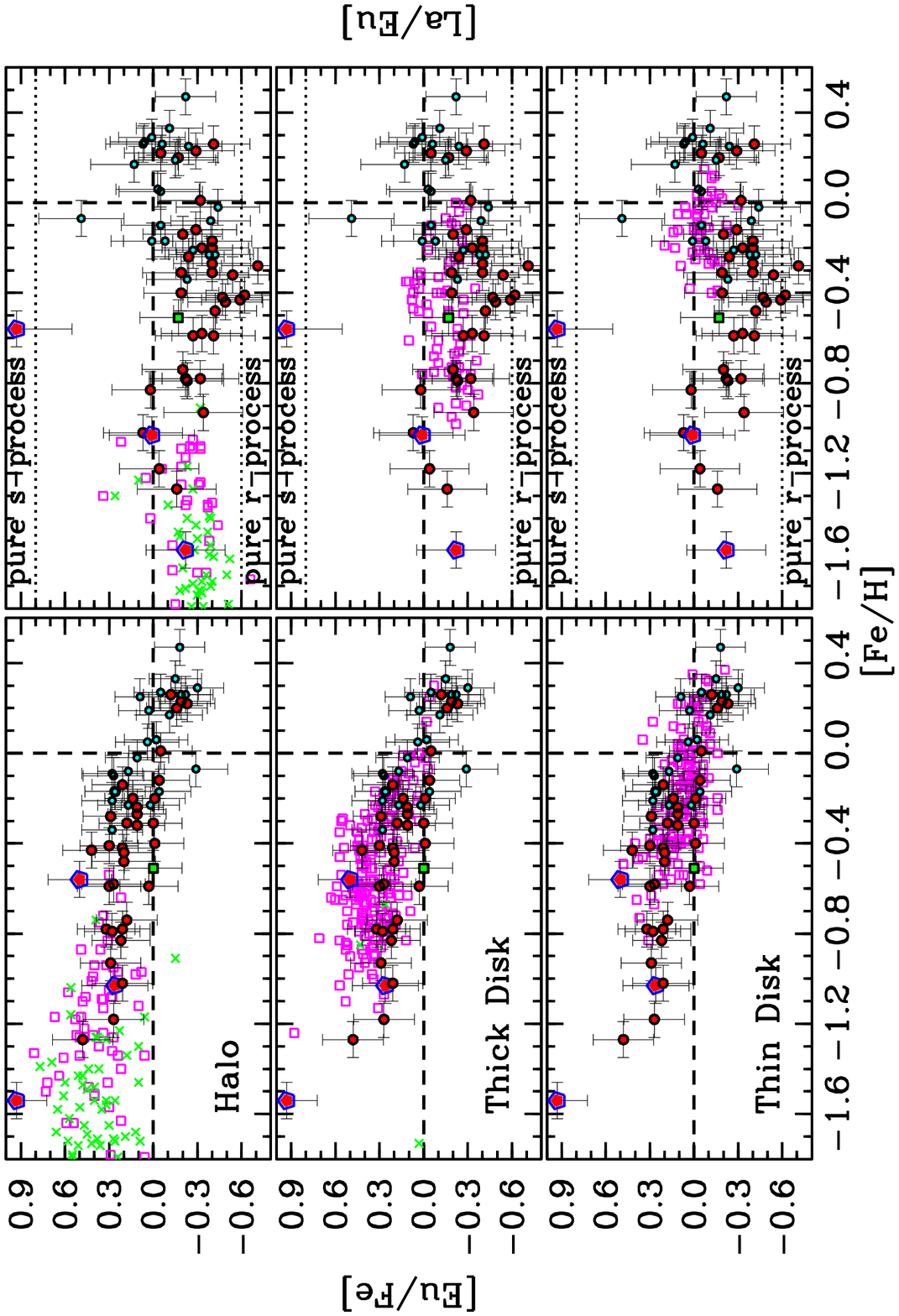}
\caption{Literature [Eu/Fe] and [La/Eu] data are plotted as a function of
[Fe/H] in comparison with our bulge results.  The symbols are the same as those
in Figure \ref{f13}.  The halo data are from McWilliam et al. (1995), Ryan et
al. (1996), Fulbright (2000), Johnson (2002), Simmerer et al. (2004), and
Reddy et al. (2006).  The thin and thick disk data are from Prochaska et al. 
(2000), Reddy et al. (2003), Bensby et al. (2005), and Brewer \& Carney 
(2006).}
\label{f17}
\end{figure}

In Figure \ref{f17} we plot [Eu/Fe] and [La/Eu] as a function of [Fe/H] for 
Plaut's field and the halo, thick disk, and thin disk.  As was mentioned in
$\S$4.1.2, the [Eu/Fe] distributions are indistinguishable between the 
halo, thick disk, and bulge for metallicities at which they overlap.  While 
the thin disk appears to also exhibit 
a similar [Eu/Fe] distribution, this is mostly a consequence of mixing 
results from the literature.  Individual, self--consistent studies (e.g., 
Bensby et al. 2005) tend to find that the thin disk stars are under--abundant
in [Eu/Fe], relative to the thick disk, by at least 0.1 dex, and we find that
the bulge [Eu/Fe] distribution is more similar to the halo and thick disk than
the thin disk.  We have also taken the position that the decrease in [Eu/Fe]
at [Fe/H]$\sim$--0.4, at least for the bulge, is more a reflection of the 
metallicity dependent yields and singular production source (core--collapse
SNe) of Eu than an indicator of Fe production from Type Ia SNe (see also
McWilliam et al. 2010).

As the right panels of Figure \ref{f17} indicate, the halo and disk data 
form a trend of slowly increasing [La/Eu] ratios with increasing metallicity.
This contrasts with the generally flat [La/Eu] trends for the Plaut and Baade's
window giants (see also Figure \ref{f12}), but we note that a large portion 
of the bulge and thick disk [La/Eu] data overlap.  In the metallicity range
over which the bulge and thin disk overlap, the thin disk [La/Eu] ratios
are, on average, at least 0.2 dex higher than the bulge values.  The enhanced
[La/Eu] ratios for the inner disk clump stars seem compatible with this trend.
We interpret this as evidence that the main s--process played a much larger 
role in especially the thin disk's chemical enrichment compared to the bulge.  
In comparison with the halo and thick disk, the enhanced [La/Eu] ratios in 
many of the metal--poor Plaut (and Baade's window) stars is not clear.  Only a
small percentage of halo and thick disk stars at [Fe/H]$\la$--1 have 
[La/Eu]$>$--0.20.  Therefore, we would not expect to find such a large 
percentage of bulge stars in our limited sample to have [La/Eu]$>$--0.20, 
if the bulge formed from a mixture of halo and thick disk gas.  A larger sample
of metal--poor bulge stars is needed to further address this issue.

\section{SUMMARY}

We have determined chemical abundances of Na, Al, Zr, La, Nd, and Eu 
for 39 RGB stars and an additional 23 potential red clump inner disk stars in 
Plaut's low extinction window.  We have also measured Li in one of the stars 
that was identified as being super Li--rich.  The abundances were measured from
high resolution (R$\approx$25,000), high S/N ($\sim$50--100 pixel$^{\rm -1}$)
spectra obtained with the Hydra multifiber spectrograph on the CTIO Blanco 4m
telescope.  The [Fe/H], [Si/Fe], and [Ca/Fe] abundances and the model 
atmosphere parameters were taken from the first paper in this series (Johnson
et al. 2011).  We developed a new computationally parallel version of the 
spectrum synthesis module for MOOG, and employed this code to derive all
abundances presented here.

At the metal--poor end, we find that the bulge RGB stars in Plaut's field have 
[Na/Fe]$\sim$--0.30 and [Al/Fe]$\sim$$+$0.20.  However, the [Na/Fe] ratios
increase with metallicity and reach a maximum of [Na/Fe]$\sim$$+$0.30 at 
[Fe/H]$\sim$--0.20.  The [Na/Fe] trend flattens out and remains enhanced to 
higher metallicities.  In contrast, the [Al/Fe] trend is nearly identical to
that of [$\alpha$/Fe].  In particular, the [Al/Fe] ratio remains relatively 
flat and at [Al/Fe]$\sim$$+$0.2 until [Fe/H]$\sim$--0.20, at which point there 
is a downturn in [Al/Fe] toward solar values.  Additionally, we do not find 
evidence supporting a significant [Na/Fe]--[Al/Fe] correlation for the bulge, 
as is found in most globular clusters.  However, three metal--poor stars in
our sample exhibit anomalously low [Na/Fe] and [Al/Fe] ratios.  Two of the 
stars have Na, Al, and heavy element abundance patterns that indicate a 
possible origin as first generation globular cluster stars.  The other star
shows an enhanced [La/Eu] ratio that is generally inconsistent with the 
r--process dominant composition of globular clusters.  Although the inner disk 
clump stars generally exhibit larger star--to--star abundance dispersions, 
presumably because of larger distance uncertainties, we find that they tend to 
share a similar [Al/Fe] trend with the bulge giants but have lower [Na/Fe] 
ratios, on average.

The single super Li--rich giant (log $\epsilon$(Li)=$+$3.51) discovered in our 
sample does not appear to exhibit any other unusual spectroscopic features.  
While it has [Al/Fe] and [Eu/Fe] abundances that are slightly below average, 
these values are well within the envelope of measurements for the other bulge 
giants.  The [La/Fe] and [Nd/Fe] ratios for this star are very consistent with 
similar metallicity stars.  By combining this star with other super Li--rich
bulge giants found in the literature, we estimate that the fraction of 
super Li--rich RGB stars in the bulge is $\sim$1$\%$.  This is similar to what
is found in the disk and globular clusters.

The light neutron--capture element Zr exhibits a trend of slightly decreasing 
[Zr/Fe] with increasing metallicity.  Interestingly, in the range 
--0.8$\la$[Fe/H]$\la$--0.4, roughly a third of the sample have
[Zr/Fe]$\ga$$+$0.1.  We speculate that these stars may have been systematically
enriched by the weak s-- and/or r--processes compared to the majority of bulge 
stars that have [Zr/Fe]$\la$$+$0.00.  For the heavier neutron--capture elements
La and Nd, the most metal--poor bulge stars are enhanced at
[La,Nd/Fe]$\sim$$+$0.2 with very little star--to--star scatter; however, both 
elements exhibit decreasing trends for [Fe/H]$>$--0.8.  The La and Nd 
distributions decline until reaching [La,Nd/Fe]$\sim$--0.3 in the 
highest metallicity stars.  The inner disk clump stars appear to be more 
enhanced in both La and Nd than the bulge giants.

The [Eu/Fe] abundance distribution for the bulge stars (and the clump stars at
higher metallicity) directly follows that of the [Al/Fe] and [$\alpha$/Fe] 
trends.  Eu is enhanced at [Eu/Fe]$\sim$$+$0.3 dex until [Fe/H]$\sim$--0.2, 
at which point the [Eu/Fe] trend declines toward solar and sub--solar values.
When we examine the [La/Eu] ratio as a function of metallicity, we find that
all but the most metal--poor bulge stars have [La/Eu]$\sim$--0.3.  This 
suggests that the Plaut field bulge stars formed rapidly and before the main 
s--process could become a significant pollution source.  The more enhanced
[La/Eu] ratios in some of the most metal--poor stars is intriguing, and may
be a chemical tag indicating pre--enrichment of the early bulge gas.  
Interestingly, the clump stars exhibit noticeably higher [La/Eu] ratios than
the similar metallicity bulge giants, which suggests these stars formed over
a longer timescale.

In addition to analyzing the abundance trends of the Plaut field stars, we also
compared our results with those from the literature spanning various bulge 
locations.  We generally find good agreement between the abundance trends in
different fields, and also between the dwarfs and giants.  In particular,
we find the following trends to be ``universal" among all bulge
studies: (1) increasing [Na/Fe] with [Fe/H], (2) enhanced [Al/Fe], 
(3) a rising [Na/Al] ratio with [Fe/H], (4) a mostly flat [Zr/Fe] (or [Y/Fe])
trend with a small percentage of Zr/Y--enhanced stars, (5) a trend of decreasing
[La/Fe] (or [Ba/Fe]) with increasing metallicity, (6) an $\alpha$--like 
distribution of [Eu/Fe], and (7) a flat distribution at [La/Eu]$\sim$--0.3 
with some enhancements at the metal--poor end.  Interestingly, stars in both
the b=--4$\degr$ and b=--8$\degr$ fields exhibit identical [La/Eu] 
distributions that seem to rule out the main s--process, which operates on 
long timescales, as a significant pollution source.  This result appears to 
conflict with the observation of a non--negligible population of young 
($\sim$3--4 Gyr old) microlensed dwarfs in the bulge.

Despite the general agreement, for Na there is some evidence that the 
metal--poor ([Fe/H]$\la$--0.8) inner bulge stars may be 0.1--0.3 dex more 
enhanced than the outer bulge stars.  Interestingly, this difference disappears
at [Fe/H]$\ga$--0.8, and we interpret this as a reflection of significant 
composition inhomogeneities in the early bulge.  For Al there is remarkable 
agreement that [Al/Fe]$\sim$$+$0.2 from [Fe/H]=--1.5 up to --0.2, but two 
different abundance trends emerge in the literature at higher metallicities: 
one set finds that Al remains enhanced at super--solar metallicities and the 
other set (which includes the present work) finds that Al decreases in 
abundance like Eu and the $\alpha$--elements.  At the moment there does not 
seem to be an obvious solution to this problem.

We also compare our derived abundance trends with similar literature data for 
the Galactic halo, thick disk, and thin disk.  While we find in agreement with
past studies that the bulge and thick disk share many composition similarities, 
we also find evidence of significant differences.  One of the elements showing 
the strongest difference between the bulge and thick disk is Na.  At the 
metal--poor end, the Plaut field stars have [Na/Fe] ratios that are at least 
0.3 dex lower than similar metallicity thick disk stars, and are instead more 
similar to those found in the halo.  Although there is significant overlap in 
[Na/Fe] between the thick disk and bulge at higher metallicities, the overall 
shape of the bulge Na distribution is different and extends to higher values of
[Na/Fe].  Additionally, we find that the enhanced [La/Fe] ratios
for the metal--poor bulge stars and the steep decline in [La/Fe] with
increasing metallicity does not match the thick disk trend.  Interestingly, 
the same is not true for Eu.  The bulge, halo, and thick disk all exhibit 
nearly identical [Eu/Fe] versus [Fe/H] trends.  We interpret this as evidence 
that the downturn in [Eu/Fe] at [Fe/H]$\sim$--0.4, which is common to the 
thick disk and bulge, is more a reflection of metallicity dependent r--process 
yields than a statement of the Type II/Type Ia SN ratio.  

\acknowledgements

CIJ would like to thank David Yong for several useful discussions, and would 
also like to thank Andrea Kunder, Livia Origlia, Nils Ryde, Elena Valenti, 
Thomas Bensby, Dave Arnett, and Alvio Renzini for many useful discussions at 
the Aspen Center for Physics, which is supported by the National Science 
Foundation under award No. 1066293.  We thank Andy McWilliam for many useful 
discussions and for providing data from McWilliam et al. (2010).  We would like
to thank David Reitzel for obtaining a portion of these observations.  This 
research has made use of NASA's Astrophysics Data System Bibliographic 
Services.  This material is based upon work supported by the National Science 
Foundation under award No. AST--1003201 to CIJ and award No. AST--0709479 to 
RMR.

\clearpage

\clearpage

\clearpage

\clearpage

\clearpage

\clearpage

\clearpage

\clearpage

\clearpage

\clearpage

\clearpage

\clearpage

\clearpage

\clearpage

\clearpage

\clearpage

\clearpage

\clearpage

\tablenum{1}
\tablecolumns{5}
\tablewidth{0pt}

\begin{deluxetable}{ccccc}
\tablecaption{Linelist and Reference Arcturus Abundances}
\tablehead{
\colhead{Element}	&
\colhead{Wavelength}	&
\colhead{E.P.}	&
\colhead{log gf}	&
\colhead{[X/Fe]}	\\
\colhead{}      &
\colhead{(\AA)}	&
\colhead{(eV)}	&
\colhead{}	&
\colhead{Arcturus\tablenotemark{a}}
}

\startdata
Li I	&	6707.768	&	0.000	&	$+$0.17\tablenotemark{b}	&	\nodata	\\
Na I	&	6154.226	&	2.101	&	$-$1.55	&	$+$0.09	\\
Na I	&	6160.747	&	2.103	&	$-$1.17	&	$+$0.09	\\
Al I	&	6696.023	&	3.140	&	$-$1.57	&	$+$0.38	\\
Al I    &       6698.673        &       3.140   &       $-$1.96 &       $+$0.38
\\
Zr I	&	6134.545	&	0.000	&	$-$1.34	&	$+$0.00	\\
Zr I	&	6140.450	&	0.519	&	$-$1.50	&	$+$0.00	\\
Zr I	&	6143.200	&	0.071	&	$-$1.20	&	$+$0.00
\\
La II	&	6262.290	&	0.403	&	$-$1.22\tablenotemark{b}	&	$-$0.06	\\
Nd II   &       6740.078        &       0.060   &       $-$1.77 &       $+$0.05 \\
Eu II	&	6645.060	&	1.379	&	$+$0.12\tablenotemark{b}	&	$+$0.29	\\
\enddata

\tablenotetext{a}{The assumed Arcturus iron abundance is [Fe/H]=--0.50.}
\tablenotetext{b}{The numbers listed represent the total log gf values for each
line.  The individual components can be found in Hobbs et al. (1999) for Li, 
Lawler et al. (2001a) for La, and Lawler et al. (2001b) for Eu.  Note that the 
total log gf value for Li listed above only includes the \iso{7}{Li} lines in 
Hobbs et al. (1999).}

\end{deluxetable}

\clearpage

\tablenum{2}
\tablecolumns{12}
\tablewidth{0pt}

\begin{deluxetable}{ccccccccccccc}
\rotate
\tabletypesize{\scriptsize}
\tablecaption{Model Atmosphere Parameters and Abundance Ratios}
\tablehead{
\colhead{Star}	&
\colhead{T$_{\rm eff}$}	&
\colhead{log g}	&
\colhead{[Fe/H]}	&
\colhead{V$_{\rm t}$}	&
\colhead{[Na/Fe]}      &
\colhead{[Al/Fe]}      &
\colhead{[Zr/Fe]}      &
\colhead{[La/Fe]}      &
\colhead{[Nd/Fe]}      &
\colhead{[Eu/Fe]}      &
\colhead{RGB/Field\tablenotemark{a}}	\\
\colhead{2MASS}	&
\colhead{(K)}	&
\colhead{(cgs)}	&
\colhead{}      &
\colhead{(km s$^{\rm -1}$)}      &
\colhead{}      &
\colhead{}      &
\colhead{}      &
\colhead{}      &
\colhead{}      &
\colhead{}      &
\colhead{}      &
\colhead{}      
}

\startdata
18174532$-$3353235	&	4540	&	1.45	&	$-$1.54	&	1.65	&	$-$0.31	&	$+$0.12	&	$+$0.44	&	$+$0.71	&	$+$0.76	&	$+$0.93	&	RGB/1	\\
18182918$-$3341405	&	4125	&	0.95	&	$-$1.27	&	1.60	&	$-$0.22	&	$+$0.25	&	$+$0.02	&	$+$0.32	&	$+$0.36	&	$+$0.48	&	RGB/1	\\
18175567$-$3343063	&	4425	&	1.30	&	$-$1.18	&	1.15	&	$-$0.33	&	$+$0.28	&	$-$0.02	&	$+$0.23	&	$+$0.28	&	$+$0.27	&	RGB/1	\\
18181521$-$3352294	&	4215	&	1.00	&	$-$1.03	&	1.55	&	$+$0.03	&	$+$0.58	&	$+$0.68	&	$+$0.27	&	$+$0.38	&	$+$0.26	&	RGB/1	\\
18182256$-$3401248	&	4465	&	1.10	&	$-$1.02	&	1.15	&	$-$0.58	&	$-$0.23	&	$+$0.07	&	$+$0.28	&	$+$0.27	&	$+$0.21	&	RGB/1	\\
18174351$-$3401412	&	4450	&	1.35	&	$-$0.93	&	1.25	&	$-$0.70	&	$-$0.52	&	$-$0.02	&	$-$0.05	&	$+$0.30	&	$+$0.29	&	RGB/1	\\
18172965$-$3402573	&	4155	&	1.35	&	$-$0.83	&	1.80	&	$-$0.26	&	$+$0.19	&	$+$0.33	&	$+$0.24	&	$+$0.25	&	$+$0.22	&	RGB/1	\\
18183521$-$3344124	&	4150	&	1.25	&	$-$0.79	&	1.85	&	$-$0.51	&	$-$0.19	&	$-$0.05	&	$+$0.05	&	$+$0.09	&	$+$0.28	&	RGB/1	\\
18181435$-$3350275	&	4340	&	1.20	&	$-$0.78	&	1.25	&	$-$0.40	&	$+$0.13	&	$+$0.15	&	$-$0.01	&	$+$0.06	&	$+$0.21	&	RGB/1	\\
18183876$-$3403092	&	4000	&	1.05	&	$-$0.78	&	1.80	&	$+$0.00	&	$+$0.27	&	$+$0.28	&	$+$0.00	&	$-$0.02	&	$+$0.32	&	RGB/1	\\
18174304$-$3357006	&	4090	&	1.00	&	$-$0.74	&	1.45	&	$-$0.21	&	$+$0.19	&	$+$0.37	&	$-$0.02	&	$+$0.21	&	$+$0.18	&	RGB/1	\\
18180550$-$3407117	&	4335	&	1.25	&	$-$0.59	&	1.45	&	$-$0.01	&	$+$0.27	&	$-$0.09	&	$-$0.24	&	$-$0.02	&	$+$0.03	&	RGB/1	\\
18180831$-$3405309	&	4100	&	1.00	&	$-$0.59	&	1.95	&	$+$0.14	&	$+$0.30	&	$-$0.14	&	$-$0.11	&	$-$0.01	&	$+$0.30	&	RGB/1	\\
18174941$-$3353025	&	4275	&	1.40	&	$-$0.58	&	1.95	&	$-$0.16	&	$+$0.18	&	$-$0.04	&	$-$0.06	&	$+$0.01	&	$+$0.27	&	RGB/1	\\
18174742$-$3348098	&	4130	&	1.45	&	$-$0.56	&	1.45	&	$-$0.24	&	$+$0.11	&	$+$1.11	&	$+$1.43	&	$+$1.26	&	$+$0.50	&	RGB/1	\\
18183679$-$3251454	&	4025	&	1.10	&	$-$0.51	&	2.00	&	\nodata	&	$-$0.05	&	\nodata	&	$-$0.17	&	$+$0.06	&	$+$0.00	&	RGB/2	\\
18181929$-$3404128	&	4020	&	0.90	&	$-$0.48	&	1.30	&	$+$0.01	&	$+$0.16	&	$+$0.26	&	$-$0.22	&	$+$0.08	&	$+$0.20	&	RGB/1	\\
18183802$-$3355441	&	4325	&	1.25	&	$-$0.44	&	1.75	&	$-$0.06	&	$+$0.19	&	$-$0.15	&	$-$0.29	&	$-$0.06	&	$+$0.20	&	RGB/1	\\
18181512$-$3353545	&	4270	&	1.45	&	$-$0.44	&	1.65	&	$-$0.26	&	\nodata	&	$-$0.05	&	\nodata	&	\nodata	&	\nodata	&	RGB/1	\\
18174303$-$3355118	&	4220	&	1.30	&	$-$0.43	&	1.60	&	$-$0.17	&	$+$0.21	&	$+$0.11	&	$-$0.17	&	$+$0.08	&	$+$0.42	&	RGB/1	\\
18182470$-$3342166	&	4145	&	1.10	&	$-$0.42	&	1.60	&	$+$0.14	&	$+$0.37	&	$-$0.09	&	$-$0.26	&	$-$0.06	&	$+$0.21	&	RGB/1	\\
18180285$-$3342004	&	4140	&	1.30	&	$-$0.41	&	1.45	&	$-$0.22	&	$+$0.08	&	$-$0.10	&	$-$0.32	&	$-$0.07	&	$+$0.30	&	RGB/1	\\
18174935$-$3404217	&	4360	&	1.25	&	$-$0.40	&	1.70	&	$+$0.30	&	$+$0.37	&	$+$0.14	&	$-$0.20	&	$+$0.03	&	$-$0.01	&	RGB/1	\\
18174929$-$3347192	&	4875	&	2.35	&	$-$0.34	&	1.40	&	$-$0.11	&	$+$0.15	&	$+$0.06	&	$+$0.05	&	$-$0.31	&	$+$0.28	&	Clump/1	\\
18183604$-$3342349	&	4105	&	0.80	&	$-$0.32	&	1.90	&	$+$0.32	&	$+$0.26	&	$-$0.08	&	$-$0.43	&	$-$0.18	&	$+$0.11	&	RGB/1	\\
18173554$-$3405009	&	4625	&	2.10	&	$-$0.31	&	1.60	&	$+$0.03	&	$+$0.08	&	$-$0.11	&	$-$0.19	&	$+$0.04	&	$+$0.00	&	RGB/1	\\
18180562$-$3346548	&	4220	&	1.30	&	$-$0.31	&	1.75	&	$+$0.38	&	$+$0.44	&	$+$0.06	&	$-$0.22	&	$-$0.14	&	$+$0.18	&	RGB/1	\\
18173180$-$3349197	&	4005	&	1.15	&	$-$0.28	&	1.25	&	$-$0.01	&	$+$0.09	&	$-$0.05	&	$-$0.42	&	$+$0.03	&	$+$0.29	&	RGB/1	\\
18181924$-$3350222	&	4215	&	1.35	&	$-$0.27	&	1.55	&	$+$0.20	&	$+$0.27	&	$-$0.11	&	$-$0.29	&	$-$0.33	&	$+$0.11	&	RGB/1	\\
18175593$-$3400000	&	4160	&	1.20	&	$-$0.24	&	1.40	&	$-$0.03	&	$+$0.12	&	$-$0.04	&	$-$0.13	&	$+$0.14	&	$+$0.11	&	RGB/1	\\
18182089$-$3348425	&	4700	&	2.15	&	$-$0.23	&	1.45	&	$-$0.24	&	$-$0.02	&	$-$0.27	&	$-$0.40	&	$-$0.17	&	$+$0.02	&	Clump/1	\\
18175546$-$3404103	&	4950	&	2.40	&	$-$0.23	&	1.40	&	$-$0.28	&	$+$0.08	&	$-$0.05	&	$-$0.21	&	$+$0.21	&	$+$0.17	&	Clump/1	\\
18174798$-$3359361	&	4830	&	2.30	&	$-$0.21	&	1.75	&	$-$0.11	&	$+$0.13	&	$-$0.04	&	$+$0.01	&	$+$0.24	&	$+$0.28	&	Clump/1	\\
18182430$-$3352453	&	4130	&	1.05	&	$-$0.20	&	1.40	&	$+$0.16	&	$+$0.25	&	$-$0.03	&	$-$0.34	&	$-$0.10	&	$-$0.01	&	RGB/1	\\
18180979$-$3351416	&	4350	&	1.65	&	$-$0.20	&	1.70	&	$-$0.10	&	$+$0.01	&	$-$0.38	&	$-$0.26	&	$-$0.20	&	$+$0.14	&	RGB/1	\\
18182553$-$3349465	&	4660	&	2.00	&	$-$0.17	&	1.75	&	$-$0.01	&	$+$0.07	&	$-$0.18	&	$-$0.12	&	$+$0.09	&	$-$0.04	&	Clump/1	\\
18173251$-$3354539	&	4565	&	1.85	&	$-$0.17	&	2.10	&	$+$0.46	&	$+$0.41	&	$-$0.01	&	$-$0.14	&	$-$0.18	&	$+$0.26	&	RGB/1	\\
18181710$-$3401088	&	4940	&	2.40	&	$-$0.17	&	1.45	&	$+$0.28	&	$+$0.26	&	$+$0.32	&	$+$0.27	&	$+$0.22	&	$+$0.26	&	Clump/1	\\
18180991$-$3403206	&	4345	&	1.80	&	$-$0.14	&	1.75	&	$+$0.14	&	$+$0.2	&	$-$0.16	&	$+$0.01	&	$+$0.14	&	$+$0.21	&	RGB/1	\\
18180301$-$3405313	&	4040	&	1.00	&	$-$0.12	&	2.05	&	$+$0.32	&	$+$0.17	&	$-$0.02	&	$-$0.33	&	$-$0.28	&	$-$0.04	&	RGB/1	\\
18180502$-$3355071	&	4715	&	2.35	&	$-$0.10	&	1.60	&	$-$0.04	&	$+$0.14	&	$-$0.21	&	$+$0.22	&	$+$0.15	&	$+$0.27	&	Clump/1	\\
18182740$-$3356447	&	4690	&	2.35	&	$-$0.09	&	1.55	&	$-$0.41	&	$-$0.01	&	$-$0.46	&	\nodata	&	\nodata	&	$+$0.28	&	Clump/1	\\
18182457$-$3344533	&	4530	&	2.10	&	$-$0.08	&	1.90	&	$+$0.15	&	$+$0.09	&	$-$0.05	&	$-$0.22	&	$-$0.32	&	$+$0.17	&	Clump/1	\\
18182612$-$3353431	&	4860	&	2.40	&	$-$0.07	&	1.65	&	$+$0.24	&	$+$0.19	&	$+$0.15	&	$+$0.20	&	$+$0.07	&	$-$0.29	&	Clump/1	\\
18172979$-$3401118	&	4860	&	2.40	&	$-$0.02	&	1.40	&	$+$0.02	&	$+$0.09	&	$+$0.01	&	$-$0.33	&	$+$0.17	&	$+$0.11	&	Clump/1	\\
18183930$-$3353425	&	4200	&	1.40	&	$+$0.01	&	1.75	&	$+$0.33	&	$+$0.12	&	$-$0.28	&	$-$0.37	&	$-$0.36	&	$-$0.05	&	RGB/1	\\
18174891$-$3406031	&	4805	&	2.40	&	$+$0.05	&	1.70	&	$-$0.07	&	$+$0.02	&	$-$0.05	&	$-$0.01	&	$+$0.12	&	$+$0.04	&	Clump/1	\\
18181322$-$3402227	&	4740	&	2.40	&	$+$0.06	&	1.50	&	$+$0.04	&	$+$0.08	&	$+$0.15	&	$-$0.05	&	$+$0.14	&	$-$0.02	&	Clump/1	\\
18181033$-$3352390	&	4900	&	2.50	&	$+$0.17	&	1.30	&	$-$0.06	&	$-$0.08	&	$-$0.07	&	$+$0.02	&	$+$0.18	&	$-$0.11	&	Clump/1	\\
18174000$-$3406266	&	4565	&	2.30	&	$+$0.19	&	1.85	&	$+$0.15	&	$+$0.11	&	$-$0.12	&	$-$0.12	&	$-$0.14	&	$+$0.03	&	Clump/1	\\
18180012$-$3358096	&	4090	&	1.50	&	$+$0.20	&	1.95	&	$+$0.21	&	$-$0.06	&	$-$0.18	&	$-$0.33	&	$-$0.14	&	$-$0.16	&	RGB/1	\\
18175652$-$3347050	&	4185	&	1.65	&	$+$0.22	&	1.40	&	$-$0.22	&	$-$0.10	&	$-$0.28	&	$-$0.28	&	$-$0.12	&	$-$0.23	&	RGB/1	\\
18182472$-$3352044	&	4070	&	1.45	&	$+$0.23	&	1.80	&	$+$0.40	&	$-$0.11	&	$-$0.35	&	$-$0.48	&	$-$0.43	&	$-$0.19	&	RGB/1	\\
18173706$-$3405569	&	4405	&	2.20	&	$+$0.25	&	2.10	&	$+$0.56	&	$+$0.25	&	$+$0.00	&	$-$0.15	&	$-$0.25	&	$+$0.09	&	Clump/1	\\
18182073$-$3353250	&	4500	&	2.30	&	$+$0.26	&	1.55	&	$-$0.19	&	$-$0.05	&	$-$0.32	&	$-$0.28	&	$+$0.06	&	$-$0.22	&	Clump/1	\\
18174478$-$3343290	&	4785	&	2.40	&	$+$0.26	&	1.85	&	$+$0.46	&	$+$0.19	&	$+$0.18	&	$-$0.12	&	$-$0.31	&	$-$0.19	&	Clump/1	\\
18173994$-$3358331	&	4085	&	1.45	&	$+$0.26	&	1.65	&	$+$0.44	&	$+$0.03	&	$-$0.13	&	$-$0.53	&	$-$0.31	&	$-$0.12	&	RGB/1	\\
18183369$-$3352038	&	4585	&	2.35	&	$+$0.27	&	1.55	&	$-$0.26	&	$-$0.15	&	$-$0.33	&	$+$0.01	&	$-$0.20	&	$-$0.05	&	Clump/1	\\
18183098$-$3358070	&	4570	&	2.35	&	$+$0.29	&	1.65	&	$+$0.37	&	$+$0.04	&	$-$0.04	&	$-$0.29	&	$-$0.17	&	$-$0.30	&	Clump/1	\\
18174067$-$3356000	&	4540	&	2.35	&	$+$0.33	&	1.70	&	$+$0.40	&	$+$0.11	&	$+$0.07	&	$-$0.26	&	$-$0.16	&	$-$0.15	&	Clump/1	\\
18173118$-$3358318	&	4355	&	2.00	&	$+$0.45	&	1.85	&	$+$0.19	&	$+$0.03	&	$-$0.38	&	\nodata	&	\nodata	&	\nodata	&	Clump/1	\\
18182052$-$3345251	&	4505	&	2.30	&	$+$0.47	&	1.75	&	$+$0.04	&	$-$0.12	&	$-$0.19	&	$-$0.40	&	$-$0.22	&	$-$0.18	&	Clump/1	\\
\enddata

\tablenotetext{a}{``RGB": probable RGB member; ``Clump": probable red clump
member; `1': Field 1 (l=--1$\degr$,b=--8.5$\degr$); `2': Field 2
(l=0$\degr$,b=--8$\degr$); see Johnson et al. (2011) for further details}

\end{deluxetable}

\clearpage

\tablenum{3}
\tablecolumns{7}
\tablewidth{0pt}

\begin{deluxetable}{ccccccc}
\tablecaption{Error Estimation}
\tablehead{
\colhead{Star}  &
\colhead{$\Delta$Na}	&
\colhead{$\Delta$Al}    &
\colhead{$\Delta$Zr}    &
\colhead{$\Delta$La}    &
\colhead{$\Delta$Nd}    &
\colhead{$\Delta$Eu}    
}

\startdata
\hline
\multicolumn{7}{c}{$\Delta$T$_{\rm eff}$$+$100 K}	\\
\hline
18174532$-$3353235	&	$+$0.08	&	$+$0.08	&	$+$0.20	&	$+$0.02	&	$+$0.04	&	$+$0.00	\\
18182918$-$3341405	&	$+$0.08	&	$+$0.08	&	$+$0.24	&	$+$0.01	&	$+$0.04	&	$-$0.02	\\
18175567$-$3343063	&	$+$0.08	&	$+$0.07	&	$+$0.21	&	$+$0.01	&	$+$0.03	&	$-$0.02	\\
18181521$-$3352294	&	$+$0.09	&	$+$0.09	&	$+$0.24	&	$+$0.02	&	$+$0.04	&	$-$0.02	\\
18182256$-$3401248	&	$+$0.08	&	$+$0.08	&	$+$0.21	&	$+$0.01	&	$+$0.04	&	$-$0.01	\\
18174351$-$3401412	&	$+$0.08	&	$+$0.07	&	$+$0.21	&	$+$0.01	&	$+$0.03	&	$-$0.02	\\
18172965$-$3402573	&	$+$0.08	&	$+$0.07	&	$+$0.22	&	$+$0.02	&	$+$0.04	&	$-$0.02	\\
18183521$-$3344124	&	$+$0.08	&	$+$0.07	&	$+$0.22	&	$+$0.02	&	$+$0.04	&	$-$0.02	\\
18181435$-$3350275	&	$+$0.08	&	$+$0.07	&	$+$0.21	&	$+$0.01	&	$+$0.03	&	$-$0.03	\\
18183876$-$3403092	&	$+$0.08	&	$+$0.07	&	$+$0.22	&	$+$0.02	&	$+$0.04	&	$-$0.02	\\
18174304$-$3357006	&	$+$0.09	&	$+$0.07	&	$+$0.21	&	$+$0.01	&	$+$0.04	&	$-$0.03	\\
18180831$-$3405309	&	$+$0.08	&	$+$0.07	&	$+$0.21	&	$+$0.01	&	$+$0.04	&	$-$0.02	\\
18180550$-$3407117	&	$+$0.08	&	$+$0.07	&	$+$0.21	&	$+$0.01	&	$+$0.03	&	$-$0.02	\\
18174941$-$3353025	&	$+$0.08	&	$+$0.07	&	$+$0.21	&	$+$0.01	&	$+$0.03	&	$-$0.02	\\
18174742$-$3348098	&	$+$0.08	&	$+$0.07	&	$+$0.17	&	$+$0.02	&	$+$0.04	&	$-$0.02	\\
18183679$-$3251454	&	\nodata	&	$+$0.07	&	\nodata	&	$+$0.02	&	$+$0.04	&	$-$0.02	\\
18181929$-$3404128	&	$+$0.07	&	$+$0.07	&	$+$0.19	&	$+$0.02	&	$+$0.04	&	$-$0.02	\\
18181512$-$3353545	&	$+$0.08	&	\nodata	&	$+$0.21	&	\nodata	&	\nodata	&	\nodata	\\
18183802$-$3355441	&	$+$0.08	&	$+$0.07	&	$+$0.20	&	$+$0.01	&	$+$0.03	&	$-$0.02	\\
18174303$-$3355118	&	$+$0.08	&	$+$0.07	&	$+$0.20	&	$+$0.01	&	$+$0.03	&	$-$0.02	\\
18182470$-$3342166	&	$+$0.08	&	$+$0.07	&	$+$0.19	&	$+$0.01	&	$+$0.04	&	$-$0.02	\\
18180285$-$3342004	&	$+$0.08	&	$+$0.06	&	$+$0.19	&	$+$0.02	&	$+$0.04	&	$-$0.02	\\
18174935$-$3404217	&	$+$0.07	&	$+$0.07	&	$+$0.20	&	$+$0.01	&	$+$0.03	&	$-$0.02	\\
18174929$-$3347192	&	$+$0.06	&	$+$0.06	&	$+$0.17	&	$+$0.01	&	$+$0.02	&	$-$0.02	\\
18183604$-$3342349	&	$+$0.08	&	$+$0.07	&	$+$0.19	&	$+$0.01	&	$+$0.04	&	$-$0.02	\\
18180562$-$3346548	&	$+$0.07	&	$+$0.06	&	$+$0.20	&	$+$0.01	&	$+$0.03	&	$-$0.02	\\
18173554$-$3405009	&	$+$0.08	&	$+$0.07	&	$+$0.18	&	$+$0.01	&	$+$0.03	&	$-$0.02	\\
18173180$-$3349197	&	$+$0.07	&	$+$0.06	&	$+$0.17	&	$+$0.01	&	$+$0.04	&	$-$0.02	\\
18181924$-$3350222	&	$+$0.07	&	$+$0.06	&	$+$0.19	&	$+$0.01	&	$+$0.03	&	$-$0.02	\\
18175593$-$3400000	&	$+$0.07	&	$+$0.06	&	$+$0.17	&	$+$0.01	&	$+$0.03	&	$-$0.02	\\
18182089$-$3348425	&	$+$0.07	&	$+$0.07	&	$+$0.18	&	$+$0.01	&	$+$0.03	&	$-$0.02	\\
18175546$-$3404103	&	$+$0.07	&	$+$0.06	&	$+$0.16	&	$+$0.01	&	$+$0.03	&	$-$0.01	\\
18174798$-$3359361	&	$+$0.07	&	$+$0.06	&	$+$0.17	&	$+$0.01	&	$+$0.03	&	$-$0.02	\\
18180979$-$3351416	&	$+$0.07	&	$+$0.06	&	$+$0.19	&	$+$0.01	&	$+$0.03	&	$-$0.02	\\
18182430$-$3352453	&	$+$0.07	&	$+$0.06	&	$+$0.17	&	$+$0.01	&	$+$0.04	&	$-$0.02	\\
18173251$-$3354539	&	$+$0.08	&	$+$0.07	&	$+$0.18	&	$+$0.01	&	$+$0.03	&	$-$0.02	\\
18181710$-$3401088	&	$+$0.07	&	$+$0.06	&	$+$0.17	&	$+$0.01	&	$+$0.02	&	$-$0.01	\\
18182553$-$3349465	&	$+$0.07	&	$+$0.07	&	$+$0.18	&	$+$0.01	&	$+$0.03	&	$-$0.02	\\
18180991$-$3403206	&	$+$0.07	&	$+$0.06	&	$+$0.18	&	$+$0.01	&	$+$0.04	&	$-$0.02	\\
18180301$-$3405313	&	$+$0.07	&	$+$0.06	&	$+$0.16	&	$+$0.01	&	$+$0.03	&	$-$0.02	\\
18180502$-$3355071	&	$+$0.08	&	$+$0.07	&	$+$0.17	&	$+$0.01	&	$+$0.03	&	$-$0.01	\\
18182740$-$3356447	&	$+$0.07	&	$+$0.07	&	$+$0.17	&	\nodata	&	\nodata	&	$-$0.02	\\
18182457$-$3344533	&	$+$0.08	&	$+$0.07	&	$+$0.18	&	$+$0.01	&	$+$0.03	&	$-$0.02	\\
18182612$-$3353431	&	$+$0.07	&	$+$0.06	&	$+$0.16	&	$+$0.01	&	$+$0.03	&	$-$0.01	\\
18172979$-$3401118	&	$+$0.07	&	$+$0.06	&	$+$0.17	&	$+$0.01	&	$+$0.03	&	$-$0.02	\\
18183930$-$3353425	&	$+$0.06	&	$+$0.05	&	$+$0.17	&	$+$0.01	&	$+$0.03	&	$-$0.02	\\
18174891$-$3406031	&	$+$0.07	&	$+$0.07	&	$+$0.17	&	$+$0.01	&	$+$0.03	&	$-$0.02	\\
18181322$-$3402227	&	$+$0.08	&	$+$0.07	&	$+$0.17	&	$+$0.01	&	$+$0.03	&	$-$0.01	\\
18181033$-$3352390	&	$+$0.07	&	$+$0.06	&	$+$0.16	&	$+$0.01	&	$+$0.03	&	$-$0.02	\\
18174000$-$3406266	&	$+$0.07	&	$+$0.06	&	$+$0.17	&	$+$0.01	&	$+$0.03	&	$-$0.02	\\
18180012$-$3358096	&	$+$0.05	&	$+$0.04	&	$+$0.13	&	$+$0.01	&	$+$0.03	&	$-$0.02	\\
18175652$-$3347050	&	$+$0.05	&	$+$0.04	&	$+$0.15	&	$+$0.02	&	$+$0.03	&	$-$0.02	\\
18182472$-$3352044	&	$+$0.05	&	$+$0.03	&	$+$0.14	&	$+$0.01	&	$+$0.03	&	$-$0.02	\\
18173706$-$3405569	&	$+$0.07	&	$+$0.05	&	$+$0.16	&	$+$0.02	&	$+$0.04	&	$-$0.01	\\
18173994$-$3358331	&	$+$0.04	&	$+$0.03	&	$+$0.12	&	$+$0.01	&	$+$0.03	&	$-$0.02	\\
18182073$-$3353250	&	$+$0.08	&	$+$0.06	&	$+$0.17	&	$+$0.01	&	$+$0.03	&	$-$0.02	\\
18174478$-$3343290	&	$+$0.08	&	$+$0.07	&	$+$0.17	&	$+$0.01	&	$+$0.03	&	$-$0.02	\\
18183369$-$3352038	&	$+$0.08	&	$+$0.06	&	$+$0.17	&	$+$0.02	&	$+$0.03	&	$-$0.01	\\
18183098$-$3358070	&	$+$0.06	&	$+$0.06	&	$+$0.17	&	$+$0.01	&	$+$0.03	&	$-$0.02	\\
18174067$-$3356000	&	$+$0.06	&	$+$0.06	&	$+$0.17	&	$+$0.01	&	$+$0.03	&	$-$0.02	\\
18173118$-$3358318	&	$+$0.05	&	$+$0.04	&	$+$0.16	&	\nodata	&	\nodata	&	\nodata	\\
18182052$-$3345251	&	$+$0.07	&	$+$0.05	&	$+$0.17	&	$+$0.01	&	$+$0.03	&	$-$0.02	\\
\hline
\multicolumn{7}{c}{$\Delta$log(g)$+$0.3 dex}	\\
\hline
18174532$-$3353235	&	$-$0.01	&	$-$0.01	&	$-$0.03	&	$+$0.13	&	$+$0.13	&	$+$0.13	\\
18182918$-$3341405	&	$-$0.01	&	$+$0.00	&	$+$0.00	&	$+$0.13	&	$+$0.13	&	$+$0.14	\\
18175567$-$3343063	&	$-$0.01	&	$+$0.00	&	$+$0.00	&	$+$0.13	&	$+$0.13	&	$+$0.13	\\
18181521$-$3352294	&	$-$0.01	&	$+$0.00	&	$+$0.01	&	$+$0.14	&	$+$0.14	&	$+$0.15	\\
18182256$-$3401248	&	$-$0.01	&	$-$0.01	&	$+$0.00	&	$+$0.13	&	$+$0.13	&	$+$0.13	\\
18174351$-$3401412	&	$-$0.01	&	$+$0.00	&	$+$0.00	&	$+$0.13	&	$+$0.13	&	$+$0.13	\\
18172965$-$3402573	&	$-$0.01	&	$+$0.00	&	$+$0.02	&	$+$0.13	&	$+$0.13	&	$+$0.13	\\
18183521$-$3344124	&	$-$0.01	&	$+$0.00	&	$+$0.02	&	$+$0.12	&	$+$0.12	&	$+$0.12	\\
18181435$-$3350275	&	$-$0.01	&	$-$0.01	&	$+$0.02	&	$+$0.12	&	$+$0.12	&	$+$0.11	\\
18183876$-$3403092	&	$+$0.00	&	$+$0.01	&	$+$0.03	&	$+$0.13	&	$+$0.13	&	$+$0.13	\\
18174304$-$3357006	&	$+$0.00	&	$+$0.01	&	$+$0.02	&	$+$0.14	&	$+$0.14	&	$+$0.14	\\
18180831$-$3405309	&	$+$0.01	&	$+$0.02	&	$+$0.01	&	$+$0.14	&	$+$0.13	&	$+$0.15	\\
18180550$-$3407117	&	$-$0.01	&	$-$0.01	&	$+$0.02	&	$+$0.12	&	$+$0.12	&	$+$0.12	\\
18174941$-$3353025	&	$+$0.00	&	$+$0.00	&	$+$0.01	&	$+$0.13	&	$+$0.13	&	$+$0.14	\\
18174742$-$3348098	&	$+$0.00	&	$+$0.02	&	$+$0.04	&	$+$0.14	&	$+$0.14	&	$+$0.14	\\
18183679$-$3251454	&	\nodata	&	$+$0.01	&	\nodata	&	$+$0.12	&	$+$0.12	&	$+$0.12	\\
18181929$-$3404128	&	$+$0.00	&	$+$0.01	&	$+$0.04	&	$+$0.13	&	$+$0.13	&	$+$0.13	\\
18181512$-$3353545	&	$+$0.00	&	\nodata	&	$+$0.01	&	\nodata	&	\nodata	&	\nodata	\\
18183802$-$3355441	&	$-$0.01	&	$-$0.01	&	$+$0.02	&	$+$0.12	&	$+$0.12	&	$+$0.12	\\
18174303$-$3355118	&	$-$0.01	&	$+$0.00	&	$+$0.03	&	$+$0.12	&	$+$0.12	&	$+$0.12	\\
18182470$-$3342166	&	$+$0.00	&	$+$0.00	&	$+$0.03	&	$+$0.12	&	$+$0.12	&	$+$0.13	\\
18180285$-$3342004	&	$+$0.00	&	$+$0.00	&	$+$0.03	&	$+$0.12	&	$+$0.12	&	$+$0.12	\\
18174935$-$3404217	&	$-$0.01	&	$-$0.01	&	$+$0.02	&	$+$0.12	&	$+$0.12	&	$+$0.12	\\
18174929$-$3347192	&	$+$0.00	&	$+$0.00	&	$+$0.00	&	$+$0.14	&	$+$0.13	&	$+$0.14	\\
18183604$-$3342349	&	$-$0.01	&	$+$0.00	&	$+$0.03	&	$+$0.12	&	$+$0.12	&	$+$0.12	\\
18180562$-$3346548	&	$-$0.01	&	$+$0.01	&	$+$0.03	&	$+$0.12	&	$+$0.12	&	$+$0.12	\\
18173554$-$3405009	&	$+$0.00	&	$+$0.00	&	$+$0.01	&	$+$0.13	&	$+$0.13	&	$+$0.13	\\
18173180$-$3349197	&	$+$0.00	&	$+$0.01	&	$+$0.05	&	$+$0.12	&	$+$0.12	&	$+$0.11	\\
18181924$-$3350222	&	$+$0.00	&	$+$0.01	&	$+$0.03	&	$+$0.13	&	$+$0.13	&	$+$0.13	\\
18175593$-$3400000	&	$-$0.01	&	$-$0.01	&	$+$0.04	&	$+$0.11	&	$+$0.12	&	$+$0.11	\\
18182089$-$3348425	&	$-$0.01	&	$+$0.00	&	$+$0.01	&	$+$0.13	&	$+$0.13	&	$+$0.12	\\
18175546$-$3404103	&	$+$0.00	&	$+$0.00	&	$+$0.00	&	$+$0.14	&	$+$0.14	&	$+$0.14	\\
18174798$-$3359361	&	$+$0.00	&	$+$0.00	&	$+$0.01	&	$+$0.13	&	$+$0.13	&	$+$0.13	\\
18180979$-$3351416	&	$-$0.01	&	$+$0.00	&	$+$0.02	&	$+$0.12	&	$+$0.12	&	$+$0.12	\\
18182430$-$3352453	&	$+$0.01	&	$+$0.03	&	$+$0.03	&	$+$0.13	&	$+$0.13	&	$+$0.14	\\
18173251$-$3354539	&	$+$0.00	&	$+$0.00	&	$+$0.01	&	$+$0.13	&	$+$0.13	&	$+$0.14	\\
18181710$-$3401088	&	$-$0.01	&	$+$0.00	&	$+$0.00	&	$+$0.14	&	$+$0.14	&	$+$0.14	\\
18182553$-$3349465	&	$+$0.00	&	$+$0.01	&	$-$0.01	&	$+$0.14	&	$+$0.14	&	$+$0.15	\\
18180991$-$3403206	&	$-$0.01	&	$+$0.00	&	$+$0.02	&	$+$0.12	&	$+$0.13	&	$+$0.13	\\
18180301$-$3405313	&	$+$0.04	&	$+$0.04	&	$+$0.04	&	$+$0.14	&	$+$0.14	&	$+$0.15	\\
18180502$-$3355071	&	$+$0.00	&	$+$0.00	&	$+$0.00	&	$+$0.13	&	$+$0.13	&	$+$0.14	\\
18182740$-$3356447	&	$+$0.00	&	$+$0.00	&	$+$0.00	&	\nodata	&	\nodata	&	$+$0.13	\\
18182457$-$3344533	&	$+$0.00	&	$+$0.00	&	$+$0.01	&	$+$0.13	&	$+$0.13	&	$+$0.13	\\
18182612$-$3353431	&	$+$0.00	&	$+$0.00	&	$+$0.00	&	$+$0.14	&	$+$0.14	&	$+$0.14	\\
18172979$-$3401118	&	$+$0.00	&	$+$0.00	&	$+$0.00	&	$+$0.14	&	$+$0.14	&	$+$0.14	\\
18183930$-$3353425	&	$+$0.01	&	$+$0.02	&	$+$0.03	&	$+$0.13	&	$+$0.13	&	$+$0.14	\\
18174891$-$3406031	&	$+$0.00	&	$+$0.00	&	$+$0.00	&	$+$0.14	&	$+$0.13	&	$+$0.14	\\
18181322$-$3402227	&	$+$0.00	&	$+$0.00	&	$+$0.00	&	$+$0.14	&	$+$0.13	&	$+$0.14	\\
18181033$-$3352390	&	$+$0.00	&	$+$0.00	&	$-$0.02	&	$+$0.14	&	$+$0.14	&	$+$0.15	\\
18174000$-$3406266	&	$-$0.01	&	$+$0.00	&	$+$0.01	&	$+$0.13	&	$+$0.12	&	$+$0.13	\\
18180012$-$3358096	&	$+$0.04	&	$+$0.05	&	$+$0.05	&	$+$0.14	&	$+$0.14	&	$+$0.15	\\
18175652$-$3347050	&	$+$0.00	&	$+$0.02	&	$+$0.04	&	$+$0.12	&	$+$0.12	&	$+$0.12	\\
18182472$-$3352044	&	$+$0.02	&	$+$0.04	&	$+$0.05	&	$+$0.14	&	$+$0.13	&	$+$0.14	\\
18173706$-$3405569	&	$-$0.02	&	$+$0.00	&	$+$0.03	&	$+$0.12	&	$+$0.12	&	$+$0.11	\\
18173994$-$3358331	&	$+$0.02	&	$+$0.04	&	$+$0.05	&	$+$0.14	&	$+$0.14	&	$+$0.14	\\
18182073$-$3353250	&	$+$0.00	&	$+$0.00	&	$+$0.02	&	$+$0.13	&	$+$0.12	&	$+$0.13	\\
18174478$-$3343290	&	$+$0.00	&	$+$0.01	&	$+$0.00	&	$+$0.14	&	$+$0.13	&	$+$0.14	\\
18183369$-$3352038	&	$-$0.01	&	$+$0.01	&	$+$0.01	&	$+$0.13	&	$+$0.13	&	$+$0.13	\\
18183098$-$3358070	&	$-$0.01	&	$+$0.01	&	$+$0.02	&	$+$0.13	&	$+$0.13	&	$+$0.13	\\
18174067$-$3356000	&	$-$0.01	&	$+$0.01	&	$+$0.02	&	$+$0.13	&	$+$0.13	&	$+$0.13	\\
18173118$-$3358318	&	$+$0.02	&	$+$0.04	&	$+$0.03	&	\nodata	&	\nodata	&	\nodata	\\
18182052$-$3345251	&	$-$0.01	&	$+$0.01	&	$+$0.03	&	$+$0.12	&	$+$0.12	&	$+$0.13	\\
\hline
\multicolumn{7}{c}{$\Delta$[M/H]$+$0.3 dex}	\\
\hline
18174532$-$3353235	&	$-$0.02	&	$-$0.02	&	$-$0.04	&	$+$0.11	&	$+$0.11	&	$+$0.11	\\
18182918$-$3341405	&	$-$0.01	&	$-$0.01	&	$+$0.01	&	$+$0.11	&	$+$0.11	&	$+$0.10	\\
18175567$-$3343063	&	$-$0.01	&	$-$0.01	&	$-$0.02	&	$+$0.12	&	$+$0.12	&	$+$0.11	\\
18181521$-$3352294	&	$-$0.01	&	$-$0.01	&	$-$0.01	&	$+$0.12	&	$+$0.12	&	$+$0.12	\\
18182256$-$3401248	&	$-$0.02	&	$-$0.01	&	$-$0.03	&	$+$0.11	&	$+$0.12	&	$+$0.11	\\
18174351$-$3401412	&	$-$0.01	&	$-$0.01	&	$-$0.02	&	$+$0.12	&	$+$0.12	&	$+$0.11	\\
18172965$-$3402573	&	$+$0.00	&	$+$0.00	&	$+$0.01	&	$+$0.11	&	$+$0.12	&	$+$0.11	\\
18183521$-$3344124	&	$+$0.00	&	$+$0.00	&	$+$0.03	&	$+$0.11	&	$+$0.11	&	$+$0.10	\\
18181435$-$3350275	&	$+$0.00	&	$-$0.01	&	$+$0.00	&	$+$0.11	&	$+$0.12	&	$+$0.10	\\
18183876$-$3403092	&	$-$0.01	&	$+$0.00	&	$+$0.05	&	$+$0.11	&	$+$0.12	&	$+$0.10	\\
18174304$-$3357006	&	$+$0.00	&	$-$0.01	&	$+$0.03	&	$+$0.11	&	$+$0.11	&	$+$0.10	\\
18180831$-$3405309	&	$+$0.00	&	$+$0.00	&	$+$0.04	&	$+$0.12	&	$+$0.12	&	$+$0.11	\\
18180550$-$3407117	&	$-$0.01	&	$+$0.00	&	$+$0.00	&	$+$0.12	&	$+$0.12	&	$+$0.12	\\
18174941$-$3353025	&	$+$0.00	&	$+$0.00	&	$+$0.01	&	$+$0.12	&	$+$0.12	&	$+$0.12	\\
18174742$-$3348098	&	$+$0.01	&	$+$0.02	&	$+$0.06	&	$+$0.12	&	$+$0.13	&	$+$0.12	\\
18183679$-$3251454	&	\nodata	&	$+$0.02	&	\nodata	&	$+$0.13	&	$+$0.13	&	$+$0.12	\\
18181929$-$3404128	&	$+$0.01	&	$+$0.02	&	$+$0.07	&	$+$0.12	&	$+$0.13	&	$+$0.12	\\
18181512$-$3353545	&	$+$0.00	&	\nodata	&	$+$0.02	&	\nodata	&	\nodata	&	\nodata	\\
18183802$-$3355441	&	$+$0.00	&	$-$0.01	&	$+$0.01	&	$+$0.12	&	$+$0.12	&	$+$0.11	\\
18174303$-$3355118	&	$+$0.00	&	$+$0.00	&	$+$0.03	&	$+$0.12	&	$+$0.12	&	$+$0.11	\\
18182470$-$3342166	&	$+$0.00	&	$+$0.00	&	$+$0.04	&	$+$0.12	&	$+$0.12	&	$+$0.11	\\
18180285$-$3342004	&	$+$0.00	&	$+$0.00	&	$+$0.05	&	$+$0.12	&	$+$0.12	&	$+$0.11	\\
18174935$-$3404217	&	$-$0.02	&	$-$0.02	&	$+$0.01	&	$+$0.11	&	$+$0.12	&	$+$0.11	\\
18174929$-$3347192	&	$+$0.01	&	$+$0.00	&	$+$0.01	&	$+$0.12	&	$+$0.12	&	$+$0.11	\\
18183604$-$3342349	&	$-$0.02	&	$-$0.02	&	$+$0.05	&	$+$0.11	&	$+$0.12	&	$+$0.09	\\
18180562$-$3346548	&	$-$0.03	&	$-$0.03	&	$+$0.04	&	$+$0.11	&	$+$0.12	&	$+$0.10	\\
18173554$-$3405009	&	$+$0.00	&	$-$0.01	&	$+$0.02	&	$+$0.12	&	$+$0.12	&	$+$0.11	\\
18173180$-$3349197	&	$+$0.00	&	$+$0.00	&	$+$0.08	&	$+$0.11	&	$+$0.12	&	$+$0.09	\\
18181924$-$3350222	&	$-$0.02	&	$-$0.02	&	$+$0.04	&	$+$0.11	&	$+$0.12	&	$+$0.09	\\
18175593$-$3400000	&	$-$0.01	&	$-$0.01	&	$+$0.05	&	$+$0.11	&	$+$0.12	&	$+$0.09	\\
18182089$-$3348425	&	$+$0.01	&	$+$0.00	&	$+$0.02	&	$+$0.12	&	$+$0.13	&	$+$0.11	\\
18175546$-$3404103	&	$+$0.02	&	$+$0.00	&	$+$0.03	&	$+$0.12	&	$+$0.13	&	$+$0.11	\\
18174798$-$3359361	&	$+$0.01	&	$+$0.00	&	$+$0.02	&	$+$0.12	&	$+$0.13	&	$+$0.11	\\
18180979$-$3351416	&	$-$0.01	&	$-$0.01	&	$+$0.03	&	$+$0.12	&	$+$0.12	&	$+$0.10	\\
18182430$-$3352453	&	$-$0.01	&	$-$0.02	&	$+$0.05	&	$+$0.11	&	$+$0.12	&	$+$0.10	\\
18173251$-$3354539	&	$-$0.01	&	$-$0.02	&	$+$0.02	&	$+$0.12	&	$+$0.13	&	$+$0.11	\\
18181710$-$3401088	&	$+$0.00	&	$+$0.00	&	$+$0.02	&	$+$0.12	&	$+$0.13	&	$+$0.11	\\
18182553$-$3349465	&	$+$0.00	&	$+$0.00	&	$+$0.02	&	$+$0.12	&	$+$0.13	&	$+$0.11	\\
18180991$-$3403206	&	$-$0.01	&	$-$0.01	&	$+$0.04	&	$+$0.12	&	$+$0.12	&	$+$0.11	\\
18180301$-$3405313	&	$+$0.00	&	$-$0.01	&	$+$0.08	&	$+$0.12	&	$+$0.13	&	$+$0.10	\\
18180502$-$3355071	&	$+$0.00	&	$+$0.00	&	$+$0.02	&	$+$0.12	&	$+$0.13	&	$+$0.11	\\
18182740$-$3356447	&	$+$0.01	&	$+$0.00	&	$+$0.02	&	\nodata	&	\nodata	&	$+$0.11	\\
18182457$-$3344533	&	$+$0.00	&	$-$0.01	&	$+$0.02	&	$+$0.12	&	$+$0.13	&	$+$0.11	\\
18182612$-$3353431	&	$-$0.01	&	$+$0.00	&	$+$0.02	&	$+$0.13	&	$+$0.13	&	$+$0.12	\\
18172979$-$3401118	&	$+$0.01	&	$+$0.00	&	$+$0.01	&	$+$0.13	&	$+$0.13	&	$+$0.12	\\
18183930$-$3353425	&	$+$0.00	&	$+$0.01	&	$+$0.06	&	$+$0.13	&	$+$0.13	&	$+$0.12	\\
18174891$-$3406031	&	$+$0.01	&	$+$0.01	&	$+$0.02	&	$+$0.13	&	$+$0.13	&	$+$0.12	\\
18181322$-$3402227	&	$+$0.00	&	$+$0.00	&	$+$0.01	&	$+$0.13	&	$+$0.13	&	$+$0.12	\\
18181033$-$3352390	&	$+$0.00	&	$+$0.00	&	$+$0.02	&	$+$0.13	&	$+$0.13	&	$+$0.12	\\
18174000$-$3406266	&	$-$0.01	&	$+$0.01	&	$+$0.04	&	$+$0.13	&	$+$0.13	&	$+$0.12	\\
18180012$-$3358096	&	$+$0.02	&	$+$0.02	&	$+$0.07	&	$+$0.11	&	$+$0.12	&	$+$0.10	\\
18175652$-$3347050	&	$+$0.01	&	$+$0.01	&	$+$0.06	&	$+$0.10	&	$+$0.11	&	$+$0.09	\\
18182472$-$3352044	&	$+$0.02	&	$+$0.02	&	$+$0.07	&	$+$0.10	&	$+$0.10	&	$+$0.09	\\
18173706$-$3405569	&	$+$0.01	&	$+$0.01	&	$+$0.04	&	$+$0.09	&	$+$0.09	&	$+$0.08	\\
18173994$-$3358331	&	$+$0.02	&	$+$0.01	&	$+$0.06	&	$+$0.08	&	$+$0.09	&	$+$0.07	\\
18182073$-$3353250	&	$+$0.01	&	$+$0.00	&	$+$0.03	&	$+$0.08	&	$+$0.08	&	$+$0.08	\\
18174478$-$3343290	&	$+$0.00	&	$+$0.00	&	$+$0.01	&	$+$0.08	&	$+$0.08	&	$+$0.08	\\
18183369$-$3352038	&	$+$0.00	&	$+$0.00	&	$+$0.02	&	$+$0.08	&	$+$0.08	&	$+$0.07	\\
18183098$-$3358070	&	$+$0.00	&	$+$0.00	&	$+$0.02	&	$+$0.07	&	$+$0.07	&	$+$0.06	\\
18174067$-$3356000	&	$+$0.00	&	$+$0.00	&	$+$0.02	&	$+$0.05	&	$+$0.05	&	$+$0.05	\\
18173118$-$3358318	&	$-$0.01	&	$+$0.00	&	$+$0.00	&	\nodata	&	\nodata	&	\nodata	\\
18182052$-$3345251	&	$+$0.00	&	$+$0.00	&	$+$0.00	&	$-$0.01	&	$-$0.01	&	$-$0.01	\\
\hline
\multicolumn{7}{c}{$\Delta$V$_{\rm t}$$+$0.3 km s$^{\rm -1}$}	\\
\hline
18174532$-$3353235	&	$+$0.00	&	$+$0.00	&	$+$0.00	&	$-$0.03	&	$-$0.02	&	$-$0.04	\\
18182918$-$3341405	&	$-$0.01	&	$-$0.02	&	$-$0.02	&	$-$0.04	&	$-$0.02	&	$-$0.02	\\
18175567$-$3343063	&	$+$0.00	&	$-$0.02	&	$-$0.01	&	$-$0.03	&	$-$0.02	&	$-$0.02	\\
18181521$-$3352294	&	$-$0.02	&	$-$0.05	&	$-$0.08	&	$-$0.04	&	$-$0.03	&	$-$0.02	\\
18182256$-$3401248	&	$-$0.01	&	$-$0.01	&	$-$0.01	&	$-$0.05	&	$-$0.03	&	$-$0.03	\\
18174351$-$3401412	&	$+$0.00	&	$-$0.01	&	$-$0.01	&	$-$0.02	&	$-$0.02	&	$-$0.03	\\
18172965$-$3402573	&	$-$0.02	&	$-$0.04	&	$-$0.08	&	$-$0.03	&	$-$0.02	&	$-$0.01	\\
18183521$-$3344124	&	$-$0.01	&	$-$0.02	&	$-$0.04	&	$-$0.02	&	$-$0.01	&	$-$0.02	\\
18181435$-$3350275	&	$-$0.01	&	$-$0.03	&	$-$0.04	&	$-$0.03	&	$-$0.02	&	$-$0.03	\\
18183876$-$3403092	&	$-$0.04	&	$-$0.05	&	$-$0.13	&	$-$0.03	&	$-$0.01	&	$-$0.03	\\
18174304$-$3357006	&	$-$0.03	&	$-$0.06	&	$-$0.15	&	$-$0.04	&	$-$0.04	&	$-$0.03	\\
18180831$-$3405309	&	$-$0.06	&	$-$0.07	&	$-$0.05	&	$-$0.02	&	$-$0.02	&	$-$0.03	\\
18180550$-$3407117	&	$-$0.04	&	$-$0.05	&	$-$0.03	&	$-$0.02	&	$-$0.02	&	$-$0.02	\\
18174941$-$3353025	&	$-$0.03	&	$-$0.05	&	$-$0.03	&	$-$0.02	&	$-$0.01	&	$-$0.02	\\
18174742$-$3348098	&	$-$0.04	&	$-$0.05	&	$-$0.31	&	$-$0.25	&	$-$0.18	&	$-$0.05	\\
18183679$-$3251454	&	\nodata	&	$-$0.05	&	\nodata	&	$-$0.02	&	$-$0.02	&	$-$0.02	\\
18181929$-$3404128	&	$-$0.08	&	$-$0.09	&	$-$0.22	&	$-$0.04	&	$-$0.05	&	$-$0.05	\\
18181512$-$3353545	&	$-$0.03	&	\nodata	&	$-$0.05	&	\nodata	&	\nodata	&	\nodata	\\
18183802$-$3355441	&	$-$0.04	&	$-$0.06	&	$-$0.03	&	$-$0.02	&	$-$0.01	&	$-$0.03	\\
18174303$-$3355118	&	$-$0.05	&	$-$0.08	&	$-$0.09	&	$-$0.03	&	$-$0.02	&	$-$0.05	\\
18182470$-$3342166	&	$-$0.09	&	$-$0.10	&	$-$0.08	&	$-$0.03	&	$-$0.02	&	$-$0.04	\\
18180285$-$3342004	&	$-$0.05	&	$-$0.08	&	$-$0.09	&	$-$0.02	&	$-$0.02	&	$-$0.05	\\
18174935$-$3404217	&	$-$0.09	&	$-$0.09	&	$-$0.06	&	$-$0.02	&	$-$0.02	&	$-$0.02	\\
18174929$-$3347192	&	$-$0.02	&	$-$0.03	&	$-$0.01	&	$-$0.02	&	$-$0.01	&	$-$0.02	\\
18183604$-$3342349	&	$-$0.11	&	$-$0.09	&	$-$0.09	&	$-$0.02	&	$-$0.02	&	$-$0.04	\\
18180562$-$3346548	&	$-$0.12	&	$-$0.11	&	$-$0.09	&	$-$0.02	&	$-$0.01	&	$-$0.03	\\
18173554$-$3405009	&	$-$0.04	&	$-$0.05	&	$-$0.01	&	$-$0.01	&	$-$0.01	&	$-$0.01	\\
18173180$-$3349197	&	$-$0.10	&	$-$0.10	&	$-$0.21	&	$-$0.03	&	$-$0.05	&	$-$0.06	\\
18181924$-$3350222	&	$-$0.10	&	$-$0.10	&	$-$0.08	&	$-$0.02	&	$-$0.01	&	$-$0.04	\\
18175593$-$3400000	&	$-$0.09	&	$-$0.09	&	$-$0.13	&	$-$0.05	&	$-$0.05	&	$-$0.04	\\
18182089$-$3348425	&	$-$0.03	&	$-$0.04	&	$-$0.01	&	$-$0.01	&	$-$0.01	&	$-$0.02	\\
18175546$-$3404103	&	$-$0.01	&	$-$0.04	&	$+$0.00	&	$-$0.01	&	$-$0.01	&	$-$0.02	\\
18174798$-$3359361	&	$-$0.03	&	$-$0.04	&	$-$0.01	&	$-$0.02	&	$-$0.01	&	$-$0.02	\\
18180979$-$3351416	&	$-$0.06	&	$-$0.06	&	$-$0.03	&	$-$0.02	&	$-$0.01	&	$-$0.03	\\
18182430$-$3352453	&	$-$0.11	&	$-$0.11	&	$-$0.16	&	$-$0.04	&	$-$0.04	&	$-$0.04	\\
18173251$-$3354539	&	$-$0.09	&	$-$0.08	&	$-$0.02	&	$-$0.01	&	$-$0.01	&	$-$0.02	\\
18181710$-$3401088	&	$-$0.06	&	$-$0.05	&	$-$0.01	&	$-$0.03	&	$-$0.01	&	$-$0.03	\\
18182553$-$3349465	&	$-$0.05	&	$-$0.05	&	$-$0.01	&	$-$0.02	&	$-$0.01	&	$-$0.01	\\
18180991$-$3403206	&	$-$0.09	&	$-$0.09	&	$-$0.05	&	$-$0.03	&	$-$0.02	&	$-$0.03	\\
18180301$-$3405313	&	$-$0.13	&	$-$0.10	&	$-$0.15	&	$-$0.02	&	$-$0.02	&	$-$0.02	\\
18180502$-$3355071	&	$-$0.05	&	$-$0.06	&	$-$0.01	&	$-$0.03	&	$-$0.01	&	$-$0.03	\\
18182740$-$3356447	&	$-$0.02	&	$-$0.05	&	$-$0.01	&	\nodata	&	\nodata	&	$-$0.03	\\
18182457$-$3344533	&	$-$0.08	&	$-$0.07	&	$-$0.03	&	$-$0.01	&	$+$0.00	&	$-$0.02	\\
18182612$-$3353431	&	$-$0.07	&	$-$0.06	&	$-$0.01	&	$-$0.03	&	$-$0.01	&	$+$0.00	\\
18172979$-$3401118	&	$-$0.05	&	$-$0.06	&	$-$0.01	&	$-$0.01	&	$-$0.01	&	$-$0.02	\\
18183930$-$3353425	&	$-$0.14	&	$-$0.11	&	$-$0.08	&	$-$0.02	&	$-$0.01	&	$-$0.03	\\
18174891$-$3406031	&	$-$0.05	&	$-$0.05	&	$-$0.01	&	$-$0.02	&	$-$0.01	&	$-$0.02	\\
18181322$-$3402227	&	$-$0.07	&	$-$0.07	&	$-$0.03	&	$-$0.02	&	$-$0.02	&	$-$0.02	\\
18181033$-$3352390	&	$-$0.06	&	$-$0.05	&	$-$0.02	&	$-$0.03	&	$-$0.02	&	$-$0.02	\\
18174000$-$3406266	&	$-$0.11	&	$-$0.09	&	$-$0.04	&	$-$0.02	&	$-$0.01	&	$-$0.02	\\
18180012$-$3358096	&	$-$0.16	&	$-$0.11	&	$-$0.15	&	$-$0.02	&	$-$0.02	&	$-$0.02	\\
18175652$-$3347050	&	$-$0.11	&	$-$0.11	&	$-$0.15	&	$-$0.04	&	$-$0.03	&	$-$0.02	\\
18182472$-$3352044	&	$-$0.18	&	$-$0.11	&	$-$0.13	&	$-$0.02	&	$-$0.01	&	$-$0.02	\\
18173706$-$3405569	&	$-$0.16	&	$-$0.12	&	$-$0.08	&	$-$0.02	&	$-$0.01	&	$-$0.02	\\
18173994$-$3358331	&	$-$0.19	&	$-$0.14	&	$-$0.19	&	$-$0.02	&	$-$0.02	&	$-$0.03	\\
18182073$-$3353250	&	$-$0.08	&	$-$0.09	&	$-$0.05	&	$-$0.02	&	$-$0.02	&	$-$0.01	\\
18174478$-$3343290	&	$-$0.12	&	$-$0.08	&	$-$0.03	&	$-$0.02	&	$+$0.00	&	$-$0.01	\\
18183369$-$3352038	&	$-$0.07	&	$-$0.08	&	$-$0.03	&	$-$0.03	&	$-$0.01	&	$-$0.02	\\
18183098$-$3358070	&	$-$0.14	&	$-$0.10	&	$-$0.06	&	$-$0.02	&	$-$0.01	&	$-$0.01	\\
18174067$-$3356000	&	$-$0.15	&	$-$0.11	&	$-$0.08	&	$-$0.02	&	$-$0.01	&	$-$0.02	\\
18173118$-$3358318	&	$-$0.17	&	$-$0.13	&	$-$0.07	&	\nodata	&	\nodata	&	\nodata	\\
18182052$-$3345251	&	$-$0.13	&	$-$0.10	&	$-$0.07	&	$-$0.01	&	$-$0.01	&	$-$0.02	\\
\hline
\multicolumn{7}{c}{Line--to--Line Dispersion\tablenotemark{a}}	\\
\hline
18174532$-$3353235	&	0.10	&	0.10	&	0.10	&	0.07	&	0.07	&	0.07	\\
18182918$-$3341405	&	0.02	&	0.04	&	0.07	&	0.07	&	0.07	&	0.07	\\
18175567$-$3343063	&	0.04	&	0.10	&	0.10	&	0.07	&	0.07	&	0.07	\\
18181521$-$3352294	&	0.02	&	0.04	&	0.06	&	0.07	&	0.07	&	0.07	\\
18182256$-$3401248	&	0.00	&	0.07	&	0.14	&	0.07	&	0.07	&	0.07	\\
18174351$-$3401412	&	0.04	&	0.10	&	0.00	&	0.07	&	0.07	&	0.07	\\
18172965$-$3402573	&	0.01	&	0.11	&	0.00	&	0.07	&	0.07	&	0.07	\\
18183521$-$3344124	&	0.10	&	0.04	&	0.12	&	0.07	&	0.07	&	0.07	\\
18181435$-$3350275	&	0.00	&	0.10	&	0.09	&	0.07	&	0.07	&	0.07	\\
18183876$-$3403092	&	0.04	&	0.09	&	0.10	&	0.07	&	0.07	&	0.07	\\
18174304$-$3357006	&	0.02	&	0.10	&	0.09	&	0.07	&	0.07	&	0.07	\\
18180831$-$3405309	&	0.05	&	0.04	&	0.04	&	0.07	&	0.07	&	0.07	\\
18180550$-$3407117	&	0.10	&	0.11	&	0.04	&	0.07	&	0.07	&	0.07	\\
18174941$-$3353025	&	0.06	&	0.14	&	0.02	&	0.07	&	0.07	&	0.07	\\
18174742$-$3348098	&	0.10	&	0.07	&	0.28	&	0.07	&	0.07	&	0.07	\\
18183679$-$3251454	&	\nodata	&	0.08	&	\nodata	&	0.07	&	0.07	&	0.07	\\
18181929$-$3404128	&	0.03	&	0.04	&	0.04	&	0.07	&	0.07	&	0.07	\\
18181512$-$3353545	&	0.14	&	\nodata	&	0.02	&	\nodata	&	\nodata	&	\nodata	\\
18183802$-$3355441	&	0.07	&	0.07	&	0.08	&	0.07	&	0.07	&	0.07	\\
18174303$-$3355118	&	0.05	&	0.04	&	0.04	&	0.07	&	0.07	&	0.07	\\
18182470$-$3342166	&	0.07	&	0.07	&	0.01	&	0.07	&	0.07	&	0.07	\\
18180285$-$3342004	&	0.01	&	0.11	&	0.07	&	0.07	&	0.07	&	0.07	\\
18174935$-$3404217	&	0.07	&	0.05	&	0.09	&	0.07	&	0.07	&	0.07	\\
18174929$-$3347192	&	0.07	&	0.13	&	0.11	&	0.07	&	0.07	&	0.07	\\
18183604$-$3342349	&	0.10	&	0.04	&	0.10	&	0.07	&	0.07	&	0.07	\\
18180562$-$3346548	&	0.04	&	0.07	&	0.07	&	0.07	&	0.07	&	0.07	\\
18173554$-$3405009	&	0.03	&	0.04	&	0.09	&	0.07	&	0.07	&	0.07	\\
18173180$-$3349197	&	0.10	&	0.09	&	0.18	&	0.07	&	0.07	&	0.07	\\
18181924$-$3350222	&	0.18	&	0.07	&	0.04	&	0.07	&	0.07	&	0.07	\\
18175593$-$3400000	&	0.11	&	0.04	&	0.04	&	0.07	&	0.07	&	0.07	\\
18182089$-$3348425	&	0.09	&	0.10	&	0.07	&	0.07	&	0.07	&	0.07	\\
18175546$-$3404103	&	0.02	&	0.07	&	0.04	&	0.07	&	0.07	&	0.07	\\
18174798$-$3359361	&	0.05	&	0.06	&	0.21	&	0.07	&	0.07	&	0.07	\\
18180979$-$3351416	&	0.00	&	0.09	&	0.02	&	0.07	&	0.07	&	0.07	\\
18182430$-$3352453	&	0.06	&	0.07	&	0.11	&	0.07	&	0.07	&	0.07	\\
18173251$-$3354539	&	0.02	&	0.12	&	0.11	&	0.07	&	0.07	&	0.07	\\
18181710$-$3401088	&	0.08	&	0.05	&	0.10	&	0.07	&	0.07	&	0.07	\\
18182553$-$3349465	&	0.04	&	0.14	&	0.00	&	0.07	&	0.07	&	0.07	\\
18180991$-$3403206	&	0.07	&	0.04	&	0.07	&	0.07	&	0.07	&	0.07	\\
18180301$-$3405313	&	0.00	&	0.07	&	0.02	&	0.07	&	0.07	&	0.07	\\
18180502$-$3355071	&	0.08	&	0.19	&	0.16	&	0.07	&	0.07	&	0.07	\\
18182740$-$3356447	&	0.10	&	0.10	&	0.10	&	\nodata	&	\nodata	&	0.07	\\
18182457$-$3344533	&	0.04	&	0.01	&	0.04	&	0.07	&	0.07	&	0.07	\\
18182612$-$3353431	&	0.09	&	0.05	&	0.11	&	0.07	&	0.07	&	0.07	\\
18172979$-$3401118	&	0.00	&	0.11	&	0.01	&	0.07	&	0.07	&	0.07	\\
18183930$-$3353425	&	0.05	&	0.11	&	0.03	&	0.07	&	0.07	&	0.07	\\
18174891$-$3406031	&	0.05	&	0.04	&	0.00	&	0.07	&	0.07	&	0.07	\\
18181322$-$3402227	&	0.03	&	0.16	&	0.00	&	0.07	&	0.07	&	0.07	\\
18181033$-$3352390	&	0.01	&	0.06	&	0.00	&	0.07	&	0.07	&	0.07	\\
18174000$-$3406266	&	0.01	&	0.14	&	0.03	&	0.07	&	0.07	&	0.07	\\
18180012$-$3358096	&	0.01	&	0.06	&	0.04	&	0.07	&	0.07	&	0.07	\\
18175652$-$3347050	&	0.00	&	0.18	&	0.04	&	0.07	&	0.07	&	0.07	\\
18182472$-$3352044	&	0.04	&	0.11	&	0.03	&	0.07	&	0.07	&	0.07	\\
18173706$-$3405569	&	0.06	&	0.21	&	0.00	&	0.07	&	0.07	&	0.07	\\
18173994$-$3358331	&	0.07	&	0.06	&	0.09	&	0.07	&	0.07	&	0.07	\\
18182073$-$3353250	&	0.11	&	0.12	&	0.06	&	0.07	&	0.07	&	0.07	\\
18174478$-$3343290	&	0.04	&	0.18	&	0.05	&	0.07	&	0.07	&	0.07	\\
18183369$-$3352038	&	0.06	&	0.16	&	0.02	&	0.07	&	0.07	&	0.07	\\
18183098$-$3358070	&	0.01	&	0.11	&	0.00	&	0.07	&	0.07	&	0.07	\\
18174067$-$3356000	&	0.08	&	0.13	&	0.00	&	0.07	&	0.07	&	0.07	\\
18173118$-$3358318	&	0.06	&	0.04	&	0.04	&	\nodata	&	\nodata	&	\nodata	\\
18182052$-$3345251	&	0.08	&	0.11	&	0.02	&	0.07	&	0.07	&	0.07	\\
\hline
\multicolumn{7}{c}{Total Uncertainty: ($\Sigma$$\sigma$$^{\rm 2}$)$^{\rm 0.5}$}	\\
\hline
18174532$-$3353235	&	0.13	&	0.13	&	0.23	&	0.19	&	0.19	&	0.19	\\
18182918$-$3341405	&	0.08	&	0.09	&	0.25	&	0.19	&	0.19	&	0.19	\\
18175567$-$3343063	&	0.09	&	0.12	&	0.23	&	0.19	&	0.19	&	0.19	\\
18181521$-$3352294	&	0.10	&	0.11	&	0.26	&	0.20	&	0.20	&	0.21	\\
18182256$-$3401248	&	0.08	&	0.11	&	0.26	&	0.19	&	0.20	&	0.19	\\
18174351$-$3401412	&	0.09	&	0.12	&	0.21	&	0.19	&	0.19	&	0.19	\\
18172965$-$3402573	&	0.08	&	0.13	&	0.24	&	0.19	&	0.20	&	0.18	\\
18183521$-$3344124	&	0.13	&	0.08	&	0.26	&	0.18	&	0.18	&	0.17	\\
18181435$-$3350275	&	0.08	&	0.13	&	0.23	&	0.18	&	0.19	&	0.17	\\
18183876$-$3403092	&	0.10	&	0.13	&	0.28	&	0.19	&	0.19	&	0.18	\\
18174304$-$3357006	&	0.10	&	0.14	&	0.28	&	0.20	&	0.20	&	0.19	\\
18180831$-$3405309	&	0.11	&	0.11	&	0.22	&	0.20	&	0.20	&	0.20	\\
18180550$-$3407117	&	0.13	&	0.14	&	0.22	&	0.18	&	0.19	&	0.18	\\
18174941$-$3353025	&	0.10	&	0.17	&	0.21	&	0.19	&	0.19	&	0.20	\\
18174742$-$3348098	&	0.13	&	0.11	&	0.46	&	0.32	&	0.27	&	0.20	\\
18183679$-$3251454	&	\nodata	&	0.12	&	\nodata	&	0.19	&	0.20	&	0.18	\\
18181929$-$3404128	&	0.11	&	0.12	&	0.30	&	0.20	&	0.21	&	0.20	\\
18181512$-$3353545	&	0.17	&	\nodata	&	0.22	&	\nodata	&	\nodata	&	\nodata	\\
18183802$-$3355441	&	0.11	&	0.12	&	0.22	&	0.18	&	0.19	&	0.18	\\
18174303$-$3355118	&	0.11	&	0.11	&	0.23	&	0.19	&	0.19	&	0.18	\\
18182470$-$3342166	&	0.14	&	0.14	&	0.21	&	0.19	&	0.19	&	0.19	\\
18180285$-$3342004	&	0.09	&	0.15	&	0.23	&	0.19	&	0.19	&	0.18	\\
18174935$-$3404217	&	0.14	&	0.13	&	0.23	&	0.18	&	0.19	&	0.18	\\
18174929$-$3347192	&	0.10	&	0.14	&	0.20	&	0.20	&	0.19	&	0.19	\\
18183604$-$3342349	&	0.17	&	0.12	&	0.24	&	0.18	&	0.19	&	0.17	\\
18180562$-$3346548	&	0.15	&	0.15	&	0.24	&	0.18	&	0.19	&	0.17	\\
18173554$-$3405009	&	0.09	&	0.09	&	0.20	&	0.19	&	0.19	&	0.18	\\
18173180$-$3349197	&	0.16	&	0.15	&	0.34	&	0.18	&	0.19	&	0.17	\\
18181924$-$3350222	&	0.22	&	0.14	&	0.22	&	0.19	&	0.19	&	0.18	\\
18175593$-$3400000	&	0.16	&	0.11	&	0.23	&	0.18	&	0.19	&	0.16	\\
18182089$-$3348425	&	0.12	&	0.13	&	0.19	&	0.19	&	0.20	&	0.18	\\
18175546$-$3404103	&	0.08	&	0.10	&	0.17	&	0.20	&	0.21	&	0.19	\\
18174798$-$3359361	&	0.09	&	0.09	&	0.27	&	0.19	&	0.20	&	0.19	\\
18180979$-$3351416	&	0.09	&	0.13	&	0.20	&	0.18	&	0.19	&	0.17	\\
18182430$-$3352453	&	0.14	&	0.15	&	0.26	&	0.19	&	0.20	&	0.19	\\
18173251$-$3354539	&	0.12	&	0.16	&	0.21	&	0.19	&	0.20	&	0.19	\\
18181710$-$3401088	&	0.13	&	0.09	&	0.20	&	0.20	&	0.20	&	0.19	\\
18182553$-$3349465	&	0.09	&	0.17	&	0.18	&	0.20	&	0.21	&	0.20	\\
18180991$-$3403206	&	0.13	&	0.11	&	0.20	&	0.19	&	0.20	&	0.19	\\
18180301$-$3405313	&	0.15	&	0.14	&	0.24	&	0.20	&	0.21	&	0.19	\\
18180502$-$3355071	&	0.13	&	0.21	&	0.23	&	0.19	&	0.20	&	0.19	\\
18182740$-$3356447	&	0.12	&	0.13	&	0.20	&	\nodata	&	\nodata	&	0.19	\\
18182457$-$3344533	&	0.12	&	0.10	&	0.19	&	0.19	&	0.20	&	0.19	\\
18182612$-$3353431	&	0.14	&	0.10	&	0.19	&	0.21	&	0.21	&	0.20	\\
18172979$-$3401118	&	0.09	&	0.14	&	0.17	&	0.20	&	0.21	&	0.20	\\
18183930$-$3353425	&	0.16	&	0.16	&	0.20	&	0.20	&	0.20	&	0.20	\\
18174891$-$3406031	&	0.10	&	0.09	&	0.17	&	0.20	&	0.20	&	0.20	\\
18181322$-$3402227	&	0.11	&	0.18	&	0.17	&	0.20	&	0.20	&	0.20	\\
18181033$-$3352390	&	0.09	&	0.10	&	0.16	&	0.21	&	0.21	&	0.21	\\
18174000$-$3406266	&	0.13	&	0.18	&	0.18	&	0.20	&	0.19	&	0.19	\\
18180012$-$3358096	&	0.17	&	0.14	&	0.22	&	0.19	&	0.20	&	0.19	\\
18175652$-$3347050	&	0.12	&	0.21	&	0.23	&	0.18	&	0.18	&	0.17	\\
18182472$-$3352044	&	0.19	&	0.16	&	0.21	&	0.19	&	0.18	&	0.18	\\
18173706$-$3405569	&	0.18	&	0.25	&	0.19	&	0.17	&	0.17	&	0.15	\\
18173994$-$3358331	&	0.21	&	0.16	&	0.26	&	0.18	&	0.18	&	0.17	\\
18182073$-$3353250	&	0.16	&	0.16	&	0.19	&	0.17	&	0.16	&	0.17	\\
18174478$-$3343290	&	0.15	&	0.21	&	0.18	&	0.18	&	0.17	&	0.18	\\
18183369$-$3352038	&	0.12	&	0.19	&	0.18	&	0.17	&	0.17	&	0.16	\\
18183098$-$3358070	&	0.15	&	0.16	&	0.18	&	0.16	&	0.17	&	0.16	\\
18174067$-$3356000	&	0.18	&	0.18	&	0.19	&	0.16	&	0.16	&	0.16	\\
18173118$-$3358318	&	0.19	&	0.15	&	0.18	&	\nodata	&	\nodata	&	\nodata	\\
18182052$-$3345251	&	0.17	&	0.15	&	0.19	&	0.14	&	0.14	&	0.15	\\
\enddata

\tablenotetext{a}{A value of 0.07 dex (the average line--to--line dispersion 
for all elements where more than one line was measured) was set for elements
and/or stars where only a single line was available.}

\end{deluxetable}

\end{document}